\documentclass[twocolumn]{emulateapj}

\shorttitle{Spitzer Spectra and Models of Protostars in Taurus}
\shortauthors{Furlan et al.}
\submitted{To appear in ApJS}

\begin{document}

\title{Spitzer IRS Spectra and Envelope Models of Class I Protostars in Taurus}


\author{E. Furlan\altaffilmark{1,2,3},  M. McClure\altaffilmark{4}, 
N. Calvet\altaffilmark{5}, L. Hartmann\altaffilmark{5}, 
P. D'Alessio\altaffilmark{6}, 
W. J. Forrest\altaffilmark{4}, D. M. Watson\altaffilmark{4}, 
K. I. Uchida\altaffilmark{1}, B. Sargent\altaffilmark{4}, 
J. D. Green\altaffilmark{4}, T. L. Herter\altaffilmark{1}
}

\altaffiltext{1}{Center for Radiophysics and Space Research, 208 Space Sciences Building,
Cornell University, Ithaca, NY 14853; furlan@astro.ucla.edu, kuchida@astro.cornell.edu,
tlh10@cornell.edu}
\altaffiltext{2}{Current address: NASA Astrobiology Institute and Department of Physics 
and Astronomy, UCLA, 430 Portola Plaza, Los Angeles, CA 90095}
\altaffiltext{3}{NASA Postdoctoral Program Fellow}
\altaffiltext{4}{Department of Physics and Astronomy, University of Rochester, Rochester, 
NY 14627; melisma@astro.pas.rochester.edu, forrest@pas.rochester.edu, 
dmw@pas.rochester.edu, bsargent@pas.rochester.edu, joel@pas.rochester.edu}
\altaffiltext{5}{Department of Astronomy, The University of Michigan, 500 Church St.,
830 Dennison Bldg., Ann Arbor, MI 48109; ncalvet@umich.edu, lhartm@umich.edu}
\altaffiltext{6} {Centro de Radioastronom{\'\i}a y Astrof{\'\i}sica, UNAM, Apartado Postal 
3-72 (Xangari), 58089 Morelia, Michoac\'an, M\'exico; p.dalessio@astrosmo.unam.mx}


\begin{abstract}
We present {\it Spitzer} Infrared Spectrograph spectra of 28 Class I protostars
in the Taurus star-forming region. The 5 to 36 $\mu$m spectra reveal excess 
emission from the inner regions of the envelope and accretion disk surrounding 
these predecessors of low-mass stars, as well as absorption features due to silicates 
and ices. Together with shorter- and longer-wavelength data from the literature, we 
construct spectral energy distributions and fit envelope models to 22 protostars 
of our sample, most of which are well-constrained due to the availability of the IRS 
spectra. We infer that the envelopes of the Class I objects in our sample cover a 
wide range in parameter space, particularly in density and centrifugal radius, implying
different initial conditions for the collapse of protostellar cores.
\end{abstract}


\section{Introduction}

In a now widely accepted evolutionary sequence based on the shape of the 
spectral energy distribution (SED) in the infrared, there are four stages in the 
pre-main-sequence life of a low-mass star:
Class 0 objects are deeply embedded protostars with large submillimeter to 
bolometric luminosity ratios, which are very faint at near-infrared wavelengths; 
Class I objects are surrounded by infalling envelopes, but are less embedded; 
Class II objects have cleared their envelopes and are surrounded by accretion 
disks; and Class III objects have dispersed almost all of their circumstellar material 
\citep{lada84, lada87,andre93}. 

Therefore, Class I protostars are thought to be in one of the earliest evolutionary 
stages. After the gravitational collapse of a molecular cloud core starts, the
infalling material forms an envelope around the central object. Since cloud 
cores are usually rotating slowly, the material falls preferentially onto a disk,
which serves as a mass reservoir, most of which is eventually accreted by 
the protostar. 
Class I objects are thus surrounded by accretion disks, in addition to envelopes
extending to several 1000 AU. The radius of the disk is roughly equal to the
centrifugal radius ($R_c$), which defines the centrifugal barrier encountered by matter 
falling in on the equatorial plane, if the specific angular momentum is conserved; 
inside this radius the density distribution flattens, and the matter follows non-radial 
trajectories, falling onto a disk. Material tends to accumulate close to $R_c$, which
is close to the protostar at the beginning of the collapse, but increases rapidly over
time \citep{terebey84,adams86}. 

Besides accreting matter, young stars generate powerful outflows which are
launched along magnetic field lines; mass accretion onto the star and mass 
loss in the form of outflows seem to be correlated \citep[e.g.,][]{hartigan95}. 
In addition to collimated outflows, protostars might also emit a wide-angle 
wind, which is expected to open up cavities in the envelope \citep{arce02}. 
The presence of both collimated outflows and wide-angle stellar winds 
explains the fact that some objects with large cavities have well-collimated
outflows. A partially evacuated cavity also forms as a natural
by-product of the collapse of a sheet \citep{hartmann94, hartmann96};
since molecular cloud cores are rarely spherically symmetric, this scenario
might be more realistic. In some cases, both stellar winds and the
flattening of the envelope are responsible for cavities oriented along the
rotational axis of a protostar.

The spectral index $n$ ($\lambda F_{\lambda} \propto {\lambda}^n$)
from 2 to 25 $\mu$m is used for the SED classification of protostars and
pre-main-sequence stars \citep{lada87,adams87,andre94}; Class I 
objects have $n > 0$, Class II objects have $ -2 < n < 0 $, and Class III 
objects have $ -3 < n < -2 $ ($n=-3$ is the spectral index of a photosphere 
in the Rayleigh-Jeans limit).
Therefore, Class I objects are characterized by a strong infrared excess; 
the dust in the envelope absorbs radiation from the central protostar and disk 
and reemits it at infrared to mm wavelengths. The peak of the SED of this 
type of objects usually lies in the mid- to far-infrared, depending on the density
of the envelope (and thus mass and infall rate), as well as the source luminosity
\citep{kenyon93a}.
Flat spectrum sources are likely objects surrounded by envelopes and disks
and therefore protostars, but with wide cavities in their envelopes and/or
seen close to pole-on \citep[e.g., DG Tau, T Tau;][]{furlan06a, calvet94}.
Some of the flat-spectrum sources could be transition objects (Class I/II), 
consisting of a T Tauri star and its disk surrounded by remnant envelope 
material (e.g., IRAS 04154+2823). Their SEDs typically show a double peak,
one centered in the near-IR and one in the far-IR.

In addition to a rising SED in the infrared due to this excess emission, Class I 
protostars display various absorption features in their near- and mid-infrared 
spectra due to silicates and ices in the envelope. However, many parts of the 
infrared spectrum are not accessible from ground-based telescopes, which in 
addition often lack the sensitivity to observe faint objects. The {\it Spitzer Space
Telescope} \citep{werner04} opened a new window onto low-mass 
star formation; with unprecedented sensitivity, the precursors of low-mass 
stars can be studied in various evolutionary stages. 

Here we present 28 spectra from 5 to 36 $\mu$m of Class I objects in the
Taurus star-forming region observed with the Infrared Spectrograph\footnote{The 
IRS was a collaborative venture between Cornell University and Ball Aerospace 
Corporation, funded by NASA through the Jet Propulsion Laboratory and the 
Ames Research Center.} \citep[IRS;][]{houck04} on board of {\it Spitzer}. 
The 5--20 $\mu$m spectra of 5 of the 28 Class I objects were already 
introduced in \citet{watson04}; these objects are IRAS 04016+2610, 
04108+2803B, 04181+2654 B, 04239+2436, and DG Tau B.
Our IRS observations are part of a larger Infrared Spectrograph guaranteed-time 
observing program; we presented the Class II and III objects from our sample of 
150 young stellar objects in Taurus in an earlier paper \citep{furlan06a}. 

Since the Taurus-Auriga star-forming region is young (1--2 Myr) and relatively 
nearby \citep[140 pc;][]{kenyon94b, bertout99}, many of its objects have 
been fairly well-studied at wavelengths covering almost the entire electromagnetic 
spectrum. Here we use SEDs constructed with data from the literature and our IRS 
spectra to constrain envelope models and to determine which parameters 
describe the circumstellar environment of the Class I objects in our sample.

This paper is structured as follows: in \S\ \ref{obs_data_reduction} we 
lay out our {\it Spitzer} IRS observations and data reduction; in \S\ 
\ref{sample} we present the IRS spectra and SEDs of the 28 Class I
objects in our sample; in \S\ \ref{env_models} we describe the 
envelope models used to fit 22 of our Class I objects and give
detailed descriptions of the individual model fits; in \S\ \ref{discussion}
we discuss the different appearances of Class I objects and give a summary
of our models; and in \S\ \ref{conclusions} we present our conclusions.

\section{Observations and Data Reduction}
\label{obs_data_reduction}

Our Taurus targets were observed with the IRS on {\it Spitzer} during IRS 
observing campaigns 3 (2004 February 6 to 8), 4 (2004 February 27 to March 5), 
12 (2004 August 30 to 31; only 3 targets), and 24 (2005 September 12; only 
IRAS 04166+2706). All targets were observed over
the full IRS range from 5 to 40 $\mu$m by using either the two low-resolution 
IRS modules (Short-Low [SL] and Long-Low [LL], 5.2--14 $\mu$m and 14--38 
$\mu$m, respectively, $\lambda$/$\Delta\lambda$ $\sim$ 90) or the SL 
module and the two high-resolution modules (Short-High [SH] and Long-High 
[LH], 10--19 $\mu$m and 19--37 $\mu$m, respectively, 
$\lambda$/$\Delta\lambda$ $\sim$ 600), depending on the object's 
expected mid-infrared flux. Thus, roughly half of our Class I objects were observed
with SL and LL, and the other half with SL, SH, and LH. 
Our observations were carried out either in IRS staring mode, where the target 
is placed on two nod positions along the slit, at 1/3 and 2/3 of the slit length, or 
in mapping mode, with $2\times3$ step maps on the target (3 steps in the 
dispersion direction and 2 steps in the spatial direction). More than half of the 
objects (19 out of 28) were observed in mapping mode. 

Data reduction was carried out using the SMART software tool \citep{higdon04}.
Spectra were extracted after bad pixels, including so-called rogue pixels,
were fixed by interpolation and after the sky in the low-resolution spectra was
subtracted using the off-order observation (see \citealt{furlan06a} for more
details). A variable-width column extraction was performed for low-resolution
spectra, while high-resolution spectra were extracted using a full-slit extraction.
The spectra were calibrated by dividing them by a calibrator spectrum
($\alpha$ Lac (A1 V) for low-resolution and $\xi$ Dra (K2 III) for
high-resolution observations), and by multiplying the result by the calibrator's 
template spectrum \citep{cohen03}. Final spectra were obtained by averaging
the two nod positions (for staring mode observations) or the two central map
positions (for mapping mode observations). We also truncated all SH spectra below 
13 $\mu$m and rebinned all high-resolution spectra to a resolution of 
$\lambda$/$\Delta\lambda$ = 200 to achieve a more uniform representation 
of the spectra.

As mentioned in \citet{furlan06a}, we applied scalar corrections to each
LH spectrum to account for the absence of sky subtraction in the high-resolution
modules. Some minor adjustments ($\lesssim$ 10\%) were also necessary for 
SL to match it to the flux level of SH. For IRAS 04154+2823, 04158+2805, 
04325+2402, and 04368+2557, a multiplicative factor of 1.2 had to be applied 
to SL, while DG Tau B is the only Class I object in our sample to require an even 
larger scaling: a multiplication by 1.35 of SL2 (5.2--7.5 $\mu$m) and by 1.21 
of SL1 (7.4--14 $\mu$m).

The flux mismatches between short- and long-wavelength modules can be
attributed to the extended nature of our targets; even though all objects in
our sample appeared as point sources when observed with the IRS, we know
that most of our targets are somewhat extended due to the presence of an 
envelope \citep[e.g.,][]{park02}. There are two effects caused by the
extension of the envelope influencing our IRS observations.

First, the LL and LH slits, which are 10.6\arcsec\ and  11.1\arcsec\ wide, 
respectively, cover a larger area and thus more of the envelope emission than 
the SL and SH slits, which are 3.6\arcsec\ and 4.7\arcsec\ wide, respectively. 
However, the cold, outer envelope regions do not contribute much flux at the 
IRS wavelengths; at a distance of 140 pc, the SL slit encompasses all emission 
within 250 AU from the central source, while LH includes emission out to 780 AU, 
sufficient to sample almost all of the material contributing to the IRS spectrum. 
Only a few objects probably suffer from slit losses beyond about 30 $\mu$m,
evidenced by a change in slope, and thus decrease in flux level, around 30 $\mu$m;
this effect is most notable in IRAS 04108+2803 B, 04158+2805, 04181+2654 A, 
04239+2436, and 04295+2251.

\begin{figure}
\epsscale{0.55}
\plotone{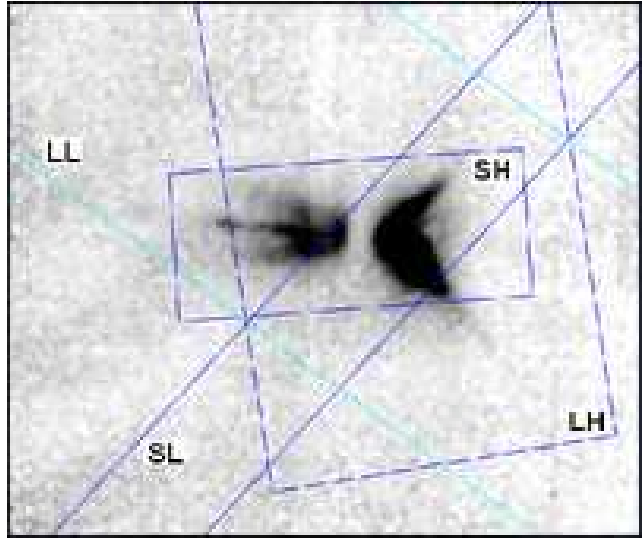}
\caption{The IRS SL, LL, SH, and LH slit positions (only one of the two nod positions 
is shown) superposed on the NICMOS $J$-band (F110W) image of DG Tau B from 
\citet{padgett99}; note that for this object the LL module was not used and is shown 
for illustrative purposes only. The slits, with their different widths and orientations, cover 
different parts of the object.  \label{DGTauB_slits}}
\end{figure}

Second, and more importantly, is the fact that protostellar envelopes are usually 
not spherically symmetric and that each of the IRS modules sample different 
regions of a protostellar system, since the slits of the various modules are not 
aligned parallel to each other and are also of different sizes (see, e.g., Fig.\ 
\ref{DGTauB_slits}). For example, a protostar seen edge-on might show a 
dark central lane representing the disk and symmetric reflection nebulae 
representing inner envelope regions; if SL is aligned along the disk, LL will
be oriented perpendicular to the dark lane and will sample more of the envelope
emission.

The larger scalar correction factors applied to some of Class I objects in our
sample (especially in the case of DG Tau B) are likely due to this second effect. 
In particular, the centering and orientation of the narrow SL slit could result in
some of the envelope emission being missed by the SL observation (see Fig.\
\ref{DGTauB_slits}).
However, in most cases only a small fraction of the warm dust contributing at 
IRS wavelengths is not included in the IRS observations, and therefore the spectra 
we obtained are believed to be a good representation of the mid-infrared emission 
of the inner envelope regions.

Our absolute spectrophotometric accuracy is of the order of 5--10\%, 
depending on the module used and the pointing uncertainty. The relative
spectral accuracy is module-dependent; spectral features above the noise level
are likely real in SL, LL2 (14--21 $\mu$m) and SH, while in LL1 (20--36 $\mu$m) 
and especially in LH artifacts from unresolved calibration issues often do not allow
reliable identifications of spectral features. Also the wavelength region where
SL and LL meet (13.5--14.5 $\mu$m) is affected by this higher uncertainty due
to possible artifacts at the order edges.

Some of the Class I objects we observed with IRS are members of multiple
systems (see Table \ref{tab_properties}). Except for GV Tau A and B, 
they either are at separations that place all members well within the IRS slits, 
or their separations are 
large enough to ensure that only one component is observed. Since the infrared
emission from protostars is dominated by envelope emission, the multiplicity
of the central source should have no impact on the mid-infrared spectrum,
except when the configuration of the system is such that it influences the 
emission from the envelope, e.g., by truncating it or causing a
departure from spherical symmetry in the illuminating radiation field. 

We encountered reduction problems in a few objects; they are listed below. \\
{\it 04108+2803B.}  This object is part of a mispointed mapping-mode 
observation; by using the spectra from the central map position for SL
and SH and scaled spectra from the third map position for LH, we 
created a continuous IRS spectrum. This stitching adds a higher uncertainty 
to the absolute flux level in our spectra, but the overall shape of the spectrum 
should be only minimally affected. \\
{\it GV Tau (A,B).} This binary is only separated by 1{\farcs}3; it was observed 
with SL, SH, and LH in mapping mode, and it was not resolved in any of the modules. 
Since this object consists of two point sources with likely some extended emission 
surrounding them, their SH and LH calibration presented some problems. The final 
spectrum was composed by using those nod and map positions for each module 
in which the flux was highest, and in addition SH was scaled to match up with SL.
Due to this procedure, the absolute flux level of this spectrum carries a higher uncertainty. 
In fact, we multiplied the entire spectrum by 0.85 to match fluxes at 11.7 and 17.9 
$\mu$m with our, yet unpublished, ground-based measurements at these wavelengths. \\
{\it HH 30.} This object was too faint in SL2; the spectrum could only be extracted 
in SL1 and in LL. \\
{\it LkHa 358.} The peak-up arrays contained bright to saturated sources during the 
SL observations, which resulted in increased noise in the SL part of the spectrum. The
LL observation included 3 sources, LkHa 358, HL Tau, and XZ Tau, but since the
observation was centered on LkHa 358, the other two sources entered only partially
in the slit. In addition, all three sources were separated sufficiently in the spatial direction 
to allow extraction of the individual spectrum of LkHa 358.

\section{The Sample}
\label{sample}

\subsection{IRS Spectra}
\label{IRS_spectra}

Our Taurus sample is based on objects analyzed by \citet{kenyon95} (from here on KH95); 
a large fraction of our Class I targets has identifiers from the {\it IRAS} Point Source 
Catalogue, since these objects were first identified as possible embedded protostars 
in the {\it IRAS} surveys and confirmed in follow-up observations \citep[e.g.,][]
{beichman86, myers87, kenyon90}. Some objects that are part of our Class I 
sample are identified as Class II objects in KH95: these objects are IRAS 04154+2823,
04158+2805, 04278+2253, CoKu Tau/1, HL Tau, IC 2087 IR, and LkHa 358.
We base our identification of these sources as Class I objects on the existence of 
extended, near-infrared nebulosity and possibly outflows around these objects, or the 
presence of a silicate absorption feature and a positive spectral index over the IRS range, 
or the detection of ice absorption features in the 5--8 $\mu$m region as well as the 
CO$_2$ ice feature at 15.2 $\mu$m. 
Therefore, this sample includes all Taurus objects we observed with the IRS 
that are thought to be at the protostellar evolutionary stage and therefore 
surrounded, to varying extents, by envelopes, even though in a few cases 
the SED slope does not lie in the typical range for Class I objects.

\begin{figure}
\centering
\includegraphics[angle=90, scale=0.62]{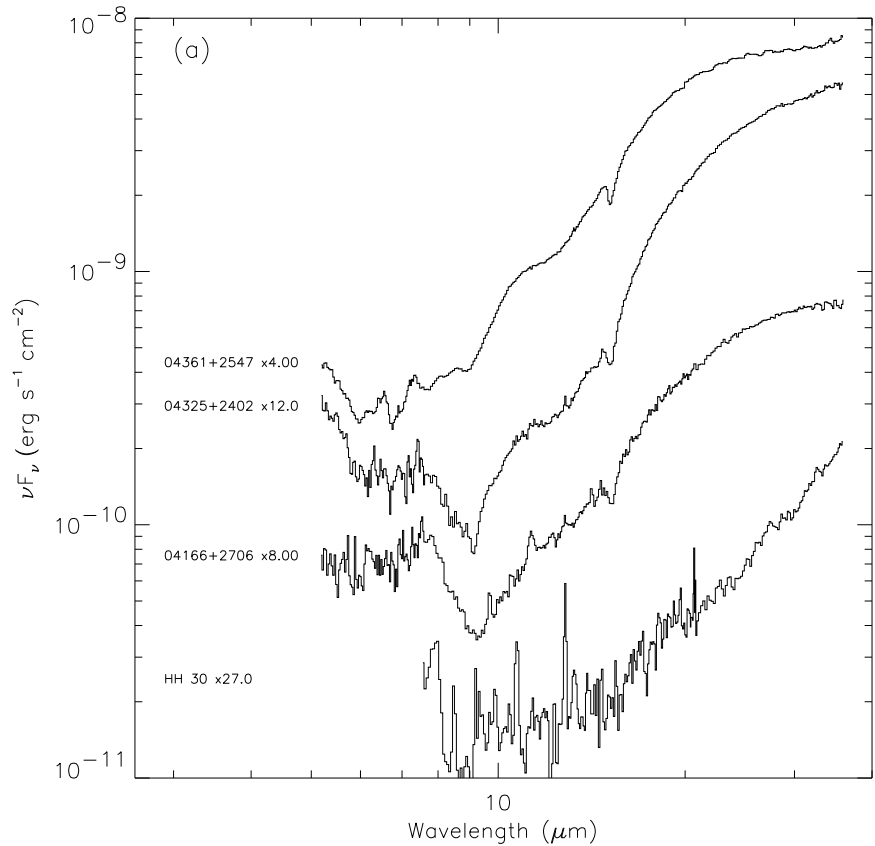}
\caption{IRS Spectra of Class I Objects in Taurus. \label{ClassI_IRS}}
\end{figure}

Figure \ref{ClassI_IRS} shows the IRS spectra of the 28 Class I objects
in our sample. Most have SEDs that are rising over the IRS spectral range,
and almost all of them show numerous absorption features in their 
mid-infrared spectra, typical of embedded protostars \citep[e.g.,][]
{alexander03, gibb04, watson04}: wide bands due to amorphous silicates 
at 9.7 and 18 $\mu$m (the latter one is much less prominent), and 
features at 6.0 $\mu$m due to H$_2$O ice, at 15.2 $\mu$m due to 
CO$_2$ ice, and at 6.8 $\mu$m due to organic compounds, possibly CH$_3$OH, 
NH$_4^+$, or strongly polar H$_2$O ice \citep{vanDishoeck04, alexander03}. 
In addition, the silicate absorption feature at 
9.7 $\mu$m often displays a wider absorption shoulder at the longer-wavelength 
end due to the broad H$_2$O ice libration band at $\sim$ 11-13 $\mu$m. 
The long-wavelength wing of the CO$_2$ ice feature also suggests that 
water ice is responsible, likely in the form of a mix with CO$_2$ ice
\citep{ehrenfreund99}. 

Some protostars show an obvious self-absorbed silicate feature at 10 $\mu$m
(e.g., IRAS 04154+2823, 04181+2654 B; see Fig.\ \ref{ClassI_IRS}d), 
where the silicate emission from inner 
disk regions is absorbed in cooler envelope regions, causing a feature that
is more or less flat, with some emission and some absorption 
characteristics \citep{mitchell81}. The self-absorbed features probably 
indicate that the optical depth of the envelope along the line of sight is 
not as large as for the cases where a clear silicate absorption feature is seen.

IRAS 04248+2612, 04264+2433, and CoKu Tau/1 (Fig.\ \ref{ClassI_IRS}e)
have a silicate emission feature,
but are surrounded by nebulosities and have an overall rising SED in the
infrared. Therefore, they are likely more evolved Class I objects seen through 
low-density envelopes, which have mostly dispersed and thus reveal the central star 
and its accretion disk. Based on the geometry of their near-infrared reflection nebulae, 
IRAS 04248+2612 and CoKu Tau/1 are likely seen close to edge-on \citep{padgett99}.

\subsection{Spectral Energy Distributions}
\label{SED_section}

Most protostars are found near or are embedded within dense molecular cloud 
cores \citep{myers87, onishi98}. Due to high extinction mostly due to their envelopes, 
protostars are usually faint or even undetectable in the optical. In the near-infrared
and in the optical, the infalling envelope often appears as a reflection 
nebula, seen in light scattered by dust grains in the inner parts of envelope cavities 
\citep[e.g.,][]{tamura91, padgett99,park02}. In addition, Class I objects are often 
extended in the sub-mm and mm \citep{chandler98,motte01}; at these wavelengths, 
the thermal emission from cold dust in the envelope is detected.

\begin{figure}
\epsscale{0.85}
\vspace*{-5mm}
\plotone{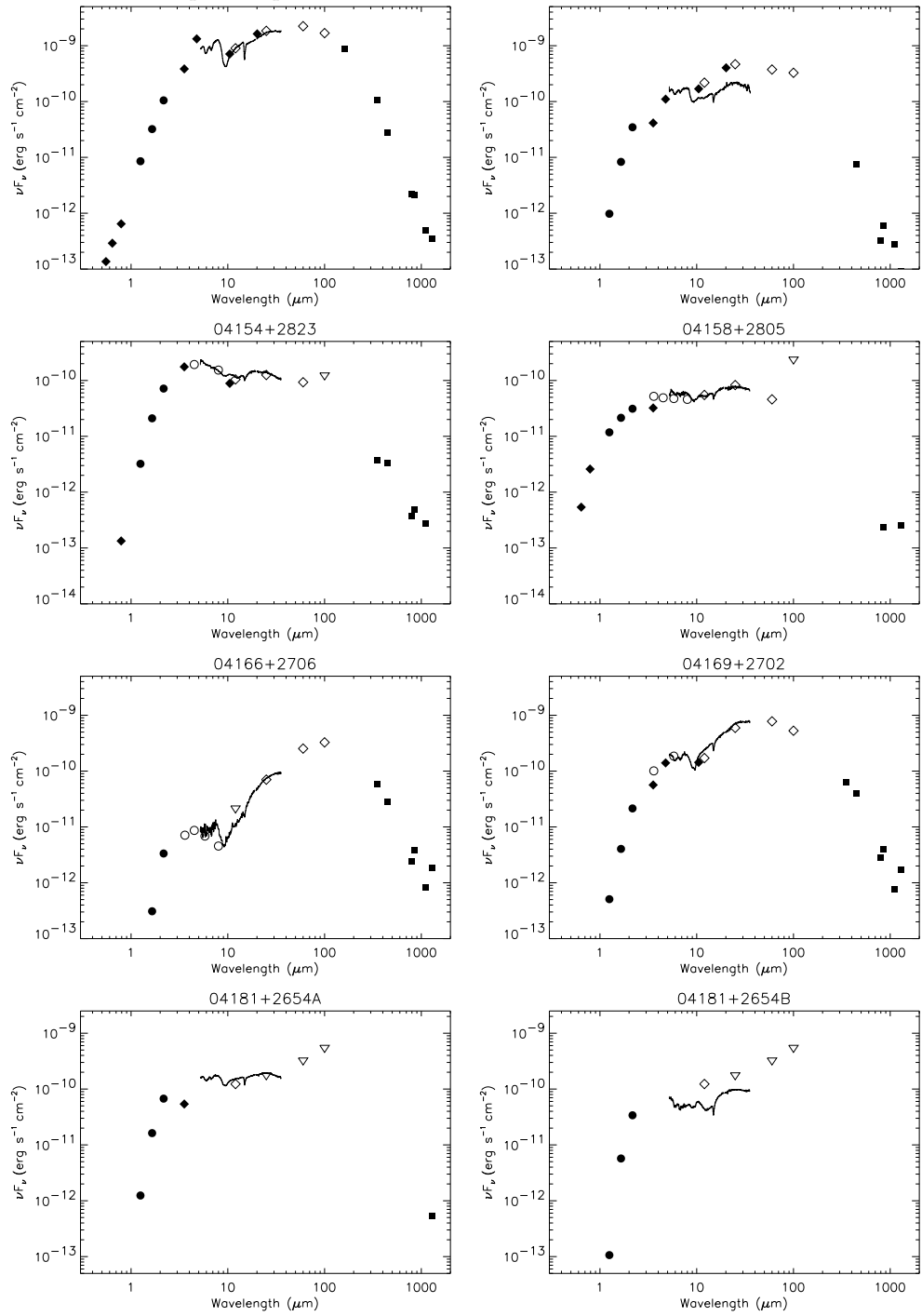}
\caption{SED plots of the Class I objects in our sample, ordered alphabetically by their 
name. Optical to mid-IR, ground-based photometry is shown as
{\it filled diamonds}, the 2MASS J, H, and K$_s$ fluxes as {\it filled circles}, the 
IRAC 3.6, 4.5, 5.8, and 8.0 $\mu$m fluxes as {\it open circles}, the {\it IRAS} 12, 25, 
and 60 $\mu$m fluxes as {\it open diamonds} or {\it open, upside down triangles}, if upper limit,
and sub-mm and mm fluxes, where available, as {\it filled squares}. The IRS spectrum
is also shown. No corrections for reddening have been applied.
\label{ClassI_SED}}
\end{figure}

To show the emission from protostellar systems at all wavelengths, in
particular how the emitted power is distributed at different wavelengths, we 
generated spectral energy distributions with data from the literature. 
In Figure \ref{ClassI_SED}, we show the SEDs of the
Class I objects in our sample, constructed with optical to mid-IR photometry
mostly from \citet{kenyon95}, the 2MASS $J$, $H$, and $K_s$ fluxes, the {\it IRAS}
12, 25, 60, and 100 $\mu$m fluxes, IRAC 3.6, 4.5, 5.8, and 8.0 $\mu$m
fluxes \citep{hartmann05,luhman06b}, sub-mm and mm fluxes, where available,
and the IRS spectrum. The {\it IRAS} fluxes are mostly from the {\it IRAS} 
Faint Source Catalog; for CoKu Tau/1, GV Tau, HL Tau, and L1551 IRS 5 we adopted 
the {\it IRAS} fluxes from \citet{weaver92}, and for the 12 and 25 $\mu$m fluxes
of DG Tau B and HH 30 the measurements from coadded {\it IRAS} scans by 
\citet{cohen87}.
The fluxes at 350, 450, 800, 850, 1100, and 1300 $\mu$m are mostly
from \citet{moriarty94}, \citet{andrews05}, \citet{young03}, and 
\citet{motte01}, while the few measurements at 160 $\mu$m were taken
by \citet{ladd91} with the Kuiper Airborne Observatory  (see Table 
\ref{tab_longwave} for a list of the long-wavelength data used in the 
SED plots). The fluxes were not corrected for reddening, since the high 
extinction of these objects is very likely intrinsic.  

The SED plots show that most of the power of a Class I object is emitted in 
the mid- to far-infrared; therefore, for many objects the IRS spectrum lies
close to the peak of the emission. For most objects, the SED is either rising
or flat over the IRS spectral range, with a deep silicate absorption feature at
10 $\mu$m as the most notable spectral feature. Only a few objects have
SEDs that seem to peak at $\sim$ 5 $\mu$m and decrease over the IRS 
spectral range; they might be more evolved protostars and will be discussed
in \S\ \ref{special_obj}.

\subsection{Notes on Individual Objects Excluded in the Modeling}
\label{notes_objects}

\subsubsection{Class I Objects} 

{\it GV Tau.} This object, also known as Haro 6-10, consists of two components 
separated by 1{\farcs}3.
Like T Tau, the GV Tau system consists of a T Tauri star (component A) and 
an embedded infrared companion (component B). The binary is surrounded by 
a circumstellar disk and an almost edge-on circumbinary envelope, with outflows 
oriented perpendicular to the disk plane \citep{menard93}. GV Tau B is likely responsible
for the outflows measured from this source \citep{chandler98, hogerheijde98}.
In the near-IR, the brightness ratio of the components varies considerably over 
the timescale of a few years \citep{leinert89,menard93,koresko97}. It is likely 
that the B component, which is more deeply embedded, varies more than the A 
component in brightness and generates the deep 10 $\mu$m silicate absorption 
feature, as well as the various ice features, observed in its mid-IR spectrum 
(see Figure \ref{ClassI_IRS}c). \\
{\it HH 30.} Optical and near-IR HST images \citep{burrows96, cotera01}
reveal that this object is seen edge-on; the central star illuminates a flared 
circumstellar disk, which obscures the protostar and is seen in scattered light. 
In addition, a narrow jet is emitted perpendicular to the disk plane. Its appearance is 
reminiscent of that of DG Tau B, which is also seen edge-on and emits a jet. 
\citet{wood98} modeled the emission of HH 30 and concluded that, in addition to 
a disk, only a tenuous envelope is required. The presence of a diffuse infalling 
envelope, as well as outflows, indicates that HH 30 is a protostar, though likely
more evolved.  \\
The {\it IRAS} fluxes of HH 30, determined by \citet{cohen87} from coadded 
{\it IRAS} scans, are highly uncertain, since they are contaminated by emission from two 
nearby sources, HL Tau and XZ Tau, which are much brighter. Our IRS spectrum
shows that the {\it IRAS} fluxes were overestimated by 1.5 to 2 orders of
magnitude (see Figure \ref{ClassI_SED}).
HH 30 is among the faintest objects in our Taurus sample; its IRS spectrum 
suggests the presence of a deep silicate absorption feature, but its SED is 
only poorly determined. 

\subsubsection{Class I/II Objects}
\label{special_obj}

As indicated in \S\ \ref{SED_section}, a few objects in our sample have
SEDs that are not typical of protostars. IRAS 04278+2253 and IC 2087 IR 
have clearly decreasing SEDs from 5 to 40 $\mu$m (see Fig.\ \ref{ClassI_IRS}d)
and would therefore be identified as Class II objects based on the slope of their 
infrared SED. 
Also IRAS 04154+2823 and LkHa 358 have SEDs that seem to peak around 
5 $\mu$m and do not display prominent 10 $\mu$m silicate absorption features. 

However, IC 2087 IR (also known as Elias 18) has ice absorption features in 
the 5--8 $\mu$m spectral region, a 10 $\mu$m silicate absorption feature, 
as well as a prominent CO$_2$ ice absorption at 15.2 $\mu$m. 
It is likely surrounded by a circumstellar disk, probably seen edge-on given 
its high visual extinction A$_V$ $\sim$ 18 \citep{shuping01}. 
\citet{bowey03} measured the 10 $\mu$m spectrum of IC 2087 IR; they noted 
a strong silicate absorption feature on a decreasing photosphere-like continuum. 
The absorption features we detect in the IRS spectrum of IC 2087 IR, and the 
fact that it is surrounded by a reflection nebulosity (IC 2087), suggest that it is 
actually a protostar, surrounded by an accretion disk and some envelope material. 
Further evidence for its earlier evolutionary stage are a molecular outflow 
\citep[e.g.,][]{heyer87} and the presence of two Herbig-Haro knots, which are 
not within the outflow, but are oriented along the axis of symmetry of the reflection
nebula \citep{gomez97}. It has been suggested that, since IC 2087 IR has a 
rather weak outflow, it could be in transition between the embedded protostellar 
and the T Tauri phase \citep{tegler95}. 

In contrast to IC 2087 IR, IRAS 04278+2253 shows no silicate absorption 
feature, only a weak CO$_2$ ice feature, and likely gas absorption features 
in the 5--8 $\mu$m wavelength range, most prominently an H$_2$O 
band centered at 5.7 $\mu$m. This feature, similar to what is observed in 
some FU Ori objects \citep{green06}, is thought to arise in the upper layers
of hot inner disk regions that are heated by accretion. IRAS 04278+2253 
is a 6{\farcs}8 binary, whose primary component is of G8 spectral type 
and has a relatively high mass accretion rate \citep[$6.6 \times 10^{-6}$ 
M$_{\odot}$ yr$^{-1}$;][]{white04}. The secondary is of K7/M0 spectral 
type \citep{white04} and thus clearly fainter than the primary. 
Given this object's relatively large mass accretion rate, it is likely still a protostar, 
but since no extended emission has been detected around it, there is likely only
little envelope material left. 

LkHa 358 forms a small cluster of stars together with HL Tau and XZ Tau 
(a flat-spectrum source, see \citealt{furlan06a}): all three objects are found 
within a radius of less than 1\arcmin. All of them are likely embedded to some 
extent; while HL Tau has an obvious absorption feature and an SED typical of 
a Class I object, LkHa 358 only has a weak absorption feature at 10 $\mu$m, 
and XZ Tau has silicate emission features at 10 and 20 $\mu$m and an overall 
flat SED in the mid-IR. Since HL Tau is the brightest source of the group, it is 
likely the dominating contribution to the IRAS source 04287+1807; therefore,
the {\it IRAS} fluxes for LkHa 358 are not known. 
The IRS spectrum shows that LkHa 358 has a silicate feature that could be 
partly self-absorbed, as well as some weak ice absorption features. The late 
spectral type of LkHa 358 (M5.5 according to \citealt{kenyon98} and 
\citealt{luhman00}) suggests that this object will end up as a very low-mass 
star, with a mass close to that of a brown dwarf. 

IRAS 04154+2823 is very faint and could be a more evolved protostar;
in fact, its 10 $\mu$m silicate feature seems to be self-absorbed,
with a broad absorption centered at 9.7 $\mu$m and emission between
10 and 12 $\mu$m, which is an indication of a tenuous envelope. Its
SED can be fitted reasonably well with a low-density envelope model (see Figure 
\ref{model_04154}). 

\section{Envelope Models}
\label{env_models}

\subsection{Model Description}
\label{model_descr}

In this paper, we generated envelope models for 22 of the 28 Class I objects we 
observed in the Taurus star-forming region using the IRS on {\it Spitzer} 
to characterize some of the envelope parameters that result in the observed 
SED shapes. These models include the contributions of the central protostar, 
the disk, and the envelope to compute the resulting SED of a protostellar system. 
We used a modeling code based on two types of models that differ in the initial
configuration: so-called TSC models \citep[after][]{terebey84}, which assume
an initially spherically symmetric molecular cloud core, and sheet-collapse 
models \citep{hartmann94, hartmann96}, which simulate the collapse of
elongated cloud cores. Both models result in flattened density distributions, 
in which the emergent SED is strongly dependent on the inclination angle of the system.
Below, we will briefly introduce these two types of models. 

\subsubsection{Envelope Component}

In the TSC models, the radiative equilibrium temperature is determined first by using
the angle-averaged density distribution for the infall region (\begin{math} \rho 
\propto r^{-3/2} \end{math} for $r \gg R_c$, \begin{math} \rho \propto 
r^{-1/2} \end{math} for $r \ll R_c$; $R_c$, the centrifugal radius, is where
material falling in on the equatorial plane lands and thus represents the outer disk
radius). In this region, departures from spherical symmetry become apparent only 
at small distances, when material falls onto a disk; at large distances, the TSC model 
can be described by spherical free-fall. 
After the temperature distribution has been calculated, the flattened, axially symmetric 
density distribution from \citet{terebey84}  is used to compute the flux emitted by 
the system \citep[][from here on KCH93]{kenyon93a}. 
The models also include isotropic scattering. The source luminosity is not set
according to the mass infall rate, which means that the disk
accretion rate can be different from the mass infall rate of the envelope. 

Outflow cavities along the poles can also be included in the models by specifying
a cavity semi-opening angle \citep{kenyon93b, calvet94}. Inside the bipolar 
cavities, material has been evacuated; outside, material will fall onto a
disk with an inner radius \begin{math} R_{min} = R_c \: \sin^2{\theta} \end{math},
where ${\theta}$ is the half opening angle of the cavity, i.e. the angle between the
polar axis and the streamline of an infalling particle at the edge of cavity at large distances,  
and $R_c$ is the centrifugal radius \citep{calvet94}. Thus, the shape of the cavities is
determined by the streamlines of the infalling material. 

The inclusion of a cavity in the density distribution of the envelope causes
departures from spherical symmetry in the temperature distribution. 
Scattered light from these cavities adds flux at near-infrared wavelengths, 
but the cavities also remove hot, inner dust. Therefore, for larger cavities, the
net result is a decrease in near- and mid-infrared flux when the inclination angle 
is larger than the opening angle of the cavity. However, when the inclination of 
the system allows the observer to see inside the cavity and thus the star and 
inner disk directly, the near-infrared emission increases \citep{calvet94}.  

Sheet-collapse models start from the collapse of sheets that are initially in 
hydrostatic equilibrium \citep{hartmann94,hartmann96}. This modification 
to the spherical infall scenario produces partially evacuated cavities, 
which, when aligned along the line of sight, reveal the inner parts of the 
envelope and disk, thus increasing the object's optical and near-infrared emission. 
Since the polar regions of the sheet collapse before the equatorial regions, polar 
cavities, which are regions where low-density material is falling in, will result 
\citep{hartmann94}. With sheet collapse, the opening angle of the cavity is 
connected to the infall and not to the outflows; eventually, outflows, but also wide-angle 
stellar winds \citep{arce02}, might easily remove the low-density material and fully 
clear out the cavities, so in the end both effects, sheet collapse and mass loss from 
a protostar, could be responsible for the formation of cavities. 

\subsubsection{Disk Component}

The disk is modeled as an optically thick, steady accretion disk irradiated as a flat disk 
by the central star ($T(r) \propto r^{-3/4}$; $T^4=T_{accretion}^4+
T_{irradiation}^4$), oriented perpendicular to the rotation axis of the envelope. 
As the outer disk regions, which emit mostly in 
the sub-mm and mm, are optically thin, and only the inner disk regions are 
optically thick, the outer disk radius was set at 10 AU to represent the optically
thick part of the disk. However, the disk is expected to extend out to $R_c$;
emission from the protostellar envelope will be obscured if it lies behind the disk 
along the line of sight. Therefore, to take this obscuration effect into account,
a second set of models was calculated with the disk radius set equal to $R_c$.
The latter set was used for the envelope and stellar emission, while the set with 
the smaller disk radius was used for the disk emission. In this way the disk 
component in our models represents only the optically thick disk regions;
the optically thin outer regions would likely add small amounts of additional flux 
at sub-mm and mm wavelengths.

This simple disk model also does not include an inner disk wall at the dust
truncation radius, which would add flux at near-IR wavelengths, in particular
for systems with higher inclination angles $i$, where the disk emission is 
reduced due to its $\cos(i)$ dependence.
In addition, since the disk model does not include an optically thin disk 
atmosphere that generates a 10 $\mu$m silicate emission feature
\citep[see, e.g.,][]{dalessio99}, the shape of the silicate feature is only
determined by the envelope. The material in the envelope will extinguish the
disk and thus generate a 10 $\mu$m absorption feature in the continuum 
emission arising from the disk, while the silicates in the envelope will produce an 
emission feature when the density is low and an absorption feature for higher densities.
Thus, the silicate feature might not be modeled realistically for Class I objects 
with low envelope densities, but should be a good approximation for more
typical, higher-density envelopes.

The inner disk radius is set by the star's magnetosphere, which channels the accretion 
flow onto the star, and is roughly at the corotation radius, at a few stellar radii from the 
stellar surface \citep{shu94}. However, in some models the inner disk radius was set
equal to the stellar radius, implying a boundary layer.
The accretion luminosity is set as the sum of the disk accretion luminosity, generated as 
matter accretes through the disk, and of the hot component luminosity, produced by 
the dissipation of the accretion shock on the stellar surface. For the cases where the
inner disk radius is equal to the stellar radius, half of the accretion luminosity is emitted
by the disk, and the other half by the boundary layer. For larger inner disk radii, the
accretion shock accounts for a larger fraction of the accretion luminosity than matter
accreting through the disk.

\subsubsection{Dust and Ices}

The dust in the envelope is composed of small silicate and graphite grains
\citep{draine84}, troilite (FeS), and water ice \citep{pollack94}, 
as well CO$_2$ ice \citep{ehrenfreund96}. The particle size distribution follows a 
power law with particle radius $a$: $n(a) da \propto a^{-3.5} da$, with maximum sizes 
of 0.3 $\mu$m. This maximum grain size corresponds to the typical value found
in the interstellar medium \citep[e.g.,][]{mathis77,weingartner01} and is motivated
by the notion that envelopes likely contain pristine material. Our data confirm this idea,
since the shape of the silicate absorption feature in most of our Class I objects indicates 
that the dust is dominated by small grains \citep[see also][]{osorio03}.
The abundance of the various components of the dust mixture is
similar to the one adopted by \citet{osorio03}, who adjusted the abundances
given in \citet{draine84} and \citet{pollack94} to fit a model to the spectrum 
of L1551 IRS 5. We do not include organics in our mixture; we assume that the 
carbon is present in the form of graphite only. The fractional abundances with 
respect to the mass of the gas for silicates, graphite, troilite, and H$_2$O ice are 
$\zeta_{sil}=0.004$, $\zeta_{gra}=0.0025$, $\zeta_{tro}=0.000768$, 
and $\zeta_{ice}=0.0005$, respectively. To get a better fit to the observed SEDs, 
these fractional mass abundances could be varied until a good match was found, 
but in order to limit the number of free parameters in the fit, we kept the abundances 
of these components constant. Only the CO$_2$ ice abundance was adjusted for each 
object to reproduce the depth of the CO$_2$ ice feature.

The opacities used in the current model should be regarded as approximations: 
the water ice opacities used in the model do not generate a feature at 6.0 $\mu$m, 
which is detected in our IRS spectra of Class I objects, and other ice opacities, 
e.g. for methanol ice, which might be responsible for the feature at 6.8 $\mu$m, 
are not included. Also, the 15.2 $\mu$m ice feature is treated as pure CO$_2$ ice, 
but in reality is likely mixed with water and other ices, making the feature broader
\citep{ehrenfreund99}. Therefore, our envelope models are not able to reproduce 
the majority of the ice features, and the most prominent ice feature, the CO$_2$ 
feature, can only be approximated in its depth, not its width. On the other hand, 
since the purpose of this study is a characterization of the overall SEDs of Class I objects, 
and thus of their envelope structure, we will neglect the fact that the ice features 
will not be reproduced well or at all. 

\subsubsection{Model Parameters}

The input parameters for the models are the total system luminosity 
($L_{star}+L_{acc}$), the fraction of the total luminosity that arises from 
the star ($\eta_{star}$), the radius of the star (assumed to be 2 R$_{\odot}$), 
the minimum disk radius, the centrifugal radius (which is essentially the outer disk radius), 
the inner and outer envelope radii, the flattening of the envelope (for sheet-collapse
models), and the reference density $\rho_1$, which is the density the envelope 
would have at a radius of 1 AU for the limit $R_c=0$.
The latter parameter, $\rho_1$, is related to the mass infall rate $\dot{M}_i$ 
and the central mass $M_{\ast}$ \citep{kenyon93a}:
\begin{equation}
\label{rho1_equ}
\rho_1 = 5.3 \times 10^{-14} \left( \frac{\dot{M}_i}
{10^{-5} \: M_{\odot} \: \mathrm{yr}^{-1}} \right)
\left( \frac{M_{\ast}}{1 \: M_{\odot}} \right)^{-1/2} 
\mathrm{g \: cm^{-3}} 
\end{equation}
For a fixed central mass, $\rho_1$ determines the mass infall rate. 
The inner radius of the envelope is determined by the dust destruction,
which, for typical dust made of silicates, is at the dust sublimation
temperature of $\sim$ 1500 K. If a cavity is included, the inner envelope
radius will be farther out.
The outer envelope radius is constrained particularly by sub-mm and mm observations, 
since this parameter affects the SED almost exclusively beyond about 100 $\mu$m; 
typical values are several 1000 AU. 

For sheet-collapse models, the flattening of the envelope is described by 
$\eta=R_{out}/H$, where $R_{out}$ is the outer radius of the envelope 
and H is the scale height of the original sheet, an isothermal, infinite, flat layer 
in hydrostatic equilibrium, and is defined by 
\begin{math} H = c_s^2/(\pi G \Sigma), \end{math}
with $\Sigma$ equal to the surface density \citep{hartmann94}. 
Over the course of the collapse of a flattened molecular cloud core, it is thought that 
$\eta$ increases, reflecting an increase in the size of the cavity.

\begin{figure}[hb]
\epsscale{0.9}
\plotone{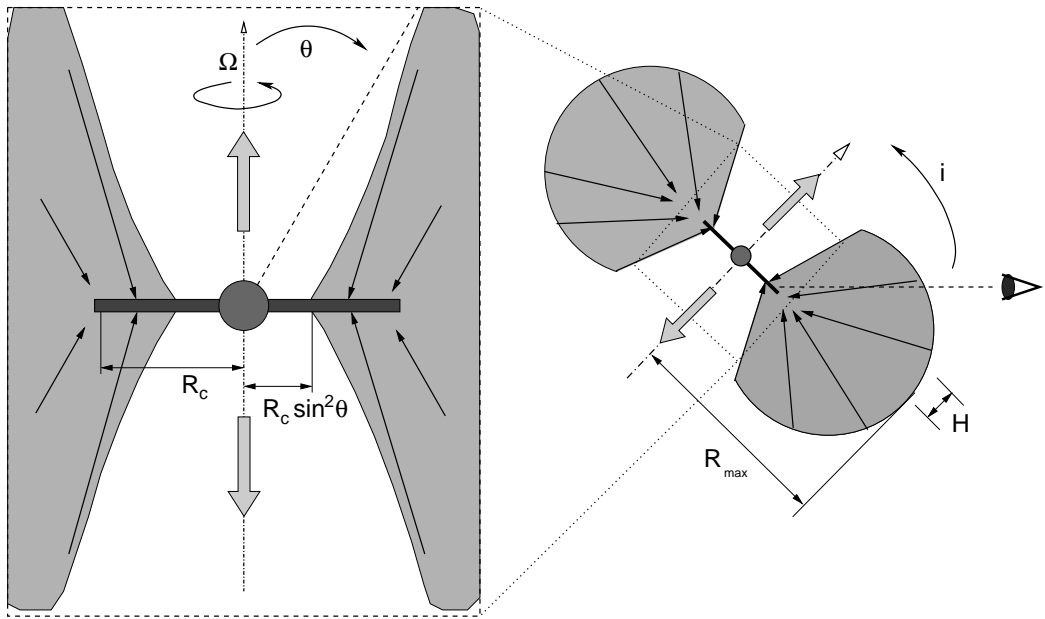}
\caption{A sketch of an envelope model as described in \S\ \ref{model_descr}. 
 \label{env_sketch}}
\end{figure}

The emergent SED is computed with the disk radius (set equal to 10 AU for
the disk component, and to $R_c$ for the stellar and envelope components), 
the inclination angle to the line of sight, and the opening angle of the cavity
as parameters. Scattering is also included, which causes an increase in 
flux at near-infrared wavelengths ($\lesssim$ 3 $\mu$m).
The parameters that affect a modeled SED the most are $\rho_1$, $R_c$,
and the inclination angle. The higher the density, the deeper the silicate
absorption feature and the longer the wavelength at which the SED peaks.
A high inclination angle will also increase the silicate feature depth, in addition 
to decreasing the emission below 8 $\mu$m and increasing the
emission in the far-infrared. The centrifugal radius, besides representing the 
outer disk radius, also defines the region where infalling material piles up;
thus, for smaller $R_c$ high-density material can be found closer to
the central protostar, resulting in more mid- to far-infrared emission and
a deeper silicate absorption feature.
A sketch of our model envelope, with important model parameters identified,
is shown in Figure \ref{env_sketch}.

From previous studies \citep[e.g.,][]{kenyon93b}, typical values for
young protostars in the Taurus star-forming region are an $R_c$ of 
$\sim$ 100 AU, and mass infall rates a few times 10$^{-6}$ 
M$_{\odot}$ yr$^{-1}$. Observations exclude centrifugal radii larger than 
about 1000 AU and mass infall rates larger than 10$^{-6}$ M$_{\odot}$
yr$^{-1}$. From equation \ref{rho1_equ}, it follows that typical values
for $\rho_1$ are of the order of 10$^{-14}$ to 10$^{-13}$ g~cm$^{-3}$.
\citet{hartmann96} find that characteristic values of $\eta$ are between
2.5 and 3.5 for the main infall phase of the envelope, resulting in large 
opening angles for the cavities. Over time, $\eta$ should increase, up to
a maximum value of $\sim$ 4, and the cavities become larger.
Inclinations are usually larger than 60\degr, but this could be a selection 
effect, since the SEDs of low-inclination protostars with large cavities would 
resemble the SEDs of T Tauri stars and therefore not be classified as Class I 
objects \citep{kenyon93b}.

In the following section, we will introduce our envelope models for 22 Class I
objects of our sample. In Figures \ref{model_04016} to \ref{model_L1551}
(excluding Figures \ref{04108_2803AB} and \ref{04325_slits}), 
we show the SED of each object together with the best-fit model (determined by eye) 
and its components, the star ({\it dotted gray line}), the disk ({\it dash-dotted gray line}), 
both extinguished by the envelope, and the envelope ({\it long-dashed gray line}); 
the sum of all components is represented by the thick, solid, gray line. All model 
protostars were placed at a distance of 140 pc.

\subsection{Models of Individual Class I Objects}

\subsubsection{04016+2610}
\citet{padgett99} imaged this object in the near-IR with HST; it is surrounded 
by a nebulosity on one side, with the protostar at its base. This might indicate 
that we are observing IRAS 04016+2610 through the opening of a large cavity in 
the envelope. A bipolar molecular outflow is oriented along the axis of symmetry 
of the nebula \citep[e.g.,][]{gomez97, hogerheijde98}; its inclination angle was
estimated to be 60\degr\ by \citet{hogerheijde98}.

\begin{figure}
\includegraphics[angle=90, scale=0.6]{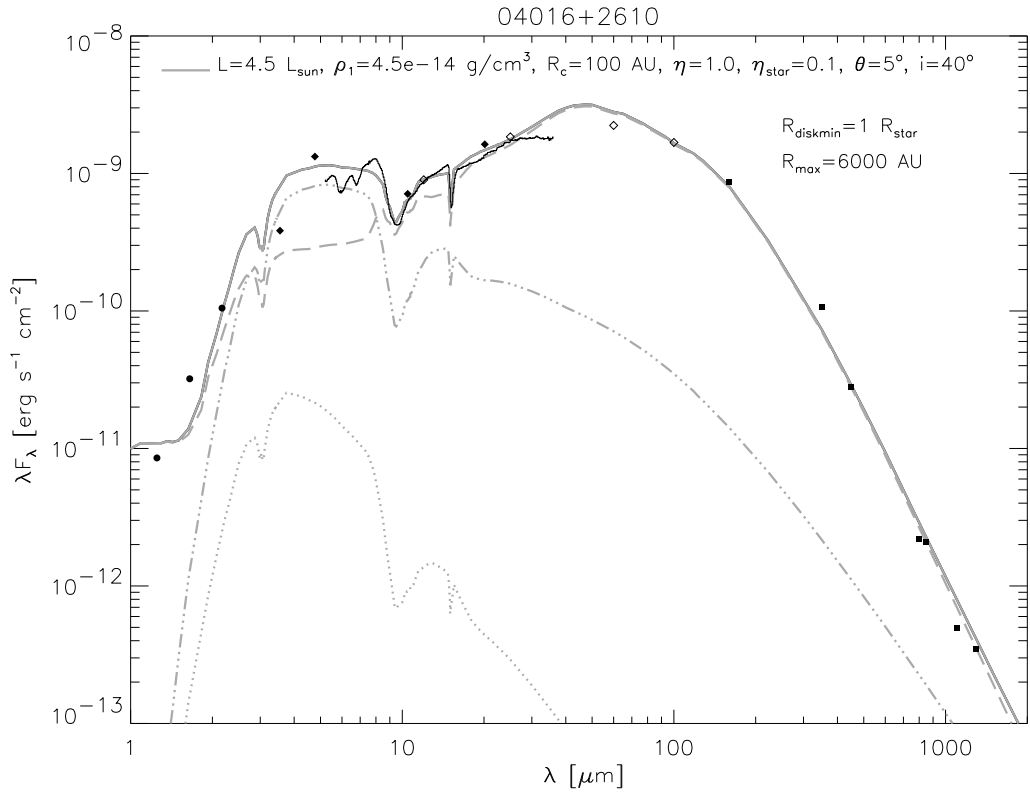}
\caption{IRS spectrum and photometric data of 04016+2610, and an envelope
model fit with L=4.5 L$_{\odot}$, $\rho_1=4.5 \times 10^{-14}$ g cm$^{-3}$,
$R_c$=100 AU, $\eta$=1.0, $\eta_{star}$=0.1, $\theta$=5\degr, 
i=40\degr, an inner disk radius of 1 stellar radius, and an outer envelope radius of 
6000 AU. The CO$_2$ ice abundance set to $1.5 \times 10^{-4}$. The gray lines 
represent the different model components: the envelope ({\it long-dashed line}), the 
star ({\it dotted line}) and the disk ({\it dash-dotted line}), both extinguished by the envelope, 
and the sum of all components ({\it solid gray line}). \label{model_04016}}
\end{figure}

\begin{figure}
\epsscale{0.65}
\plotone{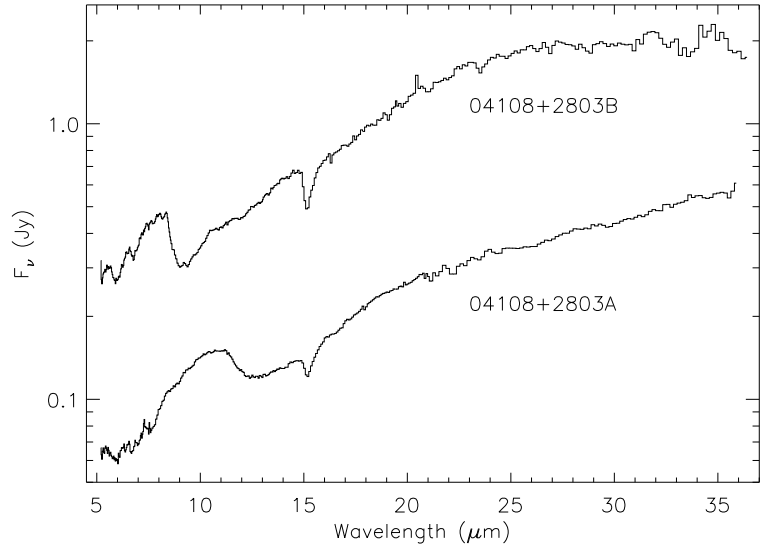}
\caption{IRS spectra of both components of the 21\arcsec\ binary IRAS 
04108+2803, plotted on the same scale. While observing 04108+2803A, the 
bright B component partly entered the LL slit, and therefore the spectrum of
component A beyond 14 $\mu$m contains an increasing amount of emission 
from component B; the spectrum of A is dominated by emission from B beyond 
about 20 $\mu$m. \label{04108_2803AB}}
\end{figure}

\begin{figure}
\includegraphics[angle=90, scale=0.6]{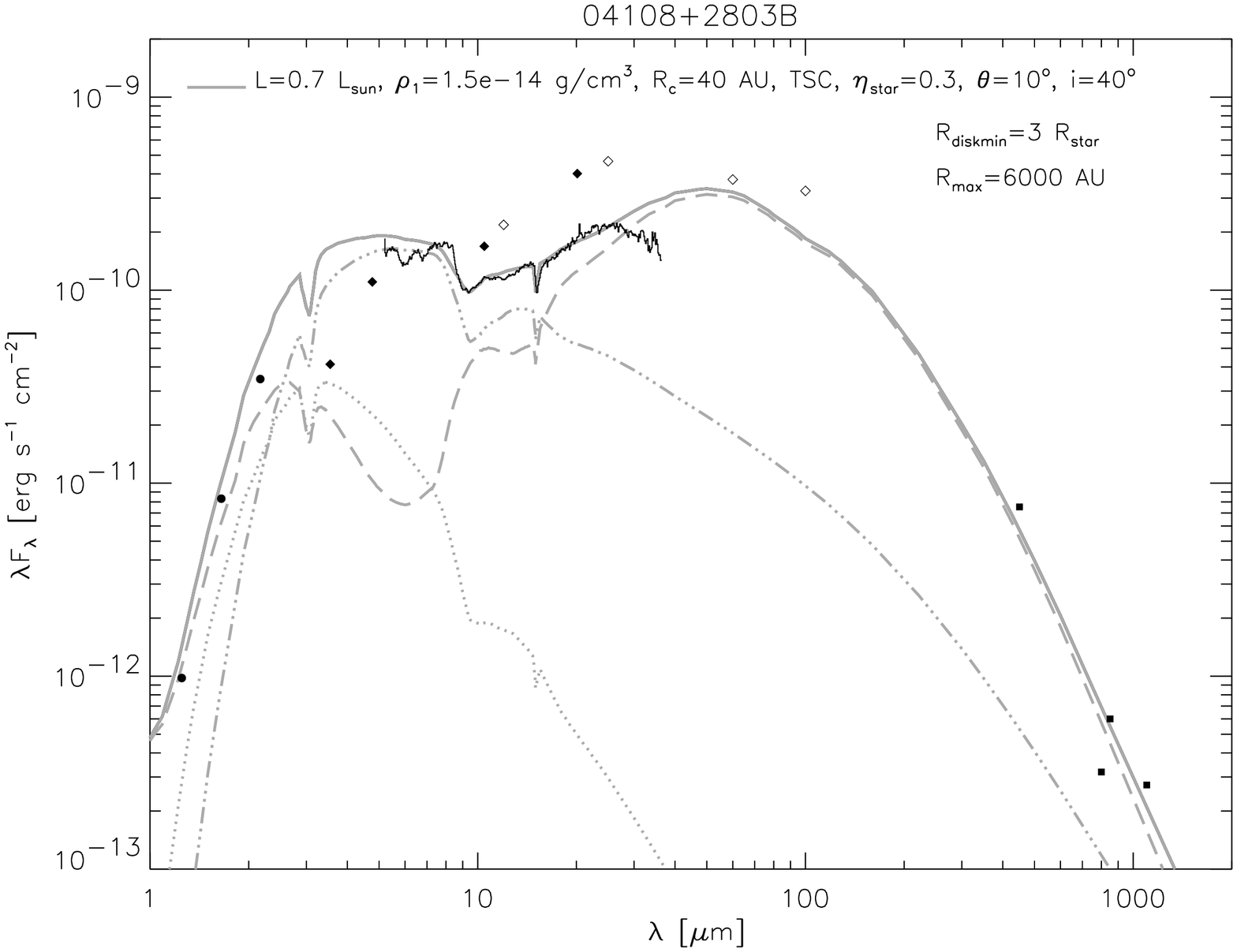}
\caption{IRS spectrum and photometric data of 04108+2803B, and an envelope
model fit with L=0.7 L$_{\odot}$, $\rho_1=1.5 \times 10^{-14}$ g cm$^{-3}$,
$R_c$=40 AU, initial TSC density distribution, $\eta_{star}$=0.3, $\theta$=10\degr, 
i=40\degr, an inner disk radius of 3 stellar radii, and an outer envelope radius of 6000 AU. 
The CO$_2$ ice abundance was set to $1.0 \times 10^{-4}$. The gray lines represent
the model components as in Figure \ref{model_04016}. \label{model_04108B}}
\end{figure}

\begin{figure}
\includegraphics[angle=90, scale=0.6]{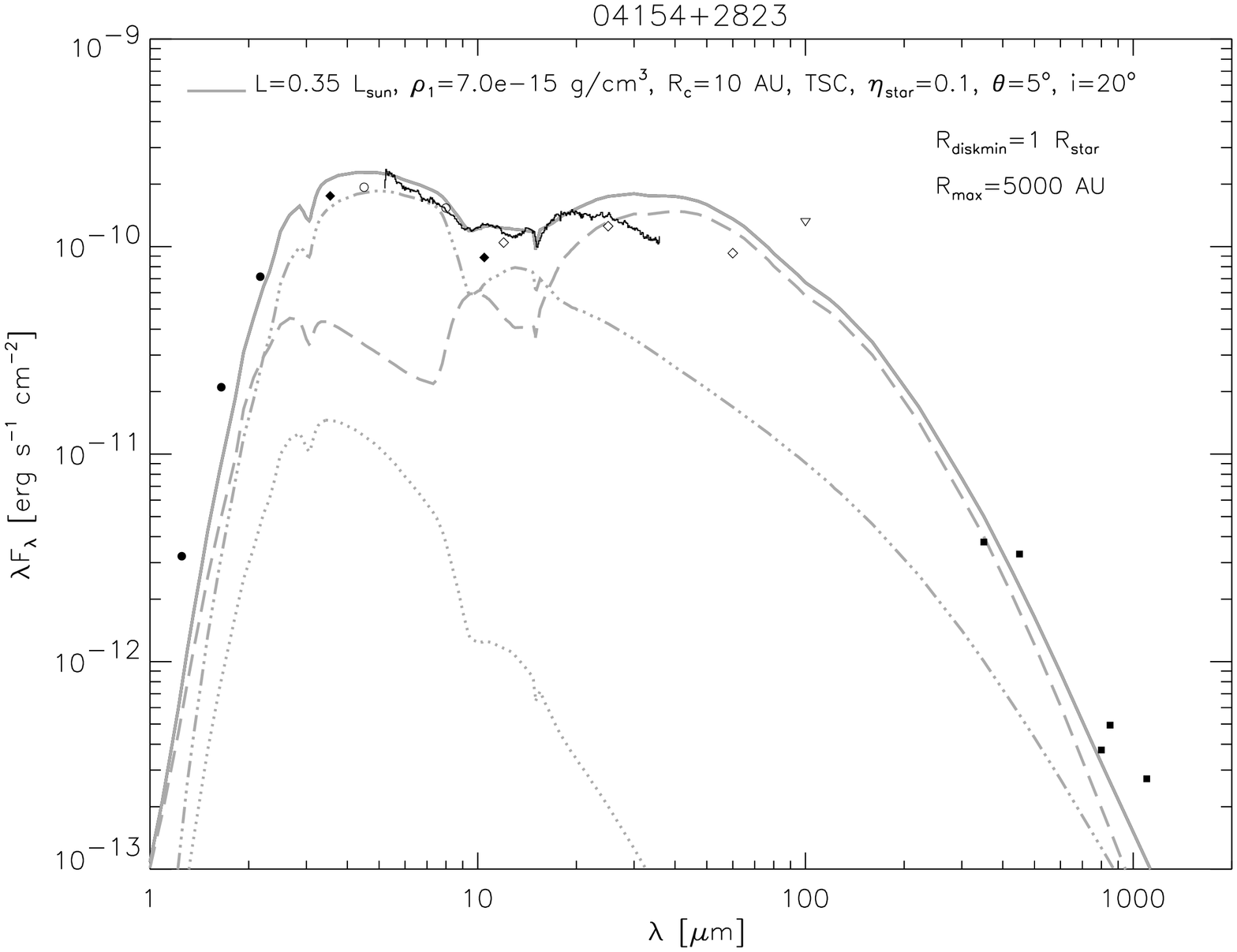}
\caption{IRS spectrum and photometric data of 04154+2823, and an envelope
model fit with L=0.35 L$_{\odot}$, $\rho_1=7.0 \times 10^{-15}$ g cm$^{-3}$,
$R_c$=10 AU, initial TSC density distribution, $\eta_{star}$=0.1, $\theta$=5\degr, 
i=20\degr, an inner disk radius of 1 stellar radius, and an outer envelope radius of 5000 AU. 
The CO$_2$ ice abundance was set to $1.0 \times 10^{-4}$. The gray lines represent 
the model components as in Figure \ref{model_04016}. \label{model_04154}}
\end{figure}

\begin{figure}
\includegraphics[angle=90, scale=0.6]{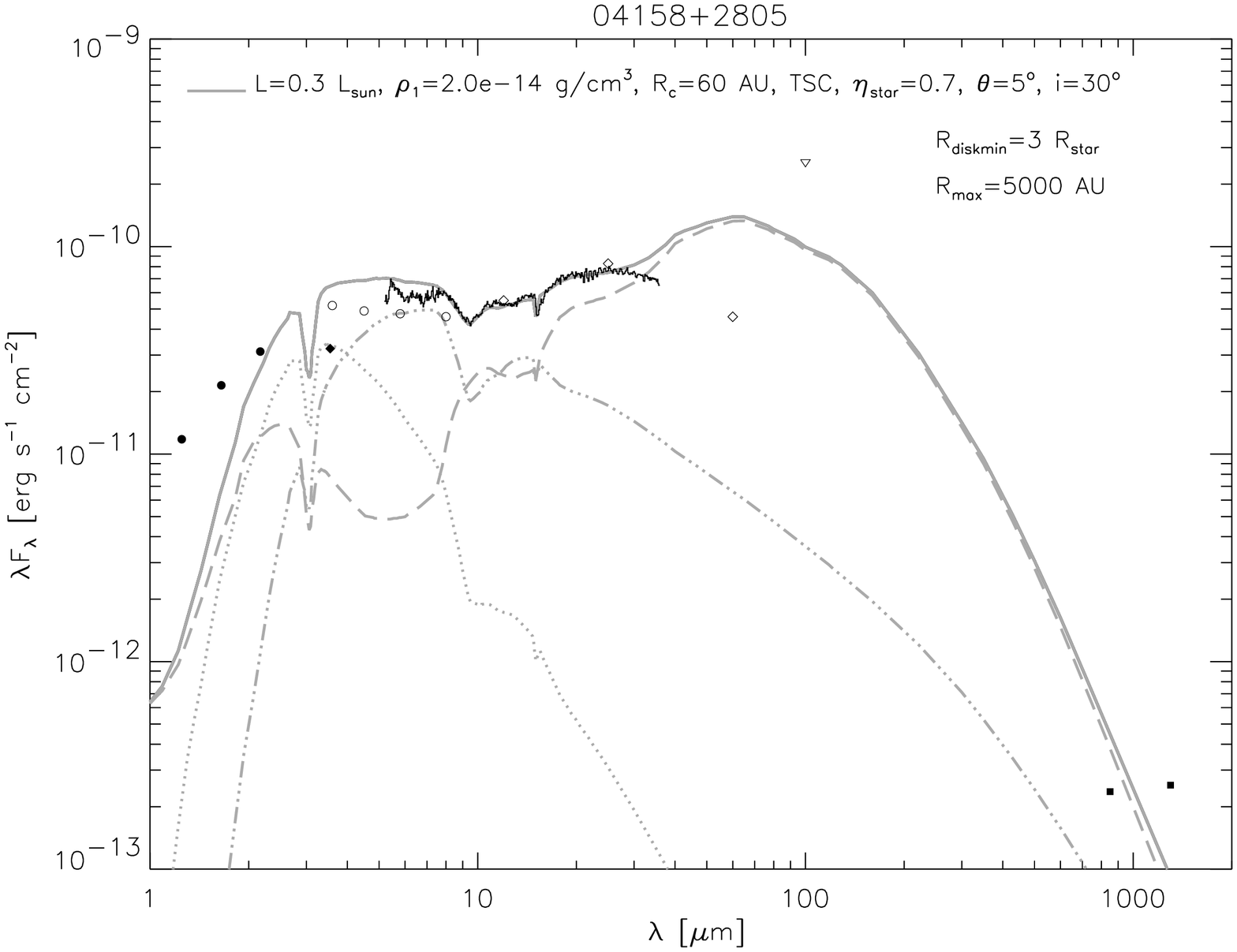}
\caption{IRS spectrum and photometric data of 04158+2805, and an envelope
model fit with L=0.3 L$_{\odot}$, $\rho_1=2.0 \times 10^{-14}$ g cm$^{-3}$,
$R_c$=60 AU, initial TSC density distribution, $\eta_{star}$=0.7, $\theta$=5\degr, 
i=30\degr, an inner disk radius of 3 stellar radii, and an outer envelope radius of 5000 AU. 
The CO$_2$ ice abundance was set to $4.0 \times 10^{-5}$. The gray lines represent
the model components as in Figure \ref{model_04016}. \label{model_04158}}
\end{figure}

\begin{figure}
\includegraphics[angle=90, scale=0.6]{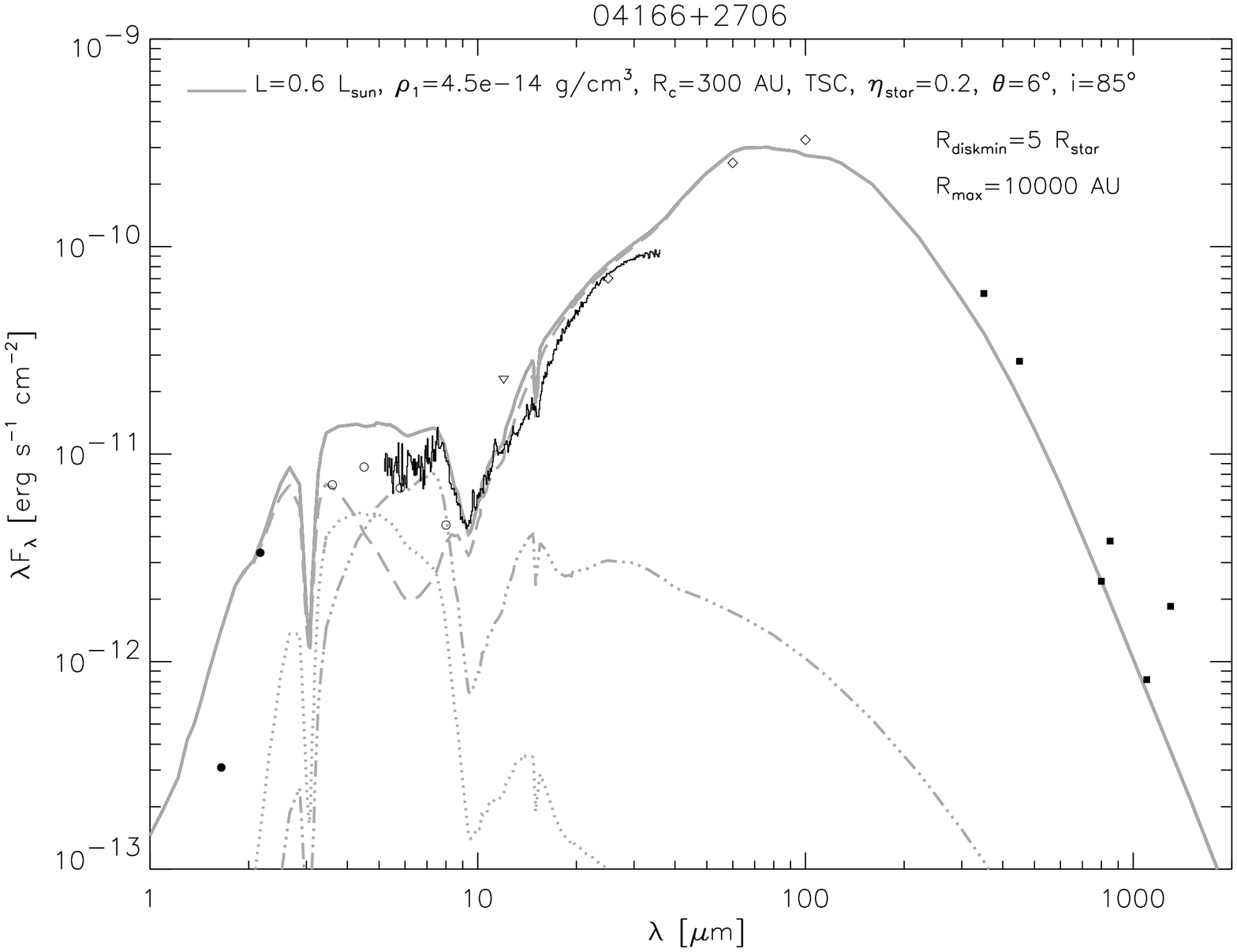}
\caption{IRS spectrum and photometric data of 04166+2706, and an envelope
model fit with L=0.6 L$_{\odot}$, $\rho_1=4.5 \times 10^{-14}$ g cm$^{-3}$,
$R_c$=300 AU, initial TSC density distribution, $\eta_{star}$=0.2, $\theta$=6\degr, 
i=85\degr, an inner disk radius of 5 stellar radii, and an outer envelope radius of 10000 AU. 
The CO$_2$ ice abundance was set to $2.0 \times 10^{-5}$. The gray lines represent
the model components as in Figure \ref{model_04016}. \label{model_04166}}
\end{figure}

\begin{figure}
\includegraphics[angle=90, scale=0.6]{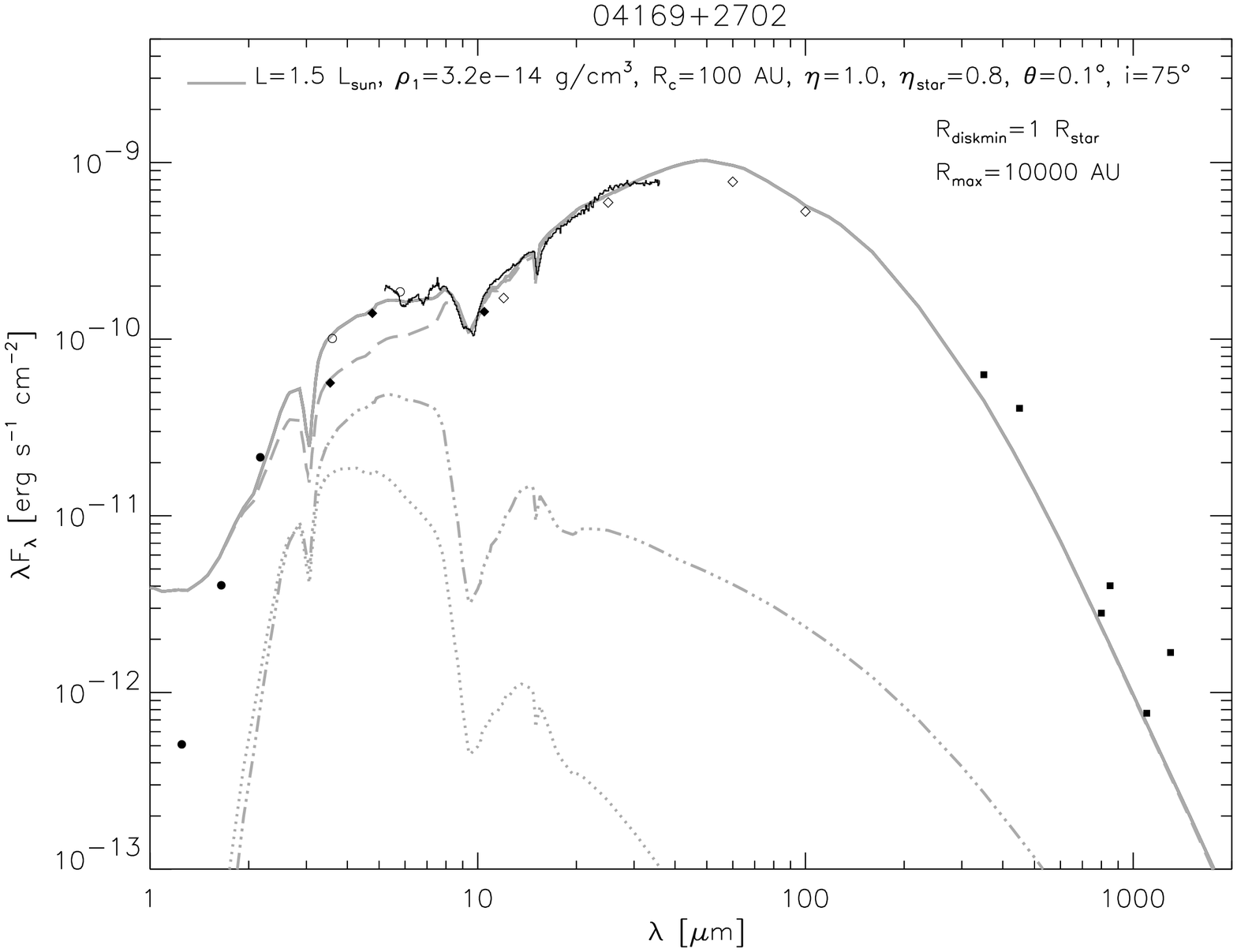}
\caption{IRS spectrum and photometric data of 04169+2702, and an envelope
model fit with L=1.5 L$_{\odot}$, $\rho_1=3.2 \times 10^{-14}$ g cm$^{-3}$,
$R_c$=100 AU, $\eta$=1.0, $\eta_{star}$=0.8, $\theta$=0.1\degr, i=75\degr,
an inner disk radius of 1 stellar radius, and an outer envelope radius of 10000 AU. 
The CO$_2$ ice abundance was set to $6.0 \times 10^{-5}$. The gray lines represent 
the model components as in Figure \ref{model_04016}. \label{model_04169}}
\end{figure}

\begin{figure}
\includegraphics[angle=90, scale=0.6]{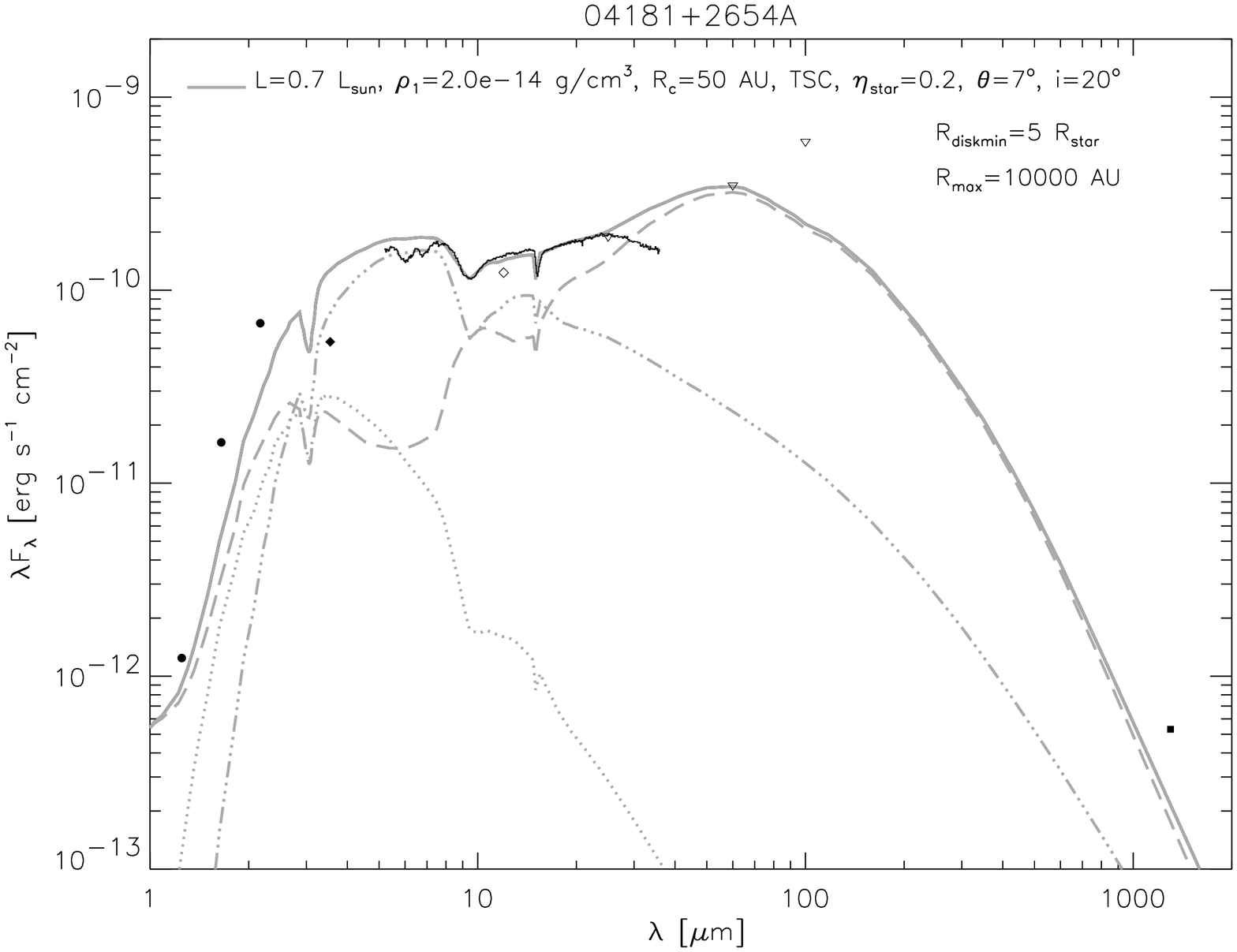}
\caption{IRS spectrum and photometric data of 04181+2654A, and an envelope
model fit with L=0.7 L$_{\odot}$, $\rho_1=2.0 \times 10^{-14}$ g cm$^{-3}$,
$R_c$=50 AU, initial TSC density distribution, $\eta_{star}$=0.2, $\theta$=7\degr, 
i=20\degr, an inner disk radius of 5 stellar radii, and an outer envelope radius of 10000 AU.
The CO$_2$ ice abundance was set to $7.0 \times 10^{-5}$. The gray lines represent
the model components as in Figure \ref{model_04016}. \label{model_04181A}}
\end{figure}

\begin{figure}
\includegraphics[angle=90, scale=0.6]{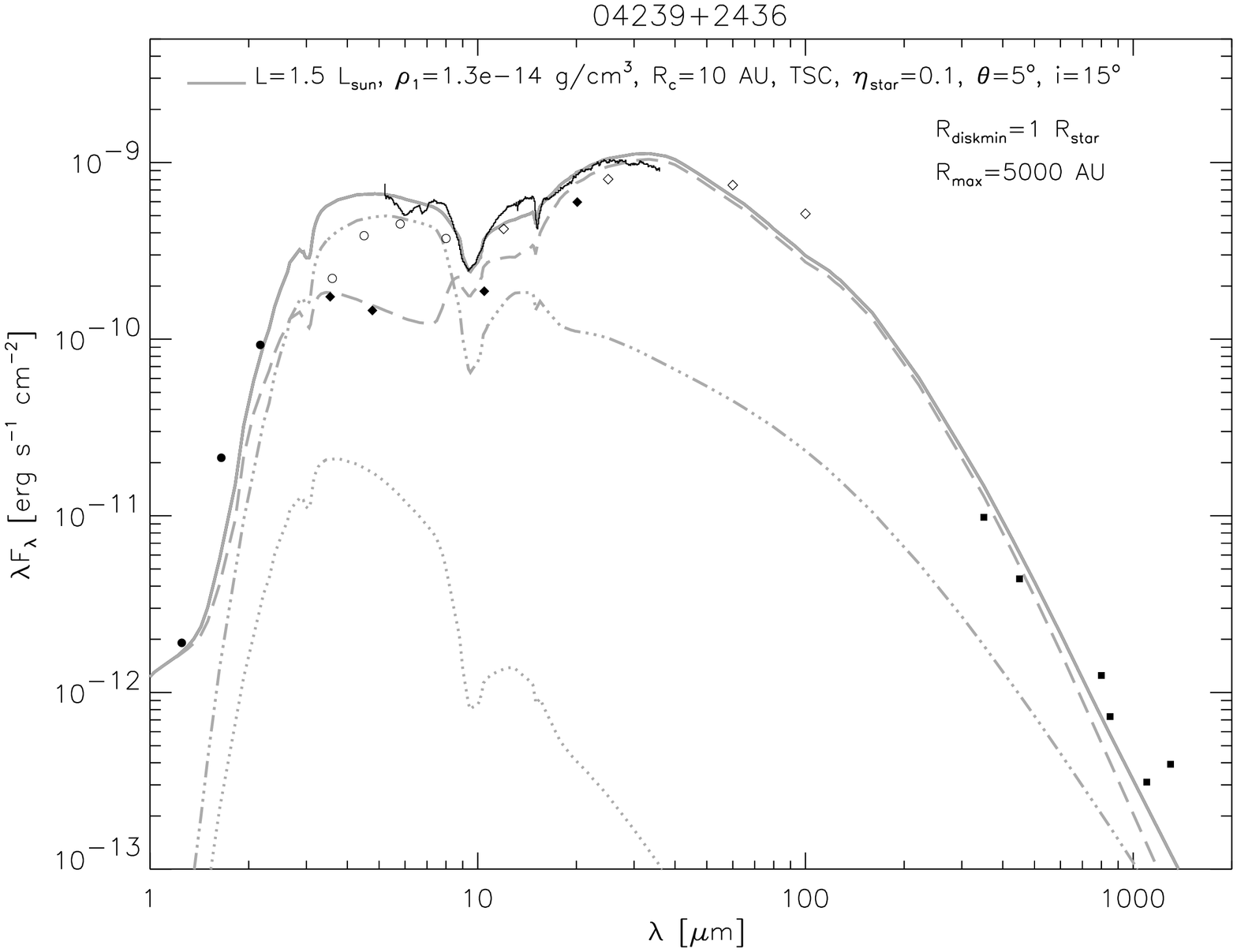}
\caption{IRS spectrum and photometric data of 04239+2436, and an envelope
model fit with L=1.5 L$_{\odot}$, $\rho_1=1.3 \times 10^{-14}$ g cm$^{-3}$,
$R_c$=10 AU, initial TSC density distribution, $\eta_{star}$=0.1, $\theta$=5\degr, 
i=15\degr, an inner disk radius of 1 stellar radius, and an outer envelope radius of 5000 AU. 
The CO$_2$ ice abundance was set to $1.0 \times 10^{-4}$. The gray lines represent 
the model components as in Figure \ref{model_04016}. \label{model_04239}}
\end{figure}

\begin{figure}
\includegraphics[angle=90, scale=0.6]{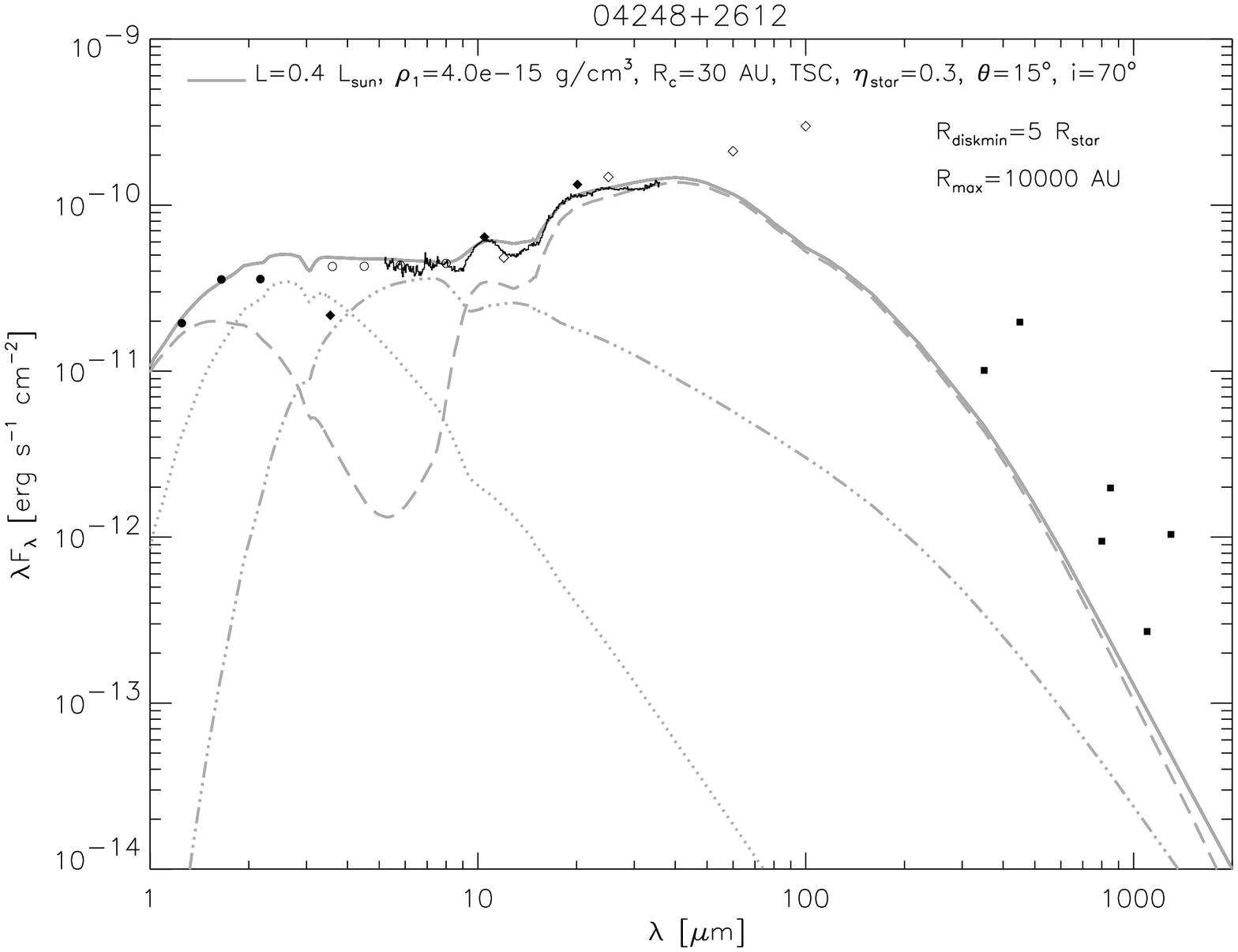}
\caption{IRS spectrum and photometric data of 04248+2612, and an envelope
model fit with L=0.4 L$_{\odot}$, $\rho_1=4.0 \times 10^{-15}$ g cm$^{-3}$, 
$R_c$=30 AU, initial TSC density distribution, $\eta_{star}$=0.3, $\theta$=15\degr,
i=70\degr, an inner disk radius of 5 stellar radii, and an outer envelope radius of 10000 AU.  
The CO$_2$ ice abundance was set to $5.0 \times 10^{-5}$. The gray lines represent 
the model components as in Figure \ref{model_04016}.
\label{model_04248}}
\end{figure}

\begin{figure}
\includegraphics[angle=90, scale=0.6]{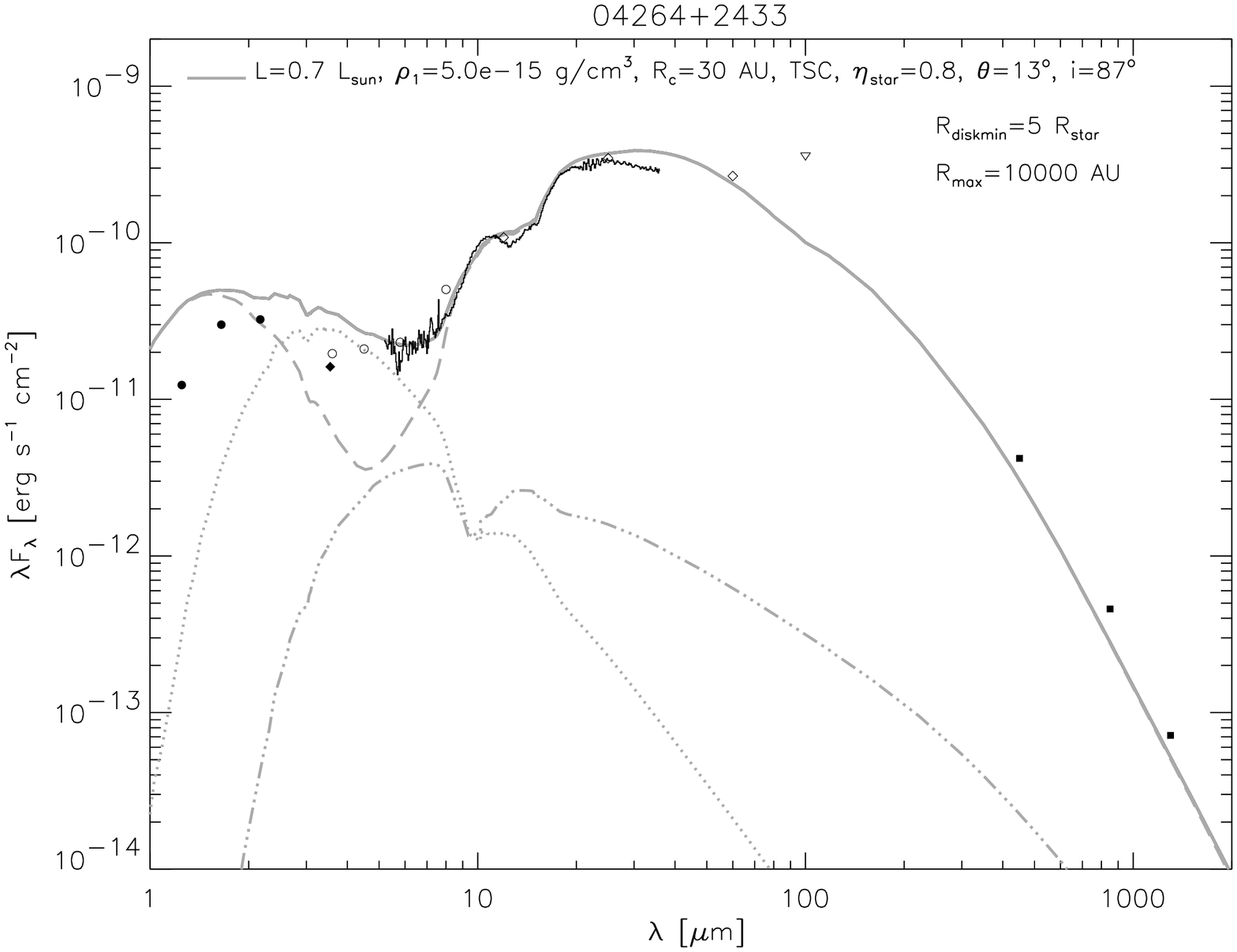}
\caption{IRS spectrum and photometric data of 04264+2433, and an envelope
model fit with L=0.7 L$_{\odot}$, $\rho_1=5.0 \times 10^{-15}$ g cm$^{-3}$, 
$R_c$=30 AU, initial TSC density distribution, $\eta_{star}$=0.8, $\theta$=13\degr, 
i=87\degr, an inner disk radius of 5 stellar radii, and an outer envelope radius of 10000 AU.  
The CO$_2$ ice abundance was set to $5.0 \times 10^{-5}$. The gray lines represent 
the model components as in Figure \ref{model_04016}. \label{model_04264}}
\end{figure}

In Figure \ref{model_04016} we show an envelope model for IRAS 04016+2610
whose parameters were adjusted to yield a good fit of the near-IR to mm SED. 
A sheet-collapse model with $\eta=1$ had to be adopted; TSC models
did not yield enough flux on the short- and long-wavelength sides of the
10 $\mu$m silicate feature. The IRS spectrum is fitted reasonably well between
about 8.5 and 25 $\mu$m; the discrepancies at the shorter wavelength side
might be resolved by adjusting the abundances of the dust and ices in the model, 
while the lower IRS fluxes beyond 25 $\mu$m could be an aperture effect, with the 
IRS slits truncating the regions contributing to the 20--40 $\mu$m region.
 
The model shown in Figure \ref{model_04016}, with a density of $4.5 \times 
10^{-14}$ g cm$^{-3}$, an $R_c$ value of 100 AU, and an inclination angle
of 40\degr, is one of several that yielded a comparable good fit. In particular, 
values for $\rho_1$ down to 3.5 $\times 10^{-14}$ g cm$^{-3}$ and
for $R_c$ down to 70 AU, as well as inclination angles within 30\degr--50\degr,
would yield acceptable fits. The lower values for $\rho_1$ and $R_c$ would
match previous modeling results of KCH93, who derived 3.2 $\times 10^{-14}$ 
g cm$^{-3}$ and 70 AU, respectively, for these two parameters. They would also
be roughly consistent with results from \citet{whitney97}, who derived an even smaller
$R_c$ of 50 AU and an inclination angle of 57\degr. On the other hand, our best-fit
model agrees very well with the results of \citet{eisner05} and of \citet{gramajo07}. 
A much smaller $R_c$ would result in an SED that is too 
narrow and contain too much flux in the 20--50 $\mu$m region, while a large $R_c$ 
would require larger inclination angles to match the 10 $\mu$m feature and 
longer-wavelength IRS flux, but it would also decrease the emission below 8 $\mu$m. 
\citet{stark06} require a larger centrifugal radius of 300 AU, combined with a larger
value for $\rho_1$ and an inclination angle of 65\degr, to reproduce the near-IR 
images of IRAS 04016+2610.

Our derived range for the inclination angle, 30\degr--50\degr, is somewhat smaller
than the value of 60\degr\ derived from observations \citep{hogerheijde98,padgett99} 
and modeling (KCH93). In addition, our derived cavity semi-opening angle of 5\degr\ is 
smaller than expected from the images; for example, \citet{stark06} derive a value 
of 25\degr\ for $\theta$. If we adopted a larger cavity, the emission from 
near-IR wavelengths to about 20 $\mu$m would decrease and therefore not fit the shape 
of most of the IRS spectrum. 
Finally, our adopted luminosity of 4.5 L$_{\odot}$ is larger than the bolometric luminosity 
of 3.7 L$_{\odot}$ (KH95), but necessary to yield enough flux. 

\subsubsection{04108+2803 B}
This protostar is part of a binary; it is separated by about 21\arcsec\ from 
component A, which is less embedded, displays a silicate emission feature, 
but is about a factor of 5 fainter in the mid-infrared (see Figure \ref{04108_2803AB}). 
Therefore, component B clearly dominates in the mid-IR; while 04108+2803 A 
probably is a reddened T Tauri star, 04108+2803 B is a Class I object. Both 
sources have no optical counterparts \citep[e.g.,][]{tamura91}.

According to \citet{young03}, IRAS 04108+2803 B is a Class I object based
on its bolometric temperature of 179 K, but would be classified as a Class 0
object based on its submillimeter luminosity. This IRAS source is not resolved in the 
450 and 850 $\mu$m maps of \citet{young03}; since it is also very faint
at 1.3 mm \citep[39 mJy;][]{motte01} and does not have an outflow, it
is likely an evolved protostar with a smaller envelope.

The envelope model for IRAS 04108+2803 B requires relatively low density 
(${\rho_1}=1.5 \times 10^{-14}$ g cm$^{-3}$), a small centrifugal 
radius (R$_c=40$ AU), a moderate-size cavity ($\theta=$ 10\degr), and 
an inclination angle of 40\degr\ (Figure \ref{model_04108B}). The outer 
envelope radius was set at 6000 AU, which is larger than the projected binary 
separation of 3000 AU, but necessary to fit the sub-mm and mm emission. 
This suggests that component A is likely obscured by this envelope; alternatively,
IRAS 04108+2803 B could be surrounded by a smaller envelope, but a fairly 
large circumstellar disk, which would add to the long-wavelength emission.

The IRS spectrum is fitted rather well with this envelope model; the change in slope 
of the spectrum beyond 25 $\mu$m could be an effect of slit truncation.
The fit of the 10 $\mu$m feature and of the slope of the IRS spectrum 
between 10 and 25 $\mu$m suggests that the values for the density and the 
centrifugal radius are relatively well-constrained. Previous models by KCH93 and
also by \citet{whitney97} obtained higher values for $\rho_1$ and $R_c$ 
(about double our values), but with comparable inclination angles. \citet{eisner05} 
derived a large value for $\rho_1$, too, but an $R_c$ similar to our value; their
best-fit model included a massive disk surrounded by a 500 AU envelope.

In our models, we can exclude $R_c$ values larger than about 50 AU, since they 
would result in a decrease in flux from 15 to 40 $\mu$m due to an overall shallower 
SED shape. A reference density up to $2.0 \times 10^{-14}$ g cm$^{-3}$ would 
still fit the observed SED, if combined with a smaller inclination angle ($\sim$ 25\degr)
to reduce the depth of the silicate absorption feature.
The cavity semi-opening angle of 10\degr\ is required to fit the IRS spectrum; a 
smaller opening angle would result in an increase in flux in the 10--25 $\mu$m 
wavelength range. The inner disk radius was set at 3 stellar radii to decrease the 
flux in the 2--8 $\mu$m region; this parameter does not affect the remaining parts
of the SED.

\subsubsection{04154+2823}
IRAS 04154+2823 might be a transition object between the Class I and II stage,
since it has an SED that is somewhat decreasing in the mid-IR, and it also does 
not display a prominent silicate absorption feature. However, its spectrum shows 
a clear CO$_2$ ice absorption feature at 15.2 $\mu$m, and its overall infrared 
SED is relatively flat, suggesting that some envelope material should be present. 
It is also very faint in the optical \citep{strom94}, possibly indicating larger
extinction along the line of sight.
It is somewhat reminiscent of IRAS 04158+2805, which is also faint and has a
relatively flat SED over the IRS spectral range.

This protostar has a low luminosity (L$_{bol}$ = 0.3--0.4 L$_{\odot}$; see
Table \ref{tab_lum_incl}); its envelope model requires a very low reference density 
of ${\rho_1}=7.0 \times 10^{-15}$ g cm$^{-3}$ and a very small centrifugal 
radius of 10 AU to reproduce the mid-IR spectrum as well as the sub-mm measurements 
(Figure \ref{model_04154}). However, this model overestimates the flux between 20 
and 100 $\mu$m and slightly underestimates the mm flux. A luminosity value
higher than about 0.4 L$_{\odot}$ would result in an increase in the long-wavelength 
flux, which would fit the mm data better, but yield even more flux between 20 
and 60 $\mu$m. An increase in $\rho_1$ would have a similar effect; in addition,
values of $\rho_1$ larger than about $8.0 \times 10^{-15}$ g cm$^{-3}$ can 
be excluded given the constraint of the weak silicate absorption feature. 
The low inclination angle ($i=20$\degr) is also well-constrained by the 10 $\mu$m 
feature and the shape of the SED.

\subsubsection{04158+2805}
According to \citet{menard01} who imaged this source in the optical, IRAS 
04158+2805 is likely surrounded by an edge-on disk and a bipolar nebula; a 
jet is oriented perpendicular to the disk plane, in a geometry similar to HH 30. 
\citet{park02} observed that this object is not extended in the near-IR or 
at millimeter continuum wavelengths, and therefore it could be a heavily reddened 
T Tauri star and not a Class I object. Recently, \citet{andrews07} modeled
the SED of IRAS 04158+2805 with only a large disk component.
However, \citet{kenyon98} concluded from their optical spectra and optical and 
near-IR photometry that IRAS 04158+2805 is a Class I source. Unpublished HST
images at optical and near-IR wavelengths (programs 9103 and 10603, PI 
K. Stapelfeldt and D. Padgett, respectively) show an extended, one-sided 
conical nebulosity, confirming that this source is surrounded by an envelope.
We note that this object is of a late spectral type (M5--M6 according to 
\citealt{luhman06a}; M6 according to \citealt{white04}); it will probably 
become a very low-mass star, close to the hydrogen-burning mass limit.

The SED of IRAS 04158+2805 can be fitted with an envelope model with relatively
low density (${\rho_1}=2.0 \times 10^{-14}$ g cm$^{-3}$), small
centrifugal radius (R$_c=60$ AU), small cavity ($\theta=$ 5\degr) and 
an inclination angle of 30\degr\ (Figure \ref{model_04158}). The
inner disk radius was set at 3 stellar radii to decrease the amount of flux
emitted at near-IR and mid-IR wavelengths out to 8 $\mu$m. The outer
envelope radius was set at 5000 AU to somewhat decrease the emission in the 
sub-mm and mm. Nevertheless, the model overestimates the flux at wavelengths 
longer than 30 $\mu$m, but lies in between the two measurements around 1 mm. 
The discrepancy between the two long-wavelength data points could be explained 
by the large difference of the two beam sizes (see Table \ref{tab_longwave}); 
the 880 $\mu$m measurement likely misses some extended emission.

We adopted a luminosity of 0.3 L$_{\odot}$, which is 50\% larger than the 
bolometric luminosity we measured by integrating the fluxes fluxes of the SED; 
a model with $L=$ 0.2 L$_{\odot}$, a smaller density ($\sim 1.0 \times 
10^{-14}$ g cm$^{-3}$), and a centrifugal radius around 10 AU would yield 
a comparable fit, but still overestimate the emission starting at 20 $\mu$m. 
Overall, the model parameters are mainly constrained by the IRS spectrum;
the depth of the silicate feature and the shape of the IRS spectrum from 12 to 
30 $\mu$m constrain the reference density $\rho_1$ and the cavity 
semi-opening angle, respectively, reasonably well.  
With the current reference density, a much smaller centrifugal radius would 
result in even higher fluxes in the 30--60 $\mu$m region, while a value larger 
than about 90 AU would decrease emission from about 15 to 40 $\mu$m. 
A $\rho_1$ value of about 1.5 $\times 10^{-14}$ g cm$^{-3}$,
combined with an $R_c$ value of 50 AU and a slightly larger inclination angle of
40\degr\ would also yield an acceptable fit.
We can exclude an edge-on orientation based on the shape of the IRS spectrum.
Only the availability of accurate, far-IR to mm fluxes will improve the
model fit of IRAS 04158+2805.

\subsubsection{04166+2706}
This IRAS source is very faint in the near-IR; no source is detected in 2MASS, 
but both \citet{kenyon90} and \citet{park02} detected a faint near-IR source 
close to the {\it IRAS} position, which we adopted for the pointing of our IRS 
observations. The latter authors describe the appearance of IRAS 04166+2706 
as low surface brightness nebulosity. \citet{motte01} and \citet{tafalla04}
detected a dense, round core at mm wavelengths coinciding with the IRAS source. 
Thus, this Class I object is likely deeply embedded.
It also has bipolar outflows \citep{bontemps96}, which are highly collimated and 
of extremely high velocity \citep{tafalla04}, consistent with this object's early 
evolutionary stage, possibly placing it at the Class 0 stage. The high degree of 
symmetry of the red- and blueshifted gas of the outflows \citep{tafalla04} 
suggests that this object is likely seen close to edge-on.
For the SED plot, we included the H and K magnitudes measured by \citet{kenyon90}.

The SED shape of IRAS 04166+2706 further indicates that it is likely seen at a large
inclination angle, close to edge-on. In fact, an envelope model that fits it well
(see Figure \ref{model_04166}) has an inclination angle of 85\degr. Smaller
inclination angles (70\degr-80\degr) would result in models that match the
observed spectrum from 15 to 30 $\mu$m more closely, but they would 
overestimate the emission from the near-IR to 8 $\mu$m. Increasing the cavity 
semi-opening angle by just 1\degr\ from its current value of 6\degr\ would 
lower the flux between 8 and 30 $\mu$m, resulting in a better fit of the slope
of the spectrum beyond 12 $\mu$m, but a poorer fits of the silicate feature. 
An inner disk radius of 5 stellar radii was adopted to lower the flux below 8 $\mu$m.

A reference density of $4.5 \times 10^{-14}$ g cm$^{-3}$ and a large centrifugal 
radius ($R_c$ = 300 AU) are required to fit the depth of the silicate feature and also 
the width of the SED. Values of $\rho_1$ and $R_c$ down to $3.5 \times 10^{-14}$ 
g cm$^{-3}$ and 200 AU, respectively, would also yield good fits; only the silicate feature 
would be somewhat narrower. The best-fitting model of KCH93, who had fewer constraints
available, resulted in $\rho_1 = 1.0 \times 10^{-13}$ g cm$^{-3}$, $R_c$=70 AU, 
and $i=30\degr$; a high-inclination model with the same $R_c$, but $i=90\degr$ and 
$\rho_1 = 3.2 \times 10^{-14}$ g cm$^{-3}$ was also suggested as a fit. 
This underlines the importance of the IRS spectrum to discriminate between different 
models.

We note that our model does not fit the 850 and 1300 $\mu$m measurements, 
which were done with a larger beam size (see Table \ref{tab_longwave}), but it 
fits the data taken at similar wavelengths, but with $\sim$ 20\arcsec\ apertures. 
This could indicate that some extended, cold dust outside of the infall region contributes 
to the long-wavelength emission \citep[see also][]{jayawardhana01}. 

\subsubsection{04169+2702}
This object, which is not detected at optical wavelengths, is surrounded by a 
cometary reflection nebula in the near-IR and associated with Herbig-Haro 
emission knots \citep{tamura91, gomez97}. It also has bipolar outflows
\citep[e.g.,][]{moriarty92, bontemps96}, which are perpendicular to an 
elongated envelope structure, about 2200 AU $\times$ 1100 AU in size
and inclined at 60\degr\ from the line of sight \citep{ohashi97b}. 
Measurements of continuum emission at 2.7 mm did not resolve the source on
scales of 2{\arcsec}--3{\arcsec}, probably indicating that this emission arises
from a disk of radius $\lesssim$ 150 AU \citep{ohashi97b}.

The SED of IRAS 04169+2702 is fitted well by a sheet-collapse envelope model 
($\eta=1$) with moderate density ($\rho_1 = 3.2 \times 10^{-14}$ 
g cm$^{-3}$), a centrifugal radius of 100 AU, basically no cavity and
a high inclination angle (Figure \ref{model_04169}). Compared to the results
obtained by KCH93, both our reference density and centrifugal radii are smaller by 
about a factor 3, and our inclination angle is larger (75\degr\ versus 30\degr).
\citet{whitney97} modeled this object with an $R_c$ of just 10 AU, a reference 
density half our value, a cavity semi-opening angle of 11\degr, and an inclination
angle of 45\degr.
However, our model is better constrained. Both $\eta$ and the 
cavity semi-opening angle are determined by the shape of the IRS spectrum
beyond about 12 $\mu$m. In particular, a TSC model would yield less flux from 
about 10 to 100 $\mu$m given the same envelope parameters.
The large $R_c$ value is required to fit the long-wavelength part of the IRS spectrum, 
while the value for the reference density is constrained by the shape of the silicate 
absorption feature and the peak of SED. The 10 $\mu$m feature and the SED
shape also constrain the high inclination angle, which has to be larger than 70\degr\
to yield a good fit.

The luminosity used for the model ($L=1.5$ L$_{\odot}$) is twice as high 
as the bolometric luminosity of 0.8 L$_{\odot}$ measured by KH95, but close 
to the value we calculated by integrating the fluxes of the SED, 1.2 L$_{\odot}$,
and to the value of 1.4 L$_{\odot}$ determined by \citet{kenyon90}.
We can exclude a lower luminosity from our model fit, since it would not 
yield enough emission to match the observations.  
As for IRAS 04166+2706, our model does not fit the long-wavelength emission
measured with large aperture sizes, but it matches the observations taken with
smaller apertures. Again, we suggest that some cold dust just outside the infalling
envelope is responsible for the extended mm emission.

\subsubsection{04181+2654 A} 
This IRAS source is a member of a wide binary system; component B lies to
the northwest of component A, at a projected distance of 31\arcsec, which 
corresponds to about 4300 AU at the distance of Taurus. In addition, the IRAS 
source 04181+2655 lies somewhat less than 1\arcmin\ to the northwest of 
04181+2654 B. A bipolar outflow is centered on IRAS 04181+2655, but no 
outflow has been detected from the IRAS sources to the south \citep{bontemps96},
which could be the result of a low inclination angle along the line of sight of these
two objects.

Of the two components of the 04181+2654 binary, the B component seems 
to be more deeply embedded, since it is fainter than A by a factor of 12 in the 
$J$ band, but only by about a factor of 2 in the $K$ band. Over the mid-IR 
spectral range, component B is fainter than component A by a factor of 2 to 3. 
The silicate feature of IRAS 04181+2654 B is seen partly in self-absorption, 
indicating that it is oriented along the line of sight such that part of the silicate 
emission originating in the inner disk is absorbed in the envelope. 

\citet{motte01} measured the 1.3 mm flux of IRAS 04181+2654 A, which 
is dominated by emission from an extended envelope. 
According to \citet{park02}, both A and B are extended at mm continuum 
wavelengths, but not in the near-IR. They suggest that both sources are protostars; 
therefore, they are likely surrounded by an envelope which could  be truncated in
its outer parts. Images at 0.9 $\mu$m obtained by \citet{eisner05} reveal a faint
nebulosity around IRAS 04181+2654 A, possibly tracing scattered light from an 
envelope cavity; the B component is not detected.
 
The SED of IRAS 04181+2654 A can be fitted by an envelope model with a reference 
density of $2.0 \times 10^{-14}$ g cm$^{-3}$, a centrifugal radius of 50 AU, 
a cavity semi-opening angle of 7\degr, and an inclination angle of 20\degr\ (Figure 
\ref{model_04181A}).
In order to fit the SED shape from the near-IR to 8 $\mu$m, we set the inner disk
radius to 5 stellar radii; even with this value, the near-IR fluxes are somewhat 
underestimated. We chose an outer envelope radius of 10000 AU; such a large 
envelope would encompass component B, too, but is necessary to yield enough
flux in the mm wavelength range. The fact that the model still underpredicts the 
flux at 1.3 mm suggests that the mm measurement probably includes some 
contribution from an envelope around component B.

The density is constrained by the depth of the silicate absorption feature,
which is rather shallow; it could be slightly larger, but then require a somewhat
smaller cavity semi-opening angle to decrease emission in the 8-25 $\mu$m range. 
The slope of the spectrum from 12 to 30 $\mu$m constrains both the opening angle 
of the cavity and the value of $R_c$. A larger $R_c$ would result in a broader SED, 
but at the same time decrease the flux levels in the long-wavelength part of the IRS 
spectrum. A smaller $R_c$ would generate a steep rise of the flux levels beyond 
12 $\mu$m. KCH93 modeled both IRAS 04181+2654 A and B with an
envelope with a higher reference density, $3.2 \times 10^{-14}$ g cm$^{-3}$, 
a somewhat larger $R_c$ (70 AU versus our 50 AU), and a comparable inclination angle. 
\citet{whitney97} used the same value for $R_c$ as we did in our model, but with 
almost double the values for reference density and inclination angle.

Overall, the model for IRAS 04181+2654 A is poorly constrained, given the lack 
of sub-mm data points and the uncertainty of the 1.3 mm measurement
due to the presence of component B. We did not model that component, 
since there are no sub-mm or mm measurements available that would constrain 
the long-wavelength part of the SED.

\subsubsection{04239+2436}
In the near-infrared, a symmetric reflection nebula surrounds this IRAS source, 
which is a binary separated by 0{\farcs}3 \citep{reipurth00}. 
The brighter component (at 2.2 $\mu$m) of this system drives a large, 
asymmetric Herbig-Haro jet, HH 300 \citep{reipurth00}; the asymmetry 
suggest that the inclination angle of this object is closer to face-on than 
edge-on. Maps of this Class I object at 450, 850, 1100, and 1300 $\mu$m 
show that it is a point source with some diffuse emission from circumstellar 
material \citep{chini01, eisner05}. Therefore, IRAS 04239+2436 is extended 
from infrared to millimeter wavelengths \citep[see also][]{park02}, typical 
for Class I objects.

The SED of IRAS 04239+2436 requires a model with relatively low density 
($\rho_1 = 1.3 \times 10^{-14}$ g cm$^{-3}$) and very small centrifugal
radius (R$_c=$ 10 AU) (Figure \ref{model_04239}). A higher density would 
increase the depth of the silicate feature, while a larger centrifugal radius would 
decrease the emission in the 10 to 40 $\mu$m region and result in a shallow 
silicate absorption feature. Increasing $\rho_1$ and $R_c$ up to about 
$2.0 \times 10^{-14}$ g cm$^{-3}$ and 20 AU, respectively, would still
result in good fits if the luminosity was increased by about 20\%.
 
Previous models by KCH93 yielded a larger density, 3.2 $\times 10^{-14}$ 
g cm$^{-3}$, and a centrifugal radius of 70 AU, while models by \citet{whitney97}
resulted in values for $\rho_1$ and $R_c$ very similar to ours. \citet{eisner05}
derived a reference density almost twice our value and a centrifugal radius of 30 AU;
more recent results by \citet{gramajo07} found the same centrifugal radius as
\citet{eisner05}, but a reference density similar to ours. 
Our inclination angle of 15\degr\ is smaller than the values derived by previous
modeling efforts, which range from 30\degr\ to 63\degr; a value of up to 
$i=30\degr$ would yield an acceptable fit, albeit with somewhat decreased emission 
from the near-IR to about 25 $\mu$m. The low inclination angle is well-constrained 
by the SED, and it is also consistent with the near-IR morphology and observed
asymmetry of the Herbig-Haro jet \citep[see, e.g.,][]{reipurth00}.

With the current values for $R_c$ and $\rho_1$, the model does not reproduce 
the shape of the silicate absorption feature very well, especially at the long-wavelength 
side; this could be an effect of dust and ice composition, since the overall width of the SED
and shape of the IRS spectrum beyond 15 $\mu$m suggest such a small
$R_c$ value. 
  
\subsubsection{04248+2612} 
This IRAS source is actually a subarcsecond binary whose components are 
roughly similar in brightness. Its IRS spectrum shows a silicate emission feature 
and some very weak ice absorption features. 
In the near-IR, it is surrounded by a bipolar reflection nebula with a somewhat 
complex morphology, but likely seen close to edge-on \citep{padgett99}.  
This object is also somewhat extended at 450 and 850 $\mu$m; the extension 
parallel to the bipolar nebulosity seen in the near-IR is likely tracing dust heated 
in an outflow \citep{young03}. \citet{white04} note that IRAS 04248+2612
is the lowest-mass protostar known that drives a molecular outflow and an HH
object \citep{moriarty92,gomez97}; they suggest it could actually be a substellar
Class I object. However, a recent spectral type determination of M4--M5 for this
object \citep{luhman06a} suggests that it will probably evolve into a very 
low-mass star.

Since this Class I object displays a silicate emission feature, and in the near-IR the
central binary is detected, its envelope likely has low densities. Our model suggests 
a reference density of $4.0 \times 10^{-15}$ g cm$^{-3}$, which is the lowest
$\rho_1$ value among the protostars modeled here, indicating an advanced stage
of evolution for this protostar. Our model further includes a relatively small 
centrifugal radius of 30 AU, a cavity semi-opening angle of 15\degr, and an 
inclination angle of 70\degr\ (see Figure \ref{model_04248}). 
IRAS 04248+2612 is similar to CoKu Tau/1, which is also seen close to edge-on 
through a low-density envelope. The short-wavelength emission is fitted very well; it is 
dominated by the stellar component, which agrees with the fact that the central binary 
is detected at near-IR wavelengths. The emission from far-IR to mm wavelengths is
underestimated; however, we found no set of parameters that could reproduce
both the mid-IR and long-wavelength parts of the SED. Thus, it is likely that the 
sub-mm and mm emission is generated by an additional component, probably by
dust in a very extended region of cold dust outside the infall zone
\citep{jayawardhana01}.

Compared to modeling results by KCH93, who obtained two models of IRAS
04248+2612, one with $\rho_1$=$1.0 \times 10^{-13}$ g cm$^{-3}$ and
$R_c$=300 AU, and one with $\rho_1$=$3.2 \times 10^{-14}$ g cm$^{-3}$ and
$R_c$=70 AU, our model has a clearly lower density (by a factor of 8) and a smaller 
centrifugal radius than the lower values of KCH93. Also models by \citet{whitney97}
resulted in a larger centrifugal radius (100 AU) and reference density ($1.5 \times 
10^{-14}$ g cm$^{-3}$) than our values; however, they derived a large inclination
angle, as we did with our model. Recent modeling efforts by \citet{stark06} yielded 
a model very similar to ours.
 
Based on our fit to the IRS spectrum, especially the presence of a silicate emission 
feature, we can exclude a much higher density, and the fit to the shape of the IRS spectrum 
between 15 and 25 $\mu$m also constrains the centrifugal radius to lie around 30 AU. 
Comparable model fits could be achieved with $\rho_1$ values in the (3.0-5.0) 
$\times 10^{-15}$ g cm$^{-3}$ range and $R_c$ values between 20 and 40 AU, 
with smaller reference densities accompanied by smaller centrifugal radii. Our inclination
angle of 70\degr\ not only matches two previously published models of this object
\citep{whitney97, stark06}, but is also in good agreement with the near-IR morphology 
observed by \citet{padgett99}.

\subsubsection{04264+2433} 
The IRS spectrum of this IRAS source shows a prominent silicate emission feature, 
but also possibly some ice features in the 5--8 $\mu$m wavelength region and an 
overall rising SED in the mid-IR. Therefore, it is probably a protostar surrounded by 
envelope material. This notion is further supported by the fact that IRAS 04264+2433 
is surrounded by a reflection nebulosity and likely emitting a small bipolar HH jet 
\citep{devine99}.

Similar to IRAS 04248+2612 and CoKu Tau/1, the detection of a silicate emission 
feature in this protostar indicates that the envelope has low densities and is thus
likely at an advanced evolutionary stage. We can reproduce the SED of IRAS 
04264+2433 fairly well with a model whose reference density is $5.0 \times 
10^{-15}$ g cm$^{-3}$; in addition, the shape of the IRS spectrum between 
about 13 and 20 $\mu$m constrains the value of the centrifugal radius, found to 
be 30 AU (see Figure \ref{model_04264}). The inclination angle of 87\degr\
and cavity semi-opening angle of 13\degr\ were fine-tuned to yield a good fit over
the 5-20 $\mu$m range. The model overestimates the emission in the near-IR 
somewhat, which can be mostly be attributed to the envelope component.

A previous model of IRAS 04264+2433 by KCH93 indicated a density twice as high
as our value, a centrifugal radius of only 10 AU, and an inclination angle of 
60\degr. Despite the differences, the earlier model already gauged the properties
of this object, namely a low-density, highly inclined envelope with a small disk. 
On the other hand, a model by \citet{whitney97} yielded different model parameters:
a centrifugal radius of 50 AU, a reference density of $3.75 \times 10^{-14}$ 
g cm$^{-3}$, and an inclination angle of 32\degr; only their cavity semi-opening
angle is similar to our value.
Our model would yield a comparable fit with a centrifugal radius of 20-40 AU, but
not with smaller or larger values for $\rho_1$.

In order to get enough flux from the far-IR to the mm wavelength range,
we had to increase the luminosity by a factor of 2 compared to the bolometric
luminosity of 0.37 L$_{\odot}$ measured by KH95. On the other hand,
integrating the measured fluxes of the SED yields a bolometric luminosity of 0.5 
L$_{\odot}$; given the high inclination angle of this object, the luminosity 
determined in this way is likely an underestimate of the true value. 

\subsubsection{04295+2251} 
This object, also known as L1536 IRS, is relatively faint; based on its bolometric
temperature of 270 K, it is classified as a Class I object, but its large sub-mm 
luminosity would suggest it is a Class 0 object \citep{young03}. Given that 
IRAS 04295+2251 seemed not to be extended in the near-IR or mm, 
\citet{park02} indicated that it could be an edge-on disk; however, data 
obtained by \citet{eisner05} shows that IRAS 04295+2251 is somewhat 
extended at 1.3 mm and also at 0.9 $\mu$m, where scattered light is detected. 
In addition, \citet{moriarty92} detected an outflow from this source, implying 
an early evolutionary stage for this object.

\begin{figure}
\includegraphics[angle=90, scale=0.6]{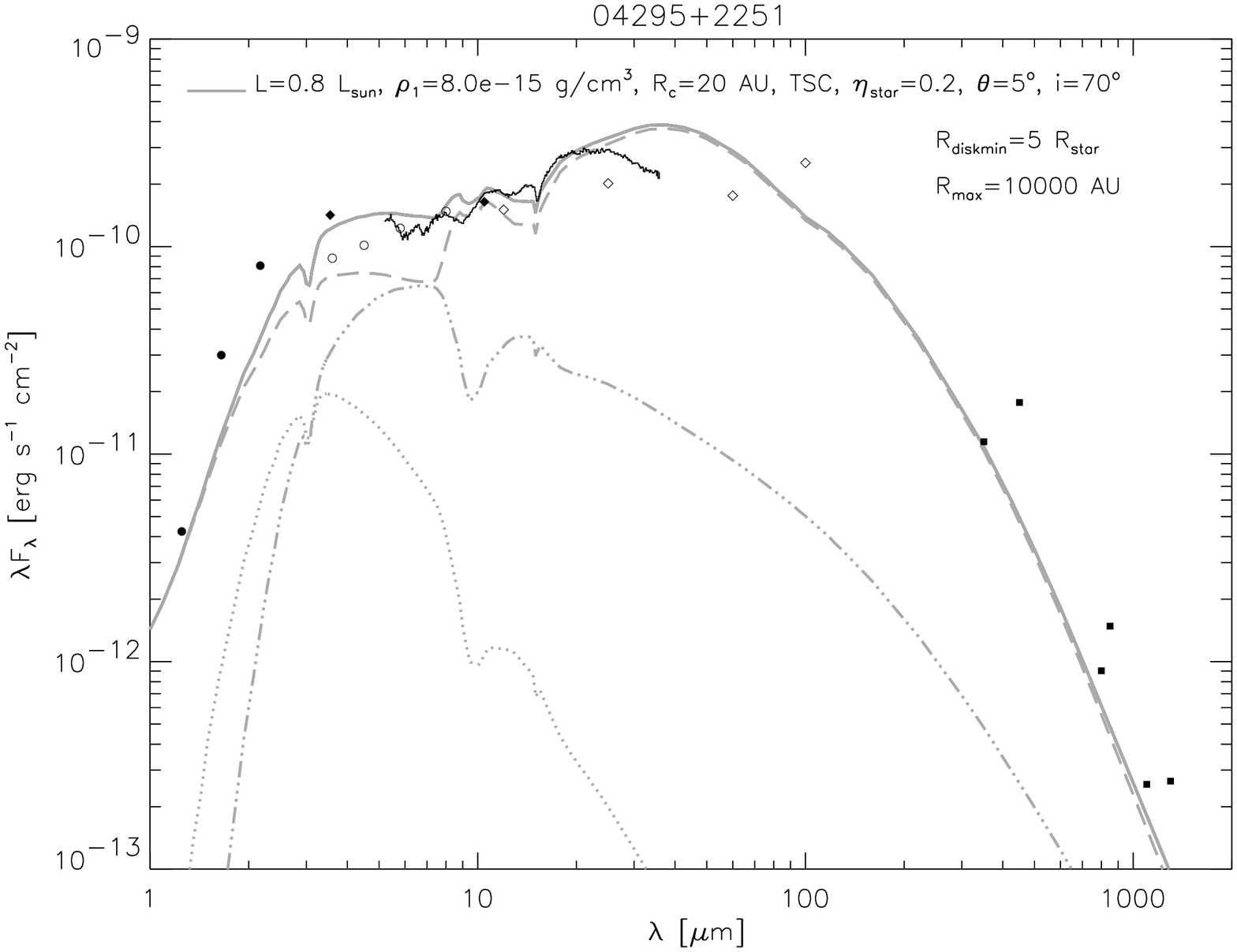}
\caption{IRS spectrum and photometric data of 04295+2251, and an envelope 
model fit with L=0.8 L$_{\odot}$, $\rho_1=8.0 \times 10^{-15}$ g cm$^{-3}$, 
$R_c$=20 AU, initial TSC density distribution, $\eta_{star}$=0.2, $\theta$=5\degr, 
i=70\degr, an inner disk radius of 5 stellar radii, and an outer envelope radius of 10000 AU.  
The CO$_2$ ice abundance was set to $1.0 \times 10^{-4}$. The gray lines represent 
the model components as in Figure \ref{model_04016}. \label{model_04295}}
\end{figure}

\begin{figure}
\includegraphics[angle=90, scale=0.6]{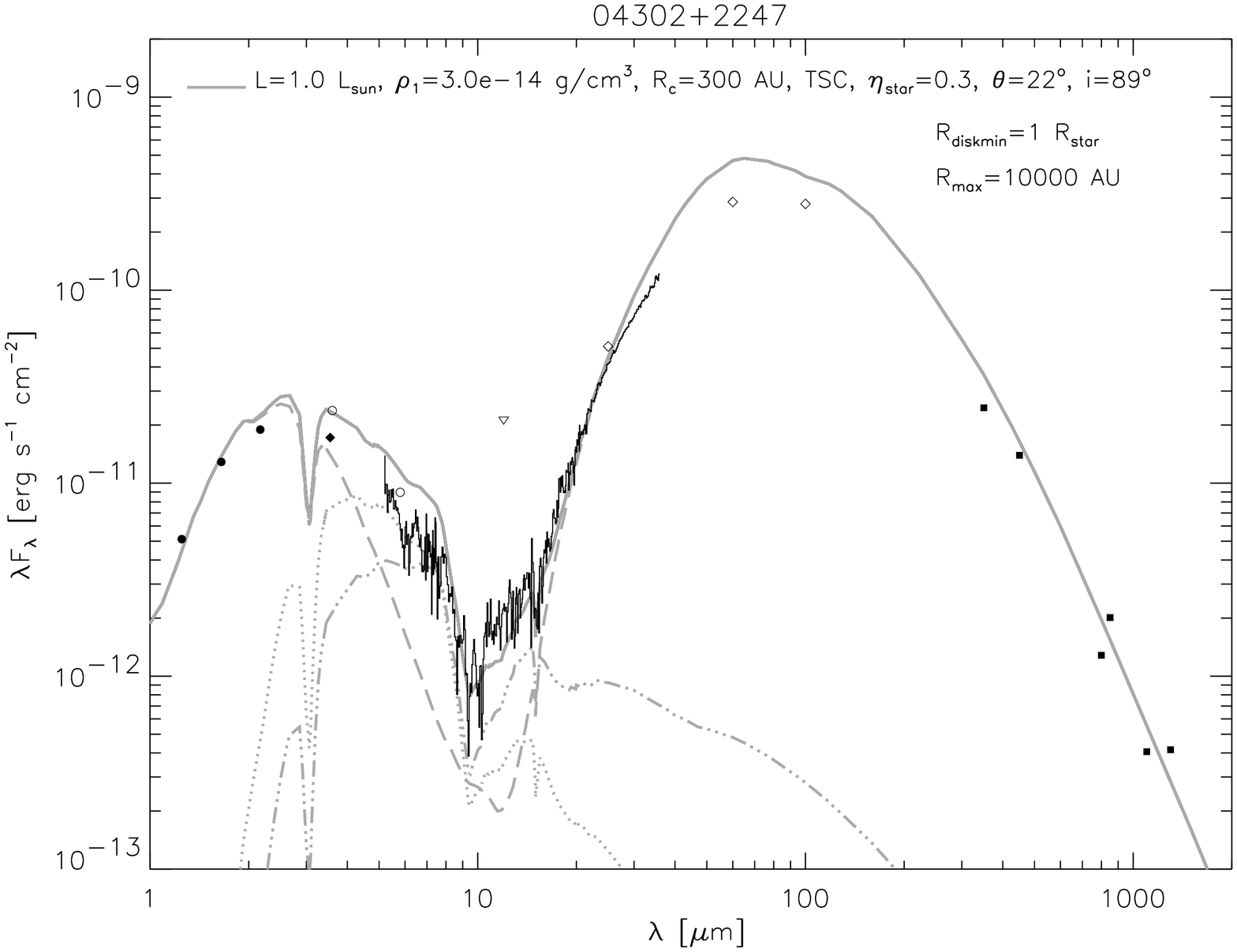}
\caption{IRS spectrum and photometric data of 04302+2247, and an envelope 
model fit with L=1.0 L$_{\odot}$, $\rho_1=3.0 \times 10^{-14}$ g cm$^{-3}$, 
$R_c$=300 AU, initial TSC density distribution, $\eta_{star}$=0.3, $\theta$=22\degr, 
i=89\degr, an inner disk radius of 1 stellar radius, and an outer envelope radius of 10000 AU. 
The CO$_2$ ice abundance was set to $4.0 \times 10^{-5}$. The gray lines represent 
the model components as in Figure \ref{model_04016}. \label{model_04302}}
\end{figure}

\begin{figure}
\epsscale{0.5}
\plotone{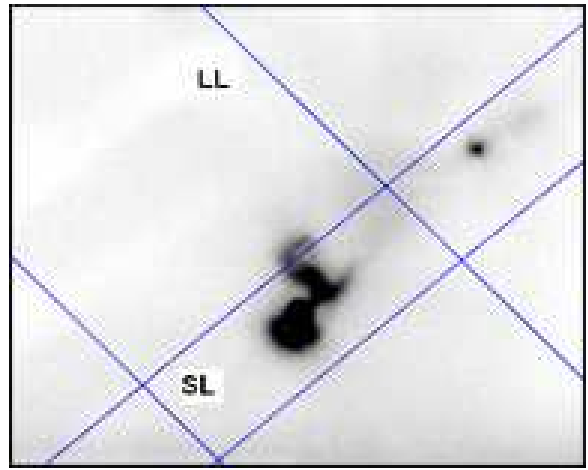}
\caption{The IRS SL and LL slit positions superposed on the NICMOS K-band (F205W) 
image of IRAS 04325+2402 by \citet{hartmann99}. The SL slit (3{\farcs}6 wide) 
runs from the lower left to upper right, while the wider LL slit extends from the upper left 
to the lower right. The full slit lengths are larger than the size of the image.\label{04325_slits}}
\end{figure}

\begin{figure}
\includegraphics[angle=90, scale=0.6]{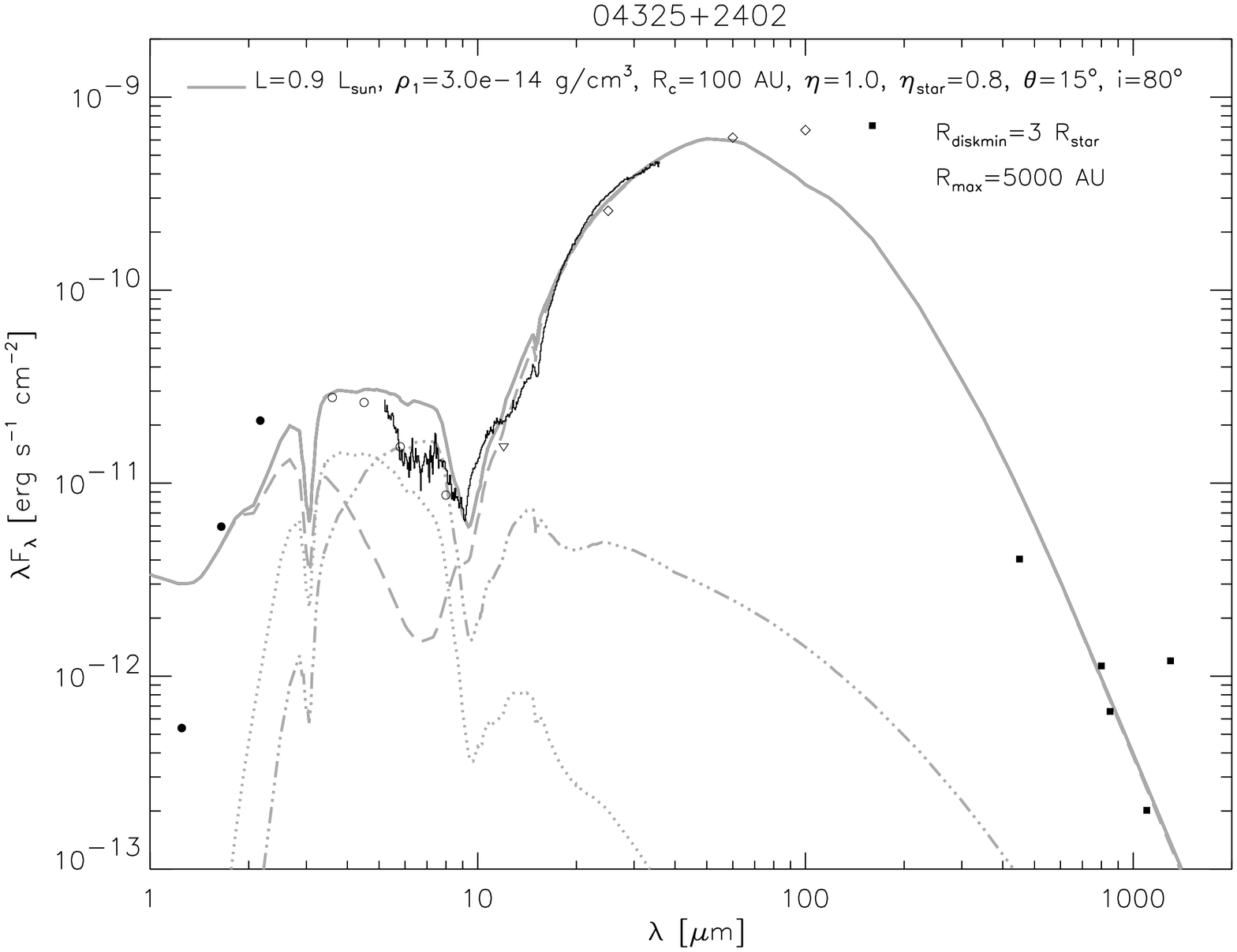}
\caption{IRS spectrum and photometric data of 04325+2402, and an envelope
model fit with L=0.9 L$_{\odot}$, $\rho_1=3.0 \times 10^{-14}$ g cm$^{-3}$,
$R_c$=100 AU, $\eta$=1.0, $\eta_{star}$=0.8, $\theta$=15\degr, i=80\degr, 
an inner disk radius of 3 stellar radii, and an outer envelope radius of 5000 AU. 
The CO$_2$ ice abundance was set to $4.0 \times 10^{-5}$. The gray lines 
represent the model components as in Figure \ref{model_04016}. \label{model_04325}}
\end{figure}

Our envelope model for IRAS 04295+2251 reproduces the general shape of 
the SED (see Figure \ref{model_04295}), but it does not represent a good 
fit to the 10 $\mu$m silicate feature and somewhat underestimates the 
near-IR and mm emission. However, at the longer wavelengths, our model 
reproduces those fluxes measured with smaller apertures, 10\arcsec-20\arcsec 
in size (see Table \ref{tab_longwave}); this suggests that a region of cold dust
outside of the infall region could be present.

A luminosity of 0.8 L$_{\odot}$ was necessary to obtain more emission 
especially at longer wavelengths, even though the source's bolometric luminosity 
is 0.5 L$_{\odot}$ (determined from integrating the fluxes of the SED). In the 
literature, L$_{bol}$ values vary from 0.44 L$_{\odot}$ (KH95) to 0.64 
L$_{\odot}$ \citep{myers87}. 
A small centrifugal radius of 20 AU was adopted to reproduce the steep rise of the SED 
past 15 $\mu$m; a value larger than 30 AU would result in a flatter SED in this wavelength 
region, while a smaller $R_c$ would decrease the near-IR emission and cause an even
larger overestimate of the far-IR flux. A reference density up to 30\% larger than our 
adopted value of $8.0 \times 10^{-15}$ g cm$^{-3}$ would still yield an acceptable 
fit, but require lower inclination angles (40\degr-50\degr) to generate only a shallow 
silicate absorption at 10 $\mu$m; these models would overestimate the emission 
between 3 and 10 $\mu$m and in the far-IR.

A previous model of IRAS 04295+2251 by KCH93 resulted in a larger centrifugal 
radius, 70 AU, and a smaller inclination angle, $i=$30\degr, but in a comparable 
reference density. We can exclude an inclination angle smaller than about 60\degr\
with the current model parameters, since it would increase the emission in the 
2-8 $\mu$m range and at far-IR wavelengths, where the model already 
overestimates the observed emission.
\citet{whitney97} modeled this object with very different model parameters,
including a larger centrifugal radius and reference density, and an inclination angle
of only 18\degr. \citet{eisner05} used the same value for $\rho_1$, 
$3.75 \times 10^{-14}$ g cm$^{-3}$, and a similar inclination angle as 
\citet{whitney97}, but an $R_c$ of 30 AU, which is comparable to our result.

The relatively low reference density, combined with the inclination of 70\degr\ to the 
line of sight and a cavity semi-opening angle of 5\degr, results in a self-absorbed silicate 
feature at 10 $\mu$m, with part absorption, part emission characteristics. However, our 
model does not reproduce the detailed shape of the observed 10 $\mu$m feature. 
This is likely due to the fact that the disk component used in the model does not include 
the optically thin disk atmosphere that would generate a 10 $\mu$m silicate emission feature, 
which would then be subject to absorption by dust in the envelope. In the current model, 
the disk component only contributes a silicate absorption feature (i.e., continuum extinguished 
by the envelope), while the dust in the envelope is responsible for the emission component 
at 10 $\mu$m (see the dash-dotted and long-dashed lines, respectively, in Figure 
\ref{model_04295}). Also the shape of the inner cavity could affect the depth of 
the silicate feature, for example by resulting in a hotter inner region for a given $\rho_1$.
It is likely that only a more complex model will yield a better fit, considering that this 
object has a relatively low-density envelope and compact morphology, but large 
sub-mm luminosity.

\subsubsection{04302+2247}
In the near-IR, this object is surrounded by two bright, symmetric nebulae separated 
by a dark lane \citep{padgett99}. Models of the very symmetric quadrupolar structure 
suggest that IRAS 04302+2247 is seen almost exactly at an inclination angle of 90\degr\ 
\citep{lucas97,wolf03}. The central source is likely obscured by a circumstellar disk, 
and the near-IR nebulosity, seen in scattered light, traces the walls of a bipolar cavity 
in the envelope \citep{lucas97, wolf03}.  Outflows oriented perpendicular to
the dark lane were detected from this object \citep{moriarty92,lucas98}. 
The edge-on orientation of IRAS 04302+2247 may explain why it is 
invisible in the optical and very faint in the mid-IR. The SED of this object peaks close 
to 100 $\mu$m and shows a pronounced 10 $\mu$m absorption feature.

The SED of IRAS 04302+2247 can be fitted with an envelope model with an
inclination angle of 89\degr, moderate density ($\rho_1 = 3.0 \times 
10^{-14}$ g cm$^{-3}$), and a centrifugal radius of 300 AU (Figure 
\ref{model_04302}). The deep silicate absorption feature and the steep 
slope of the IRS spectrum beyond 10 $\mu$m require a high inclination
angle, consistent with the near-IR images. A higher density would also 
increase the depth of the silicate feature, but it would decrease flux levels
over the 2--50 $\mu$m region, too.
The semi-opening angle of the cavity is relatively wide (22\degr), also 
roughly consistent with the results of \citet{padgett99}. The shape of the 
IRS spectrum beyond 12 $\mu$m, and the width of the SED, constrain the 
value of $R_c$, which has to be large. Previous models by KCH 93, which
had much fewer constraints, derived a centrifugal radius of only 70 AU, but
a density about a factor of 3 higher than in our model, and an inclination
angle of 60\degr. \citet{whitney97} found an even smaller value for $R_c$,
10 AU, and a reference density half as large as our value, but their results
for cavity and inclination angles agree with ours. Recent models by 
\citet{stark06} resulted in very similar parameters as for our model,
except for a somewhat smaller value for $\rho_1$.

The model emission is somewhat above the {\it IRAS} data points at 60 and 
100 $\mu$m; this is likely due to the fact that our adopted luminosity of 
1 L$_{\odot}$ is higher than the observed bolometric luminosity of 0.34 
L$_{\odot}$ (KH95), which was determined from fewer flux measurements
($\lambda \leq$ 100 $\mu$m). Our model indicates that our adopted luminosity 
might be slightly higher than the actual source luminosity.

\subsubsection{04325+2402}
This protostar is surrounded by a bipolar reflection nebula that is seen in scattered 
light in the near-IR and shows a complex morphology \citep{hartmann99}. 
It is likely a multiple system, consisting of a central, subarcsecond binary and a 
faint companion about 8\arcsec\ away; both the binary and the companion are 
surrounded by an accretion disk and are accreting from their envelopes 
\citep{hartmann99}. The subarcsecond binary, which instead could just be the 
result of scattering off a complex structure around a single star, is the likely source 
of the reflection nebula detected in the near-IR \citep{hartmann99}. 
This object is also associated with a molecular outflow \citep{moriarty92,hogerheijde98}.

Our IRS observations were centered on the central source (see Figure \ref{04325_slits}). 
Even though the SL slit was oriented such that it also included the companion about 8\arcsec\ 
away, this component did not enter our extraction window, and we also did not detect it 
separately; this is expected, since it is faint and likely seen edge-on \citep{hartmann99}, 
and even the spectrum of the bright central source is only about 40 mJy between 5 and 
10 $\mu$m, with a steeply rising continuum beyond. In LL, which is oriented close to 
perpendicular to SL, only the central source was included.

An envelope model that fits this object well requires a sheet-collapse model
($\eta=1.0$), moderate density ($\rho_1 = 3.0 \times 10^{-14}$ g cm$^{-3}$), 
a centrifugal radius of 100 AU, a cavity semi-opening angle of 15\degr, and an inclination 
angle of 80\degr\ (Figure \ref{model_04325}). Choosing a sheet-collapse model with
$\eta=1.0$ was necessary to fit the IRS spectrum in the 15-30 $\mu$m range.
The inclination angle is larger than the value of $\sim$ 60\degr\ inferred from the 
orientation of the outflow by \citet{hogerheijde98}, but the shape of the IRS spectrum
requires a high-inclination model. 
A previous model by KCH93 also determined $i=60$\degr, but a centrifugal
radius and a reference density three times as large as the values derived here.
Models by \citet{whitney97} resulted in an $R_c$ value of only 50 AU, but in
a reference density only somewhat larger than our value, and very similar values
for cavity and inclination angles (11\degr\ and 81\degr, respectively). 
Compared to previous models, the IRS spectrum provides better constraints;
a larger $R_c$ would result in a flatter mid-IR SED, while a larger density would 
generate a deeper silicate absorption feature.

The silicate feature is not well fitted with this model; it seems to be in absorption on 
the short-wavelength side, but with an emission component on the long-wavelength side. 
The binary nature of IRAS 04325+2402 could account for the peculiar shape of the 
silicate feature, with one component possibly responsible for some silicate emission.

\subsubsection{04361+2547}
This Class I object, also known as TMR 1, is deeply embedded in its dense 
environment and surrounded by a large reflection nebula, about 30\arcsec\ 
in size \citep{tamura91}. It is a close binary with a separation of 0{\farcs}31 
\citep{terebey98}; the ``companion'' detected at a projected separation of 
10\arcsec\ from this protostar, at the end of a nebulosity extending from the 
source to the southeast, is likely a background star \citep{terebey00}. 
IRAS 04361+2547 is extended in the near-IR, mid-IR and mm continuum 
\citep{motte01, park02,luhman06b}. An outflow has been detected from 
this object, but its morphology is not clearly bipolar, suggesting an inclination 
angle of about 60\degr\ \citep{hogerheijde98}.

\begin{figure}
\includegraphics[angle=90, scale=0.6]{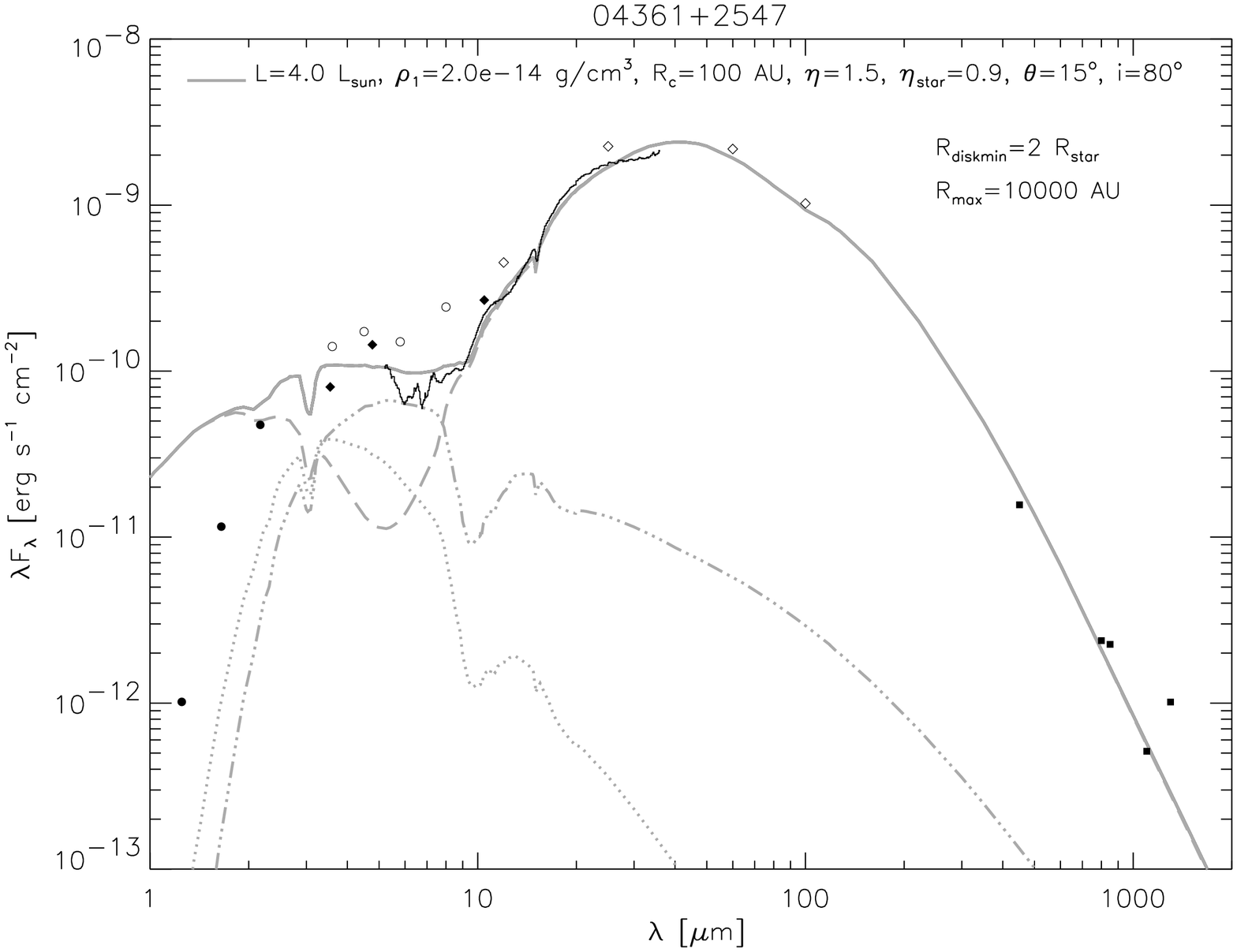}
\caption{IRS spectrum and photometric data of 04361+2547, and an envelope
model fit with L=4.0 L$_{\odot}$, $\rho_1=2.0 \times 10^{-14}$ g cm$^{-3}$,
$R_c$=100 AU, $\eta=$1.5, $\eta_{star}$=0.9, $\theta$=15\degr, i=80\degr, 
an inner disk radius of 2 stellar radii, and an outer envelope radius of 10000 AU. 
The CO$_2$ ice abundance was set to $7.0 \times 10^{-5}$. The gray lines 
represent the model components as in Figure \ref{model_04016}. \label{model_04361}}
\end{figure}

\begin{figure}
\includegraphics[angle=90, scale=0.6]{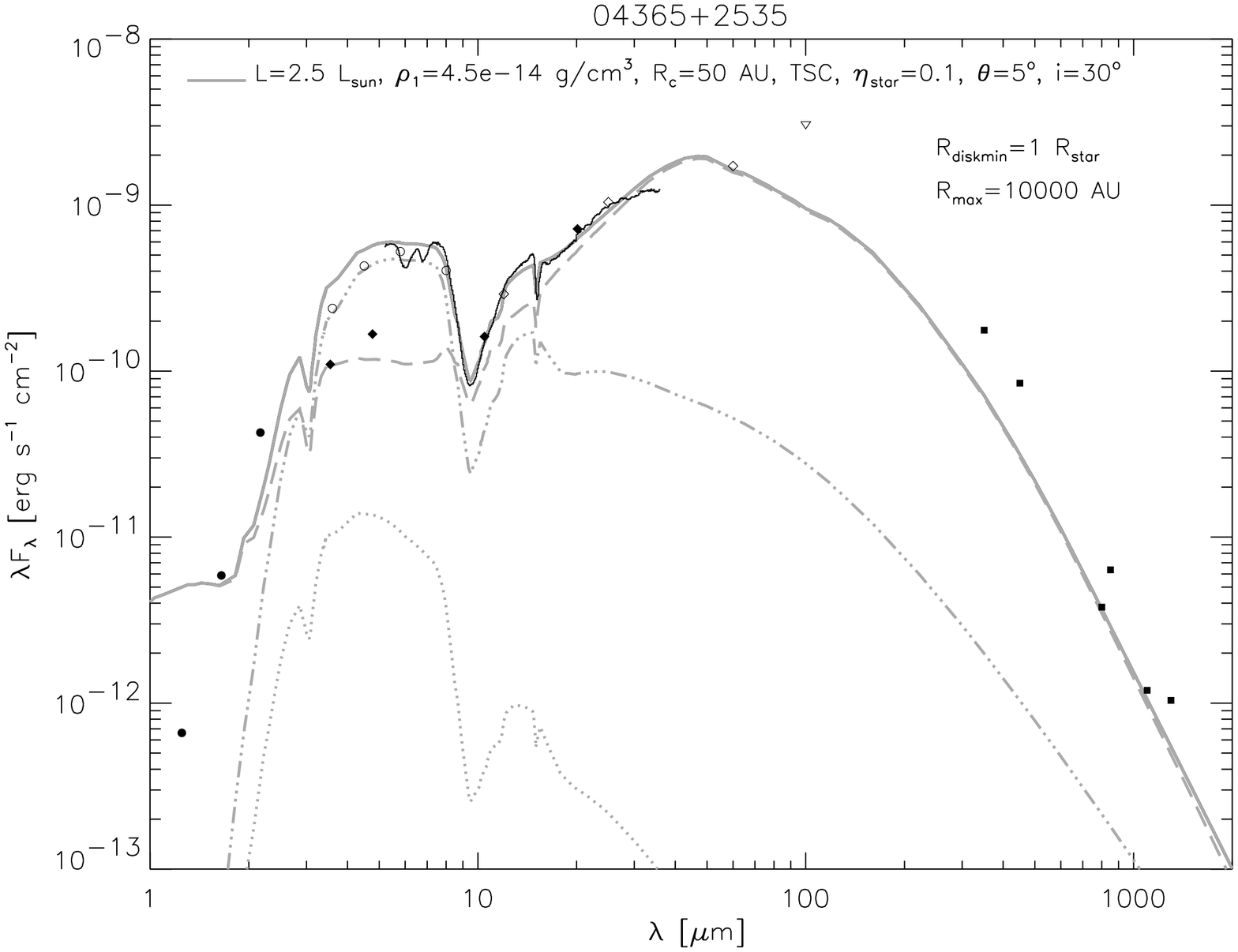}
\caption{IRS spectrum and photometric data of 04365+2535, and an envelope
model fit with L=2.5 L$_{\odot}$, $\rho_1=4.5 \times 10^{-14}$ g cm$^{-3}$,
$R_c$=50 AU, initial TSC density distribution, $\eta_{star}$=0.1, $\theta$=5\degr,
i=30\degr, an inner disk radius of 1 stellar radius, and an outer envelope radius of 10000 AU. 
The CO$_2$ ice abundance was set to $7.0 \times 10^{-5}$. The gray lines represent 
the model components as in Figure \ref{model_04016}. \label{model_04365}}
\end{figure}

The steeply rising SED of IRAS 04361+2547 indicates that this object is seen at a 
high inclination angle (Figure \ref{model_04361}). A good envelope model fit requires 
a sheet-collapse model ($\eta=1.5$), an inclination angle of 80\degr, a cavity 
semi-opening angle of 15\degr, an $R_c$ value of 100 AU, and a fairly typical 
density ($\rho_1 = 2.0 \times 10^{-14}$ g cm$^{-3}$). 
Modeling by KCH93 resulted in a reference density of 3.2 $\times 10^{-14}$ g 
cm$^{-3}$ and a centrifugal radius of only 10 AU. Also, while our model suggests 
a close to edge-on orientation, KCH93 derived an inclination angle of only 30\degr. 
Later models by \citet{whitney97} and \citet{gramajo07} confirm our result of 
a large inclination angle (and also our cavity opening angle), but they derive centrifugal 
radii about half as large as our value, and generally a somewhat larger reference 
density.

Our adopted reference density is required to fit the SED beyond 50 $\mu$m, as 
well as to produce a silicate feature at 10 $\mu$m that is mostly dominated by 
an emission component generated in the envelope (long-dashed line in
Figure \ref{model_04361}). The $R_c$ value of 100 AU was chosen to reproduce 
the SED shape from about 15 to 50 $\mu$m. A larger $R_c$ would result in a 
flatter SED, while a smaller value would result in more mid-IR flux and a pronounced 
silicate absorption feature. Choosing $\eta_{star}=0.9$ and $\theta=$ 15\degr, 
combined with the high inclination angle, depresses the emission below 10 $\mu$m. 
Our model does not reproduce the deep ice features at 6.0 and 6.8 $\mu$m, but it
traces the continuum level between 5 and 8 $\mu$m (where emission from the
disk dominates). We note that the IRS fluxes are lower than the IRAC data points;
this is likely due to the extended emission detected in IRAC \citep{luhman06b} 
and thus an aperture effect.
Below 2 $\mu$m, the near-IR scattered light component from 
the envelope is likely overestimated.
We note that even though the luminosity of our model, 4.0 L$_{\odot}$, is larger 
than the measured bolometric luminosity of 2.5 L$_{\odot}$ (see Table 
\ref{tab_lum_incl}), it is necessary to yield enough flux at all wavelengths. 

\subsubsection{04365+2535}
This IRAS source is also known as TMC 1A; it is optically invisible and thus 
deeply embedded. It is surrounded by a cometary nebulosity in the near-infrared 
\citep{tamura91, tamura96}, and it is extended both in the near-IR and at mm 
wavelengths \citep{park02}. Both \citet{chandler96} and \citet{tamura96}
measured a bipolar outflow which is somewhat conical in shape due to the
blueshifted lobe being more prominent than the redshifted one. The near-IR
nebulosity is likely light from the central source scattered by dust in the outflows
\citep{tamura96}. Also, dust continuum measured at 790 $\mu$m and 1.1 
mm is aligned with the outflow direction \citep{chandler98}. \citet{chandler96} 
concluded that the inclination angle of this object lies between 40\degr and 
68\degr, with a cavity of semi-opening angle between 15\degr\ and 21\degr.

The SED of 04365+2535 is fitted by an envelope model with relatively high density 
($\rho_1 = 4.5 \times 10^{-14}$ g cm$^{-3}$), a centrifugal radius of 50 AU, 
a small cavity, and an inclination angle of 30\degr\ (Figure \ref{model_04365}). 
Compared to models by KCH93, both our reference density and centrifugal
radius are smaller, by factors of 0.45 and 0.17, respectively, and our model
suggests a more face-on orientation as opposed to the more highly inclined
model ($i=60$\degr) of KCH93 and the inclination angle determined from the
outflows \citep{chandler96}. More recent models by \citet{whitney97} 
and \citet{gramajo07} resulted in a large inclination angle (70\degr-80\degr), 
but in values for $R_c$ and $\rho_1$ very comparable to ours.

The high density is required to fit the deep silicate absorption feature, as well as to 
shift the peak of the SED to longer wavelengths. Our model does not fit the sub-mm
and mm fluxes measured with a large aperture (40\arcsec-60\arcsec), but matches
the measurements done with apertures $\lesssim$ 20\arcsec (see Table
\ref{tab_longwave}). This suggest the presence of some extended, cold dust
beyond the infalling envelope.  
A centrifugal radius of 50 AU was chosen to fit the emission from the silicate absorption 
feature out to about 50 $\mu$m; a model with larger values for $\rho_1$ and $R_c$ 
($\sim 6.0 \times 10^{-14}$ g cm$^{-3}$ and 70 AU, respectively) would also yield 
a good fit, but underestimate the emission in the 20-30 $\mu$m range somewhat. An 
even larger centrifugal radius would decrease the emission from 15 to 40 $\mu$m.
The inclination angle of our envelope model is smaller than the value suggested by 
\citet{chandler96} and by previous models; however, we can exclude a larger 
inclination angle, since it would result in a steeper SED from 15 to 60 $\mu$m, 
as well as less flux in the 2 to 10 $\mu$m region. We can also exclude a cavity 
semi-opening angle larger than about 5\degr, since it would decrease the flux in 
the 5-15 $\mu$m wavelength range.

\subsubsection{04368+2557}
This object has a bolometric temperature of less than 70 K, thus placing it in the 
Class 0 domain, the earlier, deeply embedded protostellar stage \citep{chen95,
motte01}. It is very faint and extended in the near-IR \citep{park02}; no point 
source is detected in 2MASS. The $K$-band data point in our SED plot is from
\citet{whitney97}, who only detected some faint reflection nebulosities at the
position of the IRAS source. 
Since the central source is not detected in the near-IR, \citet{tamura96} 
conclude that this object might be seen edge-on, obscured by a disk.

Our IRS observations were centered on the position determined by interferometric 
observations at 2.7 mm by \citet{ohashi97a}, which also coincides with the 
position of the source in high-resolution 7 mm VLA maps by \citet{loinard02}. 
The latter authors resolved this IRAS source into two components separated
by 0{\farcs}17 and suggested that it is a binary in which one component is 
surrounded by an almost edge-on disk. 
In addition, this object might have a faint companion located 20\arcsec\ to the 
northwest of the main source, but it was only detected at 800 $\mu$m 
\citep{fuller96} and not confirmed at 1.3 mm \citep{motte01} and 2.7 mm 
\citep{ohashi97a}.

IRAS 04368+2557 drives molecular outflows, which are located symmetrically 
around the central source and also perpendicular to the elongated source detected 
by \citet{loinard02}. \citet{hogerheijde98} inferred an inclination angle $i>$ 65\degr\ 
from observations of the outflows. \citet{ohashi97a} observed this object at high 
resolution in the millimeter wavelength range and concluded that it is surrounded by 
an elongated envelope $\sim$ 2000 AU in radius with large bipolar cavities, seen 
edge-on.

\begin{figure}
\includegraphics[angle=90, scale=0.6]{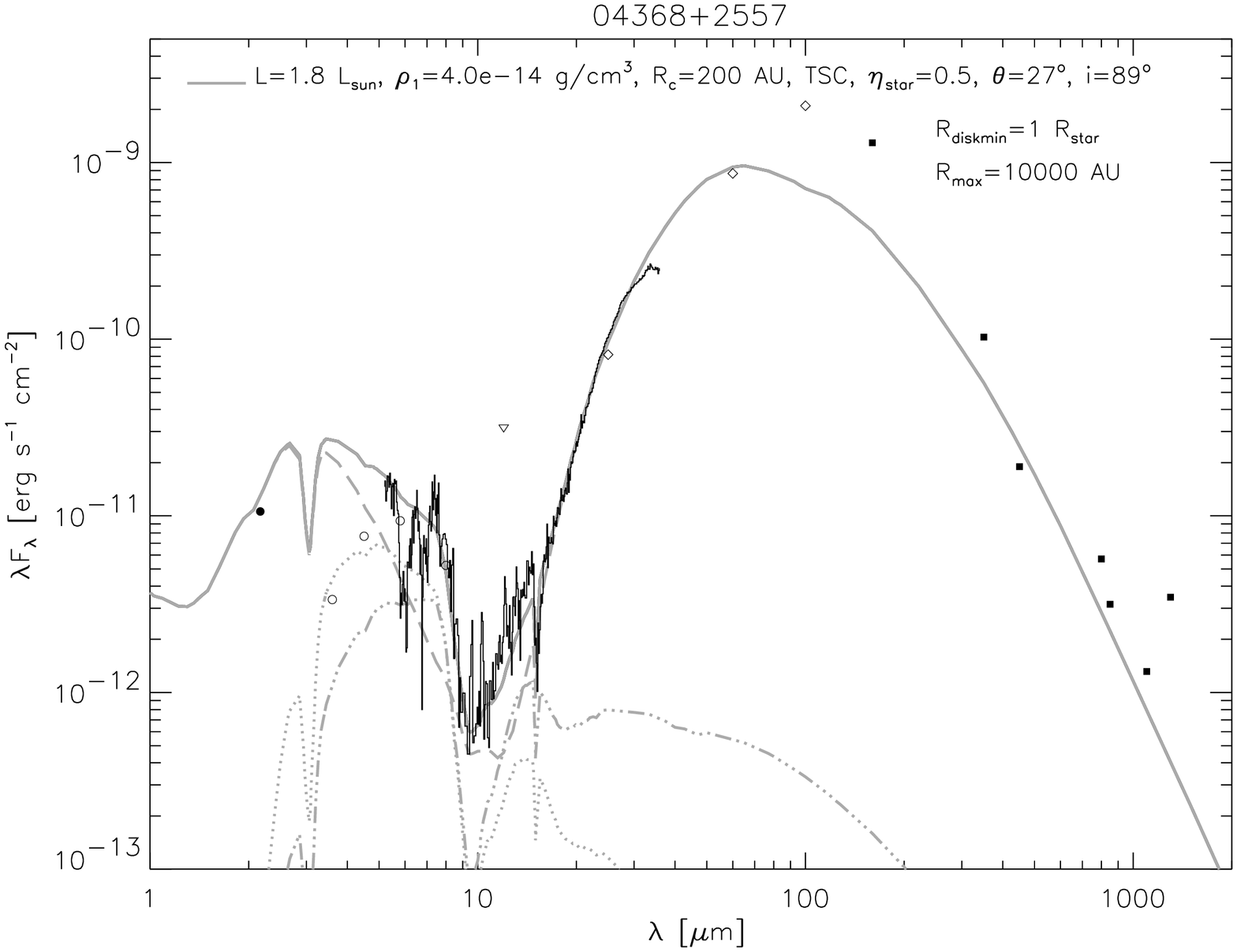}
\caption{IRS spectrum and photometric data of 04368+2557, and an envelope model 
fit with L=1.8 L$_{\odot}$, $\rho_1=4.0 \times 10^{-14}$ g cm$^{-3}$, 
$R_c$=200 AU, initial TSC density distribution, $\eta_{star}$=0.5, $\theta$=27\degr, 
i=89\degr, an inner disk radius of 1 stellar radius, and an outer envelope radius of 10000 AU. 
The CO$_2$ ice abundance was set to $1.0 \times 10^{-4}$. The gray lines represent 
the model components as in Figure \ref{model_04016}. \label{model_04368}}
\end{figure}

\begin{figure}
\includegraphics[angle=90, scale=0.6]{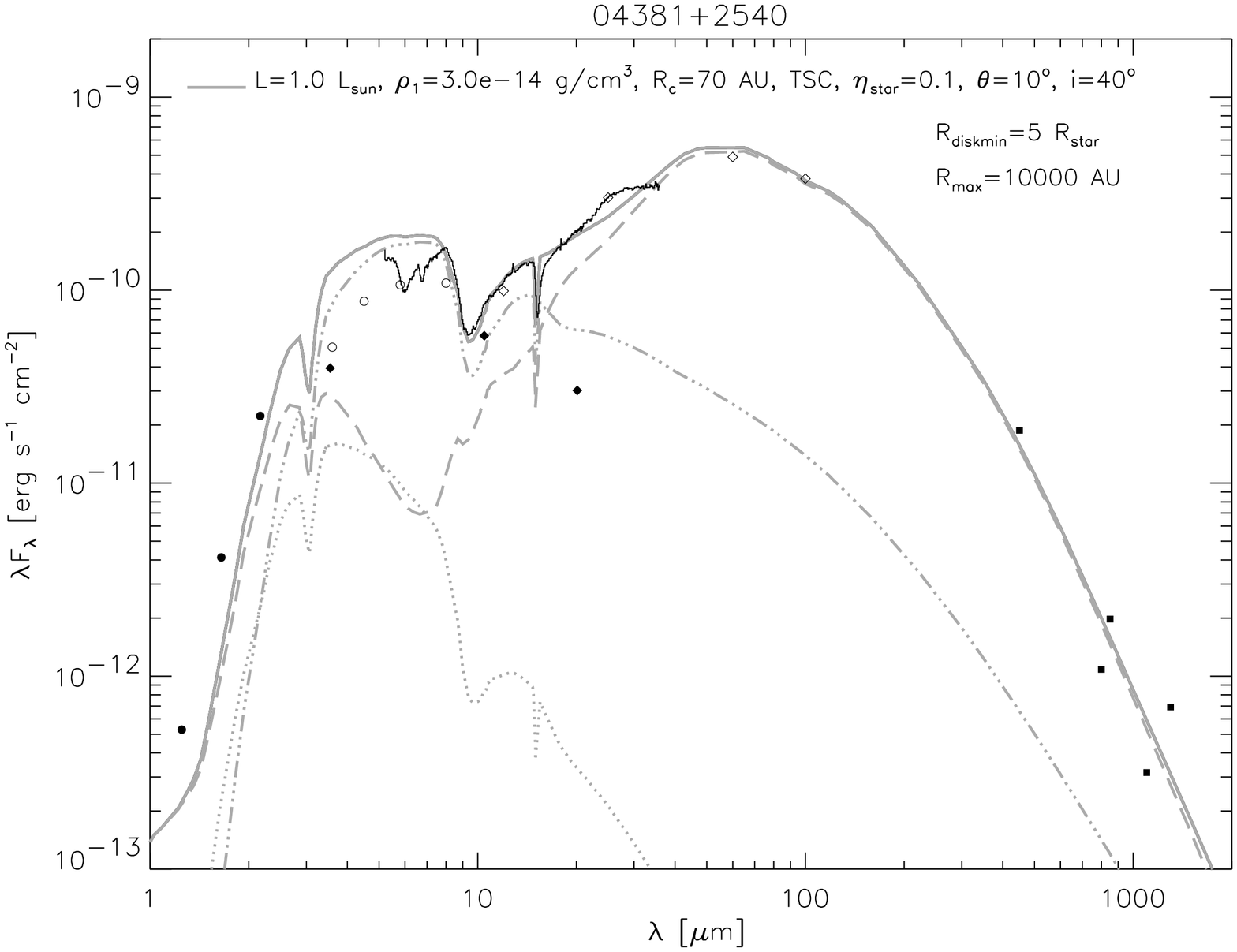}
\caption{IRS spectrum and photometric data of 04381+2540, and an envelope
model fit with L=1.0 L$_{\odot}$, $\rho_1=3.0 \times 10^{-14}$ g cm$^{-3}$,
$R_c$=70 AU, initial TSC density distribution, $\eta_{star}$=0.1, $\theta$=10\degr, 
i=40\degr, an inner disk radius of 5 stellar radii, and an outer envelope radius of 10000 AU. 
The CO$_2$ ice abundance was set to $1.2 \times 10^{-4}$. The gray lines represent
the model components as in Figure \ref{model_04016}. \label{model_04381}}
\end{figure}

The SED of IRAS 04368+2557 (Figure \ref{model_04368}) is similar to that of 
IRAS 04302+2247 (Figure \ref{model_04302}); both objects have a deep silicate 
absorption feature and SEDs that peak between 60 and 70 $\mu$m. Both are seen 
close to edge-on, but the reference density of  the IRAS 04368+2557 model fit is higher 
($\rho_1 = 4.0 \times 10^{-14}$ g cm$^{-3}$), and the centrifugal radius smaller 
($R_c$ = 200 AU). On the other hand, about 20\% changes in $\rho_1$ and $R_c$ 
yield comparable model fits, as long as the cavity semi-opening angle lies between 
20\degr\ and 30\degr, and the source is seen edge-on. We can exclude a high
density as derived by KCH93 (3.2 $\times 10^{-13}$ g cm$^{-3}$), but agree
reasonably well with their result of $R_c$ (300 AU) and inclination angle 
(60\degr-90\degr).

In the near-IR, our model reproduces the $K$-band measurement, but seems to 
overestimate the 3.6 and 4.5 $\mu$m emission, as measured by IRAC 
\citep{hartmann05}. However, this object is extended in IRAC images, and the 
photometry from the 6\arcsec\ aperture shown in Figure \ref{model_04368} 
does not capture all of the strong extended emission \citep[see also][]{hartmann05}.

As with IRAS 04302+2247, the values for $\rho_1$ and $R_c$ are fairly well 
constrained by the depth of the silicate absorption feature, the shape of the IRS 
spectrum beyond 10 $\mu$m, and the width and peak of the SED (the 100,
160, and 1300 $\mu$m measurements are likely overestimates due to aperture 
effects). The similarity between the two IRAS sources suggests that some 
Class I objects, seen at high inclination angles, could resemble Class 0 objects.

\subsubsection{04381+2540}
This IRAS source is also known as TMC 1; it is invisible in the optical and thus
deeply embedded \citep[e.g.,][]{moriarty92}. It is surrounded by a 
near-infrared nebulosity seen in scattered light \citep{tamura91,terebey06}. 
The conical shape of this near-infrared emission indicates that it likely traces an 
evacuated outflow cavity; the CO molecular outflow has a semi-opening angle 
between 13\degr\ and 19\degr\ and an inclination angle between 40\degr\ 
and 70\degr\ \citep{chandler96}.

Dust emission oriented perpendicular to outflows was detected at 790 $\mu$m 
\citep{chandler98}. \citet{young03} measured extended 450 and 850 $\mu$m 
emission from this object; based on the source's properties and model calculations, 
they confirmed that this object is a Class I source. From radial intensity profile fits 
to the 1.3 mm map, \citet{motte01} derived an outer envelope radius of 3900 AU.
Recently, \citet{apai05} found a 0{\farcs}6 companion to IRAS 04381+2540;
they suggest that this companion might be a young brown dwarf.

A good model fit for IRAS 04381+2540 (see Figure \ref{model_04381})
requires a somewhat higher luminosity than its bolometric luminosity of 0.7 L$_{\odot}$ 
(see Table \ref{tab_lum_incl}) to be able to reproduce the long-wavelength 
emission. By increasing the inner disk radius to 5 stellar radii, the emission in the 
near-IR and mid-IR out to 8 $\mu$m is decreased, resulting in a better fit. The 
centrifugal radius of 70 AU and the reference density of $3.0 \times 10^{-14}$ 
g cm$^{-3}$ are well constrained by the depth of the silicate absorption feature 
and the shape of the IRS spectrum from 15 to 30 $\mu$m. Decreasing $R_c$
to 60 AU and the inclination angle to 35\degr\ yields a nearly identical fit, but
we can exclude larger changes in these parameters, as well for the reference
density.

Previous modeling by KCH93 yielded a value of $1.0 \times 10^{-13}$ g cm$^{-3}$ 
for $\rho_1$ and 300 AU for $R_c$, and an inclination angle of 30\degr. However,
better-constrained models by \citet{whitney97} resulted in values for $\rho_1$ 
and $R_c$ similar to ours. \citet{eisner05} derived a smaller centrifugal radius
(30 AU) and also a larger reference density ( $6.75 \times 10^{-14}$ g cm$^{-3}$).

Our results for cavity semi-opening angle ($\theta=$10\degr) and inclination 
($i=$40\degr) are at the lower limits of the range suggested by 
\citet{chandler96}. Our value for $\theta$ agrees well with the modeling result 
of \citet{whitney97}, but our inclination angle is smaller; on the other hand, this latter
quantity is consistent with the result of \citet{terebey06}, but not with their cavity
semi-opening angle of 40\degr. Higher values for both parameters can be 
excluded based on the constraints placed by our IRS spectrum on the envelope models. 
Finally, we note that we adopted an outer envelope radius of 10000 AU, which is larger
than the value derived by \citet{motte01}, but necessary to fit the long-wavelength 
part of the SED.

\subsubsection{04489+3042}
Based on the M6 spectral type of IRAS 04489+3042, \citet{white04} indicated that
it could be substellar, a Class I brown dwarf. However, recent spectral type 
determinations by \citet{luhman06a} suggest that its spectral type is M3--M4, 
thus placing it into the low-mass star regime.
IRAS 04489+3042 is not extended in the near-IR or mm \citep{park02, motte01},
and no outflow has been detected \citep{gomez97}. \citet{park02} suggest it 
could be in transition between Class I and II stage, probably seen close to pole-on.

\begin{figure}
\includegraphics[angle=90, scale=0.6]{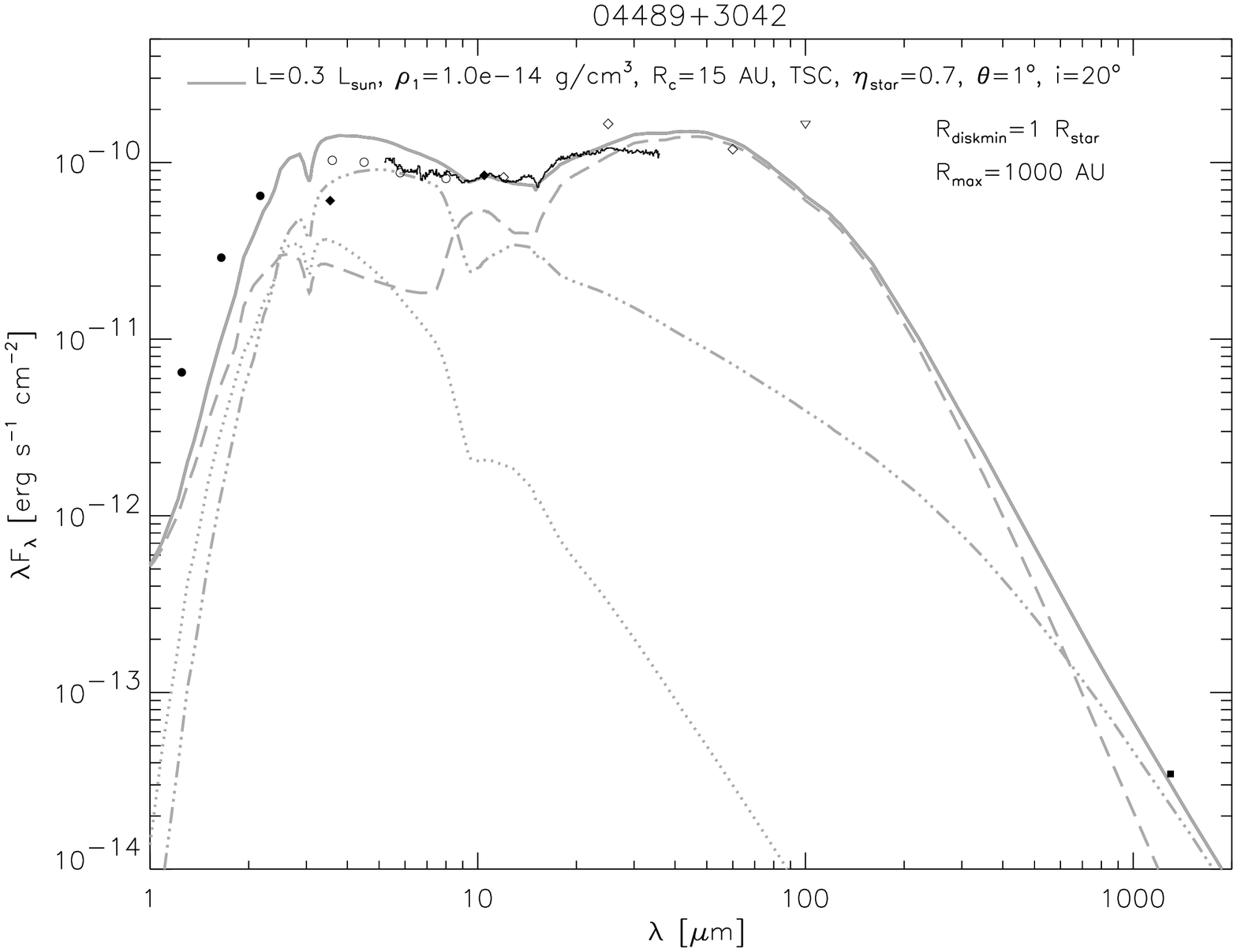}
\caption{IRS spectrum and photometric data of 04489+3042, and an envelope
model fit with L=0.3 L$_{\odot}$, $\rho_1=1.0 \times 10^{-14}$ g cm$^{-3}$,
$R_c$=15 AU, initial TSC density distribution, $\eta_{star}$=0.7, $\theta$=1\degr, 
i=20\degr, an inner disk radius of 1 stellar radius, and an outer envelope radius of 1000 AU. 
The CO$_2$ ice abundance was set to $4.0 \times 10^{-5}$. The gray lines represent 
the model components as in Figure \ref{model_04016}. \label{model_04489}}
\end{figure}

Our preliminary envelope model fit for this object (Figure \ref{model_04489})
suggests a low density ($\rho_1 = 1.0 \times 10^{-14}$ g cm$^{-3}$), 
a small centrifugal radius ($R_c$ = 15 AU), a very small cavity, and a low inclination.
Models by KCH93 yielded the same reference density as our model and a similar low 
inclination angle, but a centrifugal radius of 70 AU. On the other hand, modeling efforts
by \citet{whitney97} resulted in larger values for all these three parameters.
We can exclude a centrifugal radius larger than about 25 AU, since it would decrease 
the flux over the 10-60 $\mu$m spectral range. The low density and inclination angle 
of our model are mainly set by the very weak silicate absorption feature and roughly 
flat SED from about 5 to 15 $\mu$m.
The model does not yield a good fit of the details of the IRS spectrum below 
about 15 $\mu$m, but the overall shape of the spectrum is reproduced reasonably
well. To fit the data point at 1.3 mm with the set of model parameters described above, 
we adopted an outer envelope radius of only 1000 AU; the small envelope contributes
only little emission in the mm, where the emission from the disk dominates (see the 
dash-dotted line in Figure \ref{model_04489}). 
Since there are no sub-mm measurements available, and the IRS spectrum is relatively 
flat with no strong silicate feature, the current model parameters are only weakly constrained. 

\subsubsection{CoKu Tau/1} 
Coku Tau/1 was identified as a Class II object by KH95, but it is surrounded by 
filamentary reflection nebulae detected in the near-IR and suggested to represent 
the walls of outflow cavities \citep{padgett99}. It is seen close 
to edge-on, and, like IRAS 04248+2612, is a subarcsecond binary \citep{padgett99}.
Similar to IRAS 04248+2612 and 04264+2433, CoKu Tau/1 has a silicate emission 
feature and an infrared SED that is rising from 3 to about 20 $\mu$m (see Figure
\ref{ClassI_SED}).

The appearance of CoKu Tau/1 in the near-IR suggests that it is surrounded by
a low-density envelope and an edge-on disk. It is likely a more evolved protostar,
at a stage when the envelope has begun dissipating.
The binary and the inner disk are extinguished by the outer disk, but since in the 
near-IR the optical depth of the outer regions of the circumbinary disk decreases, 
the two stars at the center are seen at near-IR wavelengths. As in the optical, 
there is likely also some stellar light scattered by the envelope contributing to the 
near-IR emission.

\begin{figure}
\includegraphics[angle=90, scale=0.6]{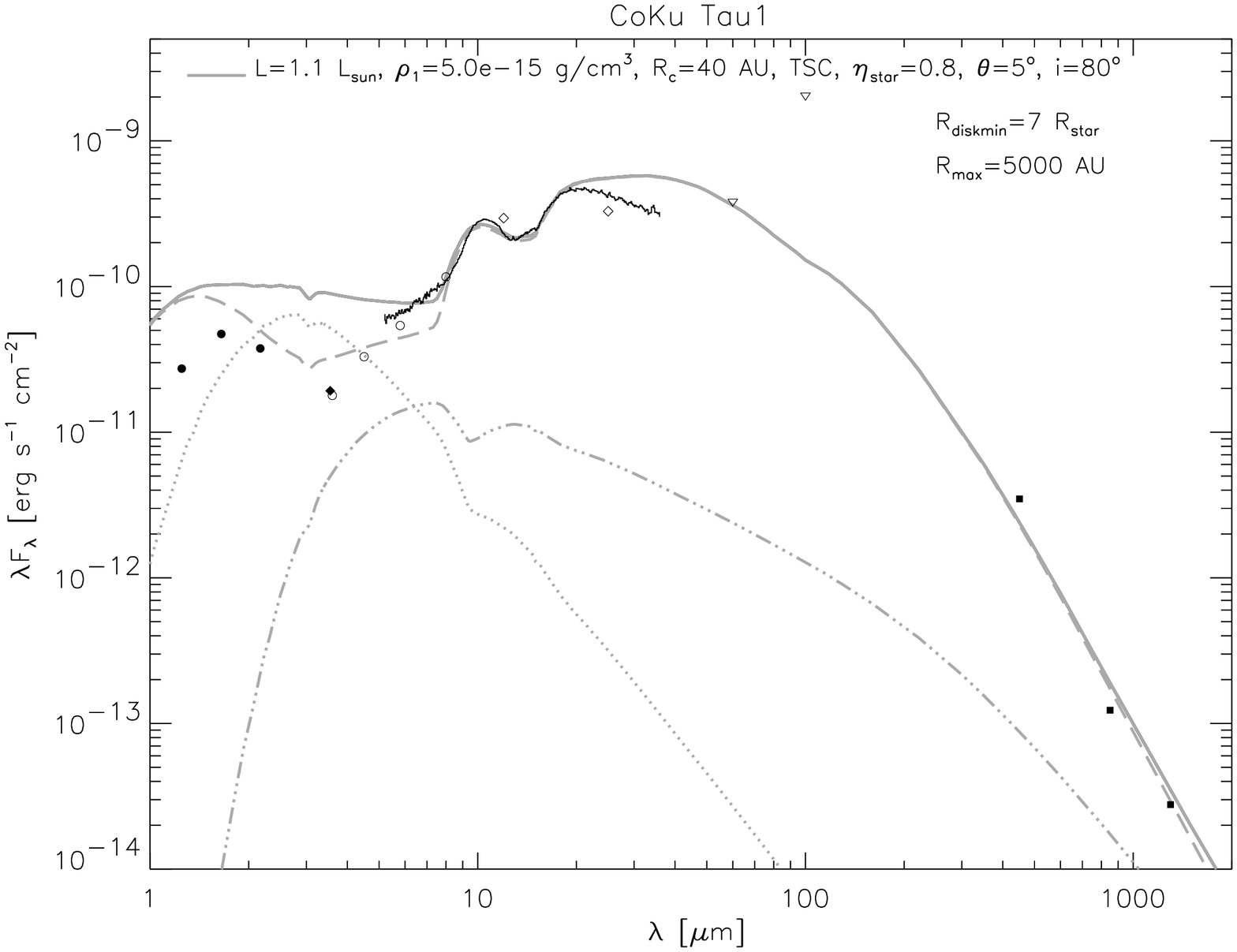}
\caption{IRS spectrum and photometric data of CoKu Tau/1, and an envelope
model fit with L=1.1 L$_{\odot}$, $\rho_1=5.0 \times 10^{-15}$ g cm$^{-3}$, 
$R_c$=40 AU, initial TSC density distribution, $\eta_{star}$=0.8, $\theta$=5\degr, 
i=80\degr, an inner disk radius of 7 stellar radii, and an outer envelope radius of 5000 AU. 
The CO$_2$ ice abundance was set to $5.0 \times 10^{-5}$. The gray lines represent
the model components as in Figure \ref{model_04016}. \label{model_CoKu_Tau1}}
\end{figure}

\begin{figure}
\includegraphics[angle=90, scale=0.58]{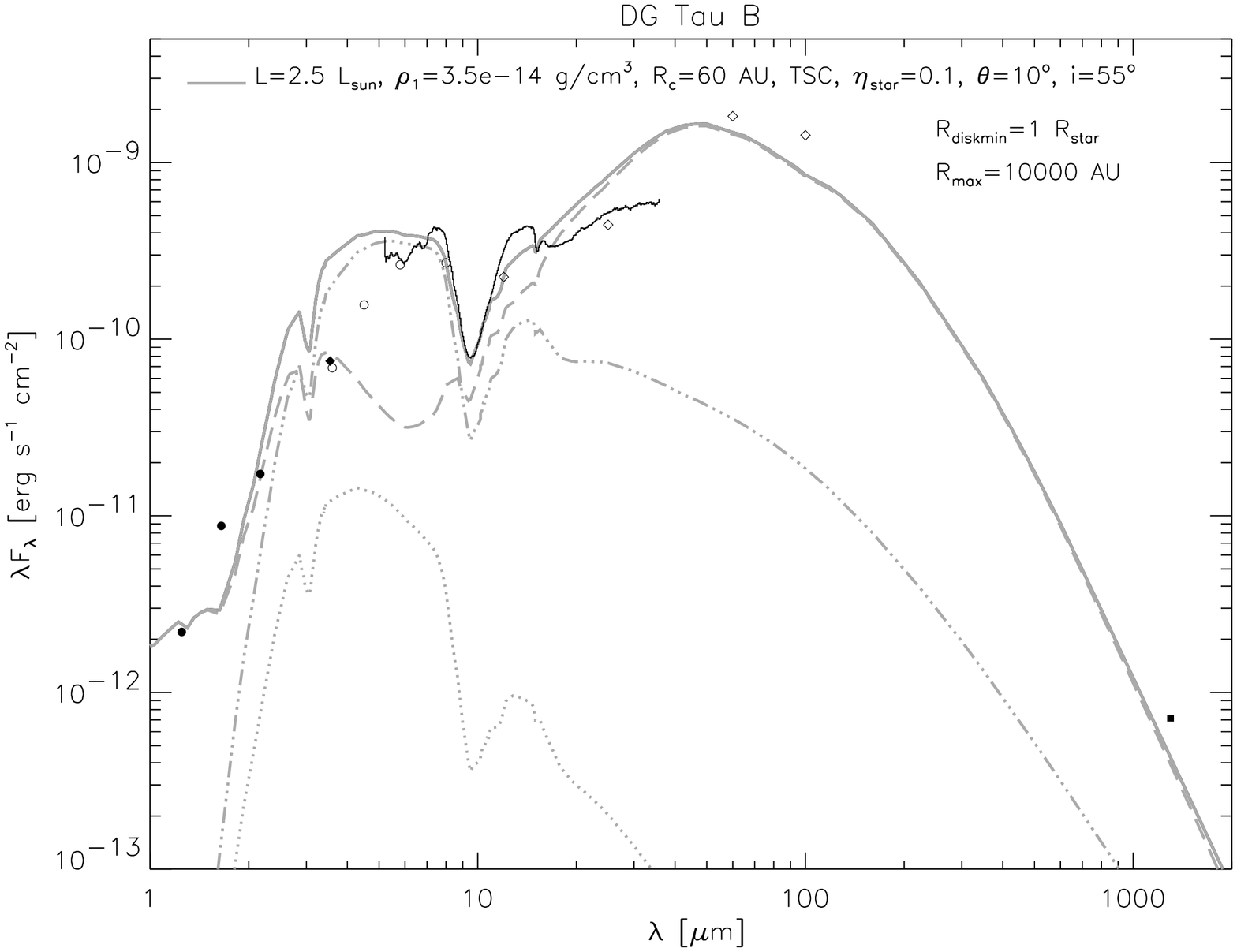}
\caption{IRS spectrum and photometric data of DG Tau B, and an envelope
model fit with L=2.5 L$_{\odot}$, $\rho_1=3.5 \times 10^{-14}$ g cm$^{-3}$,
$R_c$=60 AU, initial TSC density distribution, $\eta_{star}$=0.1, $\theta$=10\degr, 
i=55\degr, an inner disk radius of 1 stellar radius, and an outer envelope radius of 10000 AU. 
The CO$_2$ ice abundance was set to $3.0 \times 10^{-5}$. The gray lines represent 
the model components as in Figure \ref{model_04016}. \label{model_DGTauB}}
\end{figure}

\begin{figure}
\includegraphics[angle=90, scale=0.6]{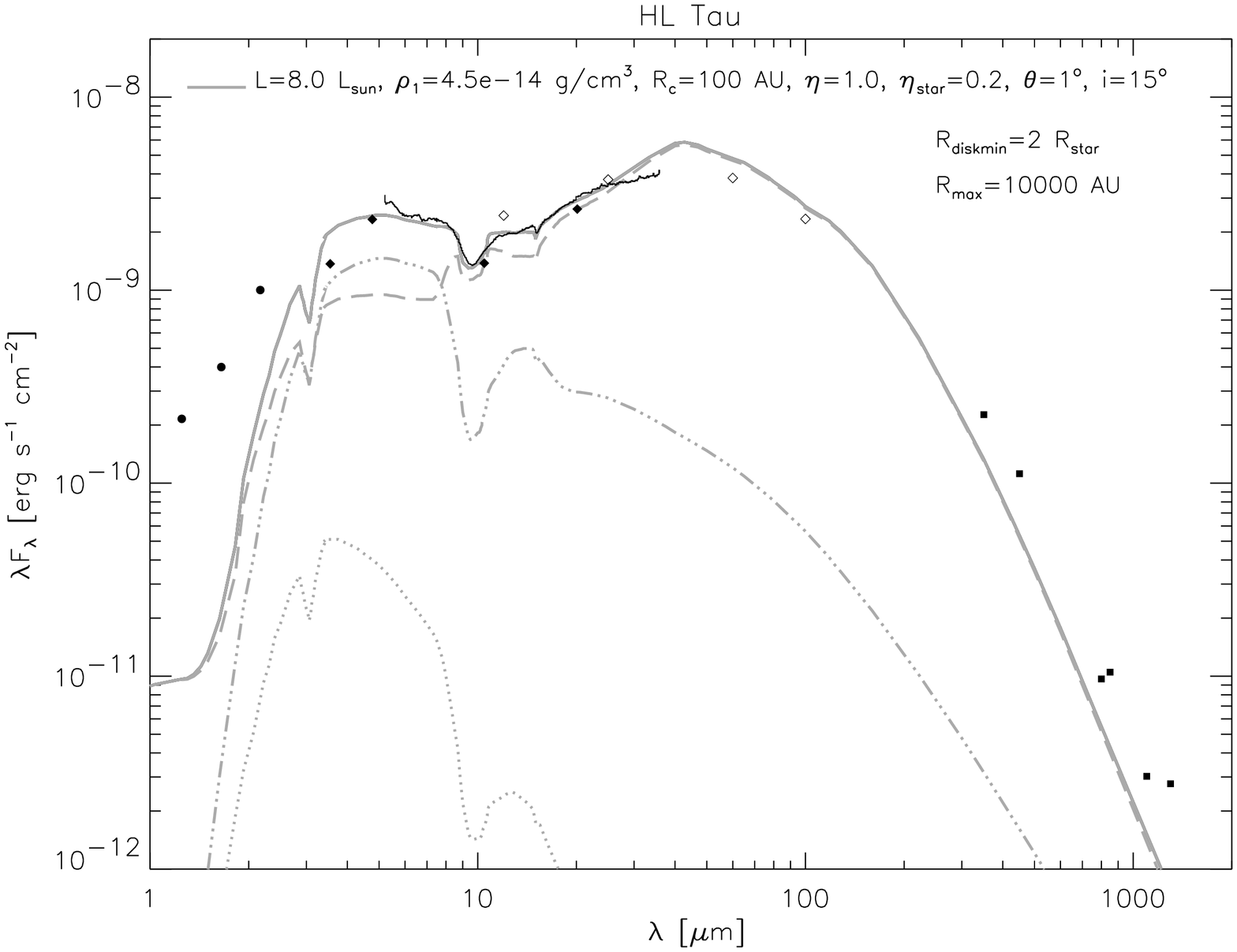}
\caption{IRS spectrum and photometric data of HL Tau, and an envelope
model fit with L=8.0 L$_{\odot}$, $\rho_1=4.5 \times 10^{-14}$ g cm$^{-3}$,
$R_c$=100 AU, $\eta$=1.0, $\eta_{star}$=0.2, $\theta$=1\degr, 
i=15\degr, an inner disk radius of 2 stellar radii, and an outer envelope radius of 10000 AU. 
The CO$_2$ ice abundance was set to $4.0 \times 10^{-5}$. The gray lines represent 
the model components as in Figure \ref{model_04016}. \label{model_HLTau}}
\end{figure}

\begin{figure}
\includegraphics[angle=90, scale=0.6]{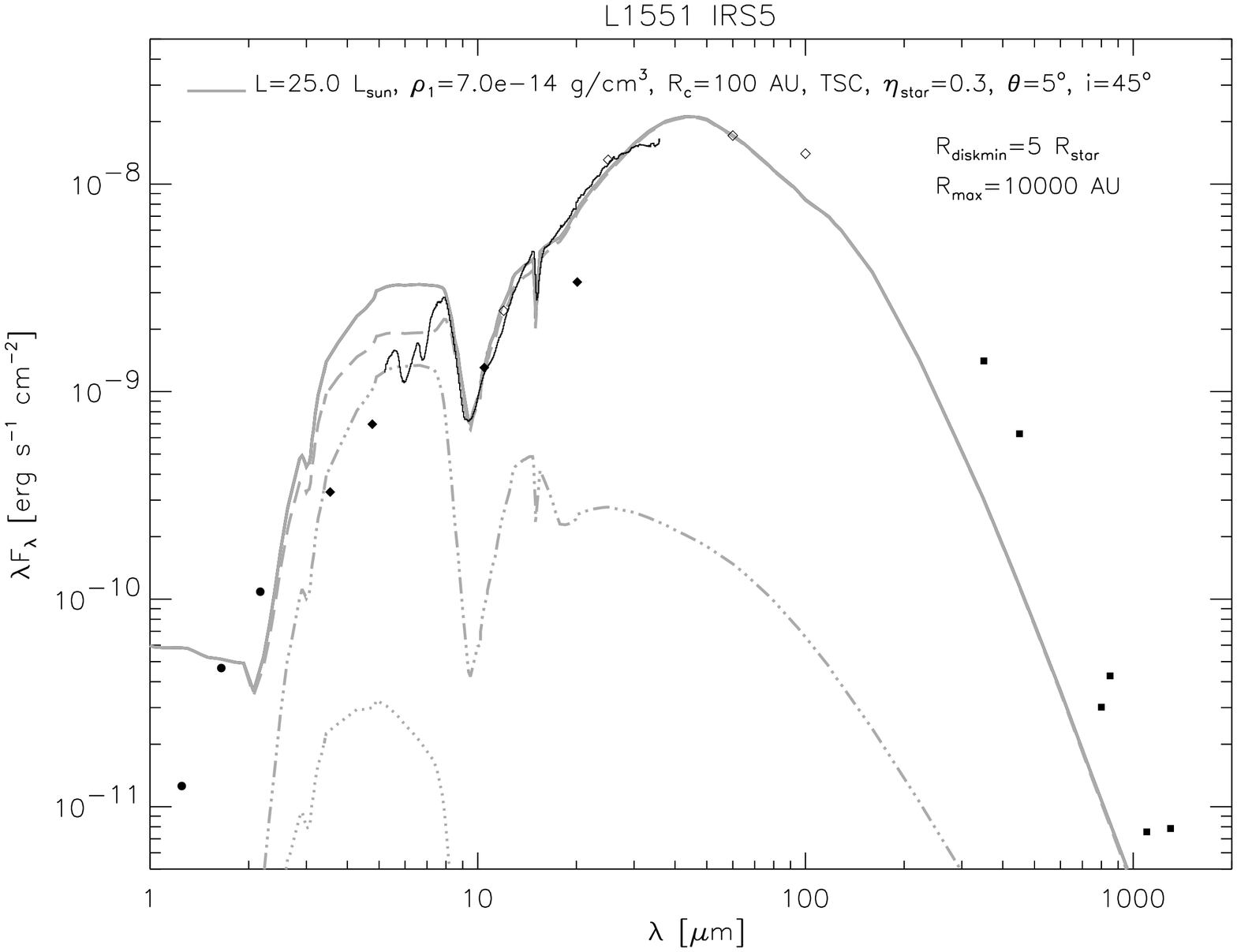}
\caption{IRS spectrum and photometric data of L1551 IRS 5, and an envelope
model fit with L=25.0 L$_{\odot}$, $\rho_1=7.0 \times 10^{-14}$ g cm$^{-3}$,
$R_c$=100 AU, initial TSC density distribution, $\eta_{star}$=0.3, $\theta$=5\degr, 
i=45\degr, an inner disk radius of 5 stellar radii, and an outer envelope radius of 10000 AU. 
The CO$_2$ ice abundance was set to $3.0 \times 10^{-4}$. The gray lines represent 
the model components as in Figure \ref{model_04016}. \label{model_L1551}}
\end{figure}

Our model of CoKu Tau/1 (see Figure \ref{model_CoKu_Tau1}) requires a low
density ($\rho_1=5.0 \times 10^{-15}$ g cm$^{-3}$), a small centrifugal
radius ($R_c$ = 40 AU), a cavity semi-opening angle of 5\degr, and a high
inclination angle ($i=80$\degr). We adopted a luminosity of 1.1 L$_{\odot}$, 
which is slightly higher than the measured bolometric luminosity of 1.0 L$_{\odot}$ 
(see Table \ref{tab_lum_incl}). The inner disk radius was set at 7 stellar radii to
decrease the emission from the near-IR to 8 $\mu$m, which is still higher than the
observed fluxes, most likely due to an overestimate of the scattered light emission 
from the envelope.

The shape of the IRS spectrum between 12 and 20 $\mu$m constrains the
centrifugal radius, while the low density is required to generate a silicate emission 
feature and cause the SED to peak around 30 $\mu$m. Decreasing $\rho_1$
by about 20\% and increasing or decreasing $R_c$ by roughly the same percentage 
yields comparable fits, provided the inclination angle lies in the 70\degr-80\degr\
range. A previous model by \citet{stark06} yielded a reference density smaller
by about a factor of three, a comparable values for $R_c$, but a somewhat smaller
inclination angle and a larger cavity opening angle. We can exclude a larger cavity, 
since it would decrease the flux of the 10 $\mu$m silicate emission feature.

As mentioned earlier, the silicate emission feature at 10 $\mu$m is generated 
by the envelope. The disk just contributes a weak silicate absorption feature, 
which results from our simple disk model that includes only continuum emission,
extinguished by the envelope. A more realistic edge-on disk would yield a somewhat 
stronger absorption feature; however, the envelope emission clearly dominates over 
that of the disk at 10 $\mu$m, a fact that also applies to IRAS 04264+2433 and, 
to a lesser extent, IRAS 04248+2612 (long-dashed gray lines in Figures 
\ref{model_04248}, \ref{model_04264}, \ref{model_CoKu_Tau1}). 
Thus, the silicate emission feature is not greatly affected by the shape of the silicate 
feature originating from the disk, and our current models of low-density envelopes 
have some validity.

\subsubsection{DG Tau B}
\label{DGTauB_model}
This object displays some similarity to HH 30; its near-IR HST images show 
an edge-on disk and a bipolar reflection nebulosity (see Fig.\ \ref{DGTauB_slits}), 
which was interpreted as the walls of an outflow cavity \citep{padgett99}. 
In the optical, the nebula is visible, too, but the star is not detected 
\citep{stapelfeldt97}. DG Tau B is also the source of a jet, which is oriented 
perpendicular to the disk and along the axis of symmetry of the reflection nebula 
\citep{eisloeffel98}. A molecular outflow, oriented along the redshifted optical
jet, has also been observed \citep{mitchell97}.

Our preliminary model for DG Tau B (see Figure \ref{model_DGTauB}) has an 
inclination angle of 55\degr\ and a cavity semi-opening angle of 10\degr;
both values are smaller than expected from the optical and near-IR images. 
However, a more highly inclined envelope would result in a steeply rising SED 
over the mid-IR range, which would not match the observed spectrum.
The centrifugal radius (and thus outer disk radius) of 60 AU is poorly
constrained; a larger value would better fit the slope of the IRS spectrum beyond
20 $\mu$m, but result in a shallower silicate feature at 10 $\mu$m.
The deep silicate  absorption feature requires a relatively high density ($\rho_1=3.5 
\times 10^{-14}$ g cm$^{-3}$), but its long-wavelength wing is not fitted 
very well. In addition, the model is not well-constrained at longer wavelengths, since
DG Tau B lacks any sub-mm measurements, and the {\it IRAS} fluxes at 60 and 
100 $\mu$m are uncertain due to the proximity of DG Tau: the far-IR fluxes contain 
the emission of both DG Tau B and DG Tau, which lies 1\arcmin\ to the 
northeast of DG Tau B and is about two to three times brighter in the mid-IR. 

Recent models by \citet{stark06} yielded quite different model parameters, except
for the reference density; they were able to fit the near-IR images of DG Tau B 
adopting a luminosity of 0.2 L$_{\odot}$, a centrifugal radius of 300 AU, a $\rho_1$
value of $3.75 \times 10^{-14}$ g cm$^{-3}$, and cavity semi-opening and
inclination angles of 30\degr\ and 73\degr, respectively. 
This set of parameters would not fit the IRS spectrum. In particular, we adopted a 
luminosity of 2.5 L$_{\odot}$; a smaller luminosity, closer to the value derived from 
integrating under the SED (1.8 L$_{\odot}$), would result in less far-IR flux and 
require a smaller inclination angle to still reproduce the mid-infrared spectrum, which
would be inconsistent with the HST images, which indicate a more edge-on orientation. 

\subsubsection{HL Tau}
Even though HL Tau is considered by some a classical T Tauri star surrounded 
by an accretion disk, it is also embedded in an envelope \citep[e.g.,][]{motte01,
white04}; this is supported by our IRS spectrum, which shows the silicate feature 
in absorption, as well as ice absorption features in the 5--8 $\mu$m range and
at 15.2 $\mu$m. HL Tau is also very bright at sub-mm and mm wavelengths, 
and a rotating, elongated, $\sim$ 2000 AU long structure has been mapped in 
$^{13}$CO \citep{beckwith90, sargent91}. \citet{stapelfeldt95} detected a 
compact, one-sided reflection nebulosity in the optical; the star is not detected. 
HL Tau drives a powerful, extended molecular outflow which was mapped in CO 
emission \citep{monin96}. \citet{close97} observed HL Tau in the near-IR 
with adaptive optics and resolved the inner accretion disk and bipolar outflow cavities. 
They concluded that HL Tau is surrounded by an accretion disk of about 150 AU 
radius, inclined at 67\degr, and an infalling envelope of about 1200 AU radius.  
However, the environment of HL Tau is quite complicated, partly due to its proximity
to XZ Tau, which is likely at the origin of an expanding shell that compresses the
nebula around HL Tau \citep{welch00}. 

The SED of HL Tau can be reproduced with an envelope model with a luminosity of 
8.0 L$_{\odot}$, which is somewhat larger than its measured bolometric luminosity 
(see Table \ref{tab_lum_incl}), a relatively high density ($\rho_1 = 4.5 \times 
10^{-14}$ g cm$^{-3}$), an $R_c$ value of 100 AU, and flattening parameter 
$\eta$ of 1.0, i.e., a sheet-collapse model (Figure \ref{model_HLTau}). 
The latter quantity was adjusted to yield more flux in the 12 to 30 $\mu$m region
compared to the TSC models. A lower density would yield a 10 $\mu$m silicate
feature that is too narrow, while a density up to $6.0 \times 10^{-14}$ g cm$^{-3}$
would still yield a good fit. An even higher density would require a larger centrifugal radius,
which would result in an SED that overestimates the flux from about 25 to 200 $\mu$m.
The $R_c$ value is constrained by the shape of the IRS spectrum; we can rule out $R_c$ 
values smaller than 100 AU and larger than about 200 AU.
These results are also roughly consistent with previous modeling results by 
\citet{calvet94}, who adopted a reference density of 3.2 $\times 10^{-14}$ 
g cm$^{-3}$ and a centrifugal radius of 200 AU.

We note that, despite the finding by \citet{close97} that the inclination
angle of HL Tau is 67\degr\ and that large cavities are present, our model 
indicates virtually no cavity and a low inclination. The model by \citet{calvet94} 
also suggests an outflow hole with a semi-opening angle of 10\degr; including 
such a cavity in our model would reduce the flux over the IRS wavelength range,
but also increase the flux somewhat in the near-IR, where our model underestimates
the emission. A larger inclination angle would decrease the short-wavelength flux
even more and generate a deeper silicate absorption feature. On the other hand, our 
value for the inclination angle is closer to the value of  30\degr\ determined by 
\citet{calvet94}, and our disk radius ($\sim$ $R_c$)  is also roughly consistent 
with the value of 150 AU found by \citet{close97}.

Our model underestimates the emission at sub-mm and mm wavelengths, despite an
outer envelope radius of 10,000 AU. Models of HL Tau by \citet{dalessio97} have 
shown that the outer disk requires higher temperatures to reproduce the SED shape 
at longer wavelengths, implying irradiation of the circumstellar disk by the infalling 
envelope. Since our model does not include this additional heating, our disk emission
is likely underestimated; thus, the ``missing'' flux at the longer wavelengths could be 
attributed in part to this additional disk component. In addition, dust outside the
infall region could contribute to the mm emission.

\subsubsection{L1551 IRS 5} 
This object is the most luminous Class I object in Taurus (L$_{bol}$ = 22-28
L$_{\odot}$; see Table \ref{tab_lum_incl}) and is therefore a well-studied 
protostellar system. It consists of a 0{\farcs}3 binary, whose components 
are each surrounded by a circumstellar disk, and of a circumbinary disk and a 
flattened envelope \citep{looney97, rodriguez98}. 
It is surrounded by an extended reflection nebula at optical and near-IR wavelengths
\citep{campbell88,tamura91,white00}, and it powers a highly collimated bipolar 
molecular outflow \citep{snell85,moriarty92}. Even though no optical outburst 
has been observed, L1551 IRS 5 is considered an FU Ori object, implying episodic
events of high disk accretion and outflow rates \citep{hartmann96a}.

The mid- and far-IR spectrum (2--200 $\mu$m) of this bright Class I object 
was obtained by \citet{white00} using ISO; they detected gas-phase emission 
lines and ice absorption features. We also detect strong ice absorption features
in our IRS spectrum, in addition to a very deep and broad silicate absorption feature 
(see Figure \ref{ClassI_IRS}c).

L1551 IRS 5 has been modeled in detail by \citet{osorio03}, taking into account
emission from the circumstellar and circumbinary disks, as well as the envelope.   
The model presented below does not include all the components of the system as 
in \citet{osorio03} and should therefore be considered as preliminary.

L1551 IRS 5 is the most luminous object in our sample (adopted luminosity of 
25~L$_{\odot}$). Its deep silicate absorption feature requires a high density 
($\rho_1 = 7.0 \times 10^{-14}$ g cm$^{-3}$), and the steep SED beyond 
15 $\mu$m a centrifugal radius of 100 AU (see Figure \ref{model_L1551}). An
$R_c$ value larger than about 150 AU would result in a decrease in flux between 
15 and 30 $\mu$m and thus degrade the fit to the IRS spectrum. Our model indicates 
that the cavity is small ($\theta=$\,5\degr), and the object is seen at an intermediate
inclination angle. Increasing $\rho_1$ would require a smaller inclination angle, but
a larger inclination is more in accordance with the observed outflows ($i$=65\degr;
\citealt{hogerheijde98}).

Compared to the models by KCH93, which suggested a reference density of 
$\sim$ $10^{-13}$ g cm$^{-3}$, a centrifugal radius between 70 and 300 AU,
and an inclination angle between 30\degr\ and 60\degr, our model has a smaller,
but still high, reference density, an $R_c$ value towards the lower end of the KCH93
range, and an inclination angle in the middle of the range determined by KCH93. 
\citet{whitney97} and, more recently, \citet{gramajo07}, modeled this object 
with a reference density of only $3.75 \times 10^{-14}$ g cm$^{-3}$ , 
$R_c \sim$ 40 AU, $\theta$=20\degr, and $i \gtrsim$ 70\degr.

Our model parameters are different from the ones found by \citet{osorio03}, who 
adopted an inclination angle of 50\degr, $\rho_1 = 4.0 \times 10^{-13}$ g cm$^{-3}$, 
$R_c= 300$ AU, and a flattening parameter $\eta=2.5$ for the envelope. We could 
obtain a comparable fit of the SED out to 30 $\mu$m by using $L=25$ L$_{\odot}$,
$\rho_1 = 2.5 \times 10^{-13}$ g cm$^{-3}$, $R_c= 100$ AU, $\eta=2.0$,
$\eta_{star}$=0.1, $\theta$=0.1\degr, and $i=$45\degr, but the far-IR emission
would be overestimated. On the other hand, the sub-mm and mm emission would 
be better matched with the sheet-collapse model. Thus, the SED of L1551 IRS 5 
can be reproduced by a TSC model, but the long-wavelength emission suggests the 
presence of an additional, cold, likely elongated, dust component that is outside of 
the infall region.

\section{Discussion}
\label{discussion}

\subsection{Identification of Class I Objects}

The mid-infrared spectra of Class I objects have quite different appearances; while some
differences are caused by individual envelope parameters like the density and size of the
infall region, the inclination angle also plays an important role. A Class I object viewed 
pole-on through a cavity in the envelope can look like a Class II object with some
additional long-wavelength excess emission (as some FU Ori objects appear; see
\citealt{green06}), while an edge-on Class I object can have 
the appearance of a Class 0 object. However, the envelope around a true protostar 
will generate excess emission at mid-IR to mm wavelengths, resulting in a larger infrared 
excess than expected from an accretion disk alone, and less excess when compared to 
the large, cool envelopes around Class 0 objects that peak at sub-mm to mm wavelengths.
In addition, if seen within a narrow range of inclination angles, an edge-on Class II object 
can appear similar to an embedded protostar \citep[see][]{dalessio99},
but it would lack additional excess emission caused by the envelope.

Observations at other wavelengths help in identifying the evolutionary state 
of an object: as noted by \citet{park02}, true Class I sources are usually 
extended in the near-IR and at sub-mm or mm wavelengths, the 
former due to light scattered by dust in inner envelope regions, the latter due to 
thermal emission by dust in the outer parts of the envelope. Furthermore, Class I 
objects usually have molecular outflows or jets detected in the optical/near-IR and
especially at mm wavelengths. 
Objects in transition between the Class I and Class II stage are expected to be 
surrounded by less envelope material, in accordance with larger cavities in their 
envelopes.

Of the 28 Class I objects presented in this paper, 7 were previously identified as Class II
objects in KH95: IRAS 04154+2823, 04158+2805, 04278+2253, CoKu Tau/1,
HL Tau, IC 2087 IR, and LkHa 358. While the SED shape of these objects might
indicate that they are T Tauri stars whose envelope has already mostly dissipated, the
presence of ice features in their mid-infrared spectrum, as well as the detection
of extended reflection nebulosities and outflows around some of these objects, 
suggests that there is still some remnant envelope material. 

IRAS 04154+2823, 04278+2253, 04489+3042, IC 2087 IR, and LkHa 358, 
as well as some of the Class II objects presented in \citet{furlan06a} 
(IRAS 04187+1927, DG Tau, DP Tau, FS Tau, HN Tau, T Tau, and XZ Tau), 
could be objects in transition between the Class I and II stage, 
when a protostar has almost completely cleared its envelope, but is  
still surrounded by some nebulosity. In some cases, the ``evolved'' protostar 
is still generating outflows, but due to envelope clearing it appears more and 
more like a classical T Tauri star. In some other cases, like T Tau, HN Tau,
and XZ Tau, a binary consists of a T Tauri star and a more deeply embedded
source, which is likely an effect of the particular viewing geometry of the system:
either one component is more aligned with the outflow cavity and thus less
embedded, or the more extinguished component is actually a T Tauri star
oriented edge-on. Therefore, as stated earlier, the inclination angle is an 
important factor in determining the appearance of Class I and II objects.

\subsection{Ices in the Envelopes of Class I Objects}

\begin{figure}
\epsscale{0.8}
\plotone{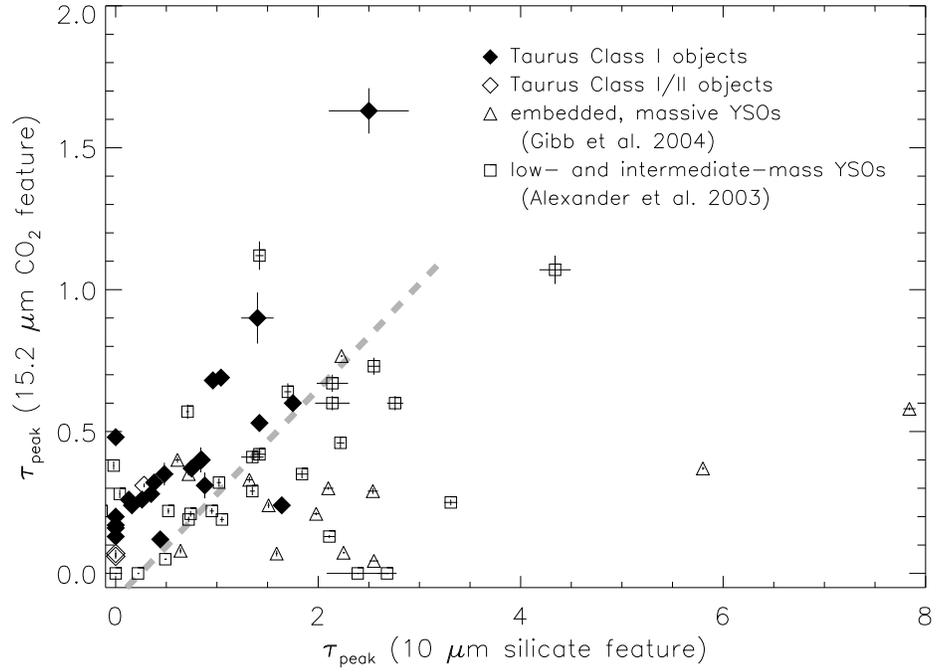}
\caption{Peak optical depth of the CO$_2$ ice feature versus the peak optical depth
of the 10 $\mu$m silicate absorption feature. The diamonds represent the Class I
objects presented in this paper, while the triangles and squares represent generally more
massive YSOs studied by \citet{gibb04} and \citet{alexander03}, respectively. 
The data points at a peak silicate optical depth of 0 are objects with either a silicate
emission feature or a feature that shows emission and absorption characteristics.
The thick, gray dashed line is drawn to guide the eye to separate the regimes of 
envelopes from that of ambient molecular cloud material (see text for details). 
\label{CO2_sil_taus}}
\end{figure}

The ice and silicate absorption features detected in the IRS spectra are also a useful 
tool to characterize Class I objects. The peak optical depth of these features reveals
their strength and can therefore be used as an indication for their origin.
To determine the optical depth of the CO$_2$ ice feature at 15.2 $\mu$m and that 
of the silicate absorption feature around 10 $\mu$m, we first used a spline fit to 
determine the underlying continuum. Then we derived the optical depth, $\tau$, 
assuming $F_{obs} = F_c e^{-\tau}$, where $F_{obs}$ and $F_c$ are the 
observed and the underlying continuum flux, respectively. The measured peak optical 
depths of the CO$_2$ and silicate absorption features are shown in Figure 
\ref{CO2_sil_taus}, together with previous measurements by \citet{gibb04} and 
\citet{alexander03}, who determined the optical depths of silicate and ice absorption 
features in ISO spectra of embedded, massive YSOs and low- and intermediate-mass 
YSOs, respectively. For the \citet{alexander03} data, we used the same conversion
as in \citet{watson04} to convert the CO$_2$ equivalent widths to peak optical depths.

As already pointed out by \citet{watson04}, the low-mass Class I objects in
Taurus generally have large peak CO$_2$ optical depths, especially considering
that their peak silicate optical depth lies below $\sim$ 2 (the object with the 
largest silicate and CO$_2$ optical depths is IRAS 04368+2557, a Class 0 object). 
Most of the objects observed by \citet{alexander03} have similarly strong 
CO$_2$ ice features for larger silicate optical depths, suggesting additional extinction 
along the line of sight towards these objects. 
It is expected that the two optical depths track each other along lines of sight through 
molecular clouds (with some dispersion due to different conditions in different clouds), 
since ice mantles are thought to grow in a similar manner onto dust grains in all 
comparably shielded regions of molecular clouds. 

The gray dashed line in Figure \ref{CO2_sil_taus} is drawn by eye to roughly 
separate the region occupied by the Taurus Class I objects, where extinction from 
the envelope dominates the optical depth, from that occupied by objects seen along 
molecular cloud lines of sight, where the extinction is mostly due to intervening molecular 
cloud material. The boundary is not well-defined, since some of the ISO observations 
did include objects in which local extinction by an envelope dominates, but it separates 
all but one of the Taurus Class I objects from the majority of the ISO targets.

Thus, we confirm the conclusions of \citet{watson04} that dust grains in protostellar 
envelopes have larger ice mantles than grains in the ambient molecular cloud, implying
ice mantle growth inside the envelopes. We defer a more detailed study of the ice
absorption features, including the detailed composition and thus derived temperature
of the ices, to a future paper (Zasowski et al. 2008, in preparation).
 
We note that the objects with a peak silicate optical depth of 0 are protostars in which
the silicate feature is either in emission or displays both an emission and an absorption
component. As discussed in \S\ \ref{IRS_spectra}, the latter type of objects is
likely seen along lines of sight that suffer from less extinction by the envelope, caused
by a more face-on orientation or a lower-density envelope.
Therefore, for these objects the optical depth of the silicate feature, taken as a measure
of dust absorption in the envelope, is probably underestimated due to contamination 
by the silicate emission arising from the disk.

The one Class I object that lies in the molecular cloud region of the plot is DG Tau/B; 
it has a deep silicate, but only a weak CO$_2$ ice absorption feature. This could
indicate that grain mantles in its envelope are less rich in CO$_2$ ice, but on the 
other hand this feature is superposed on the broad and deep silicate absorption at 
18 $\mu$m, which adds additional uncertainty to the optical depth measurement 
of the CO$_2$ feature. 

The three objects identified as ``Class I/II objects'' in Figure \ref{CO2_sil_taus}
are IRAS 04278+2253, LkHa 358, and IC 2087 IR. The former two objects have 
the lowest CO$_2$ feature strengths of our Taurus sample and  virtually no silicate 
absorption feature, which confirms that they are more evolved  Class I objects. 
IC 2087 IR has both CO$_2$ and silicate peak optical depths comparable to those 
of typical Class I objects; thus, it is still likely embedded to some extent. 
This shows that the optical depths of the silicate and ice absorption features can 
help to recognize more evolved Class I objects, but the effects of the environment 
and viewing angle also have to be taken into account.

\subsection{Summary of the Models}

Envelope models aid in the identification and characterization of Class I objects;
when models fit the SED from the near-IR to the mm, they reveal both the
large- and small-scale structure of a protostar, its accretion disk and its envelope. 
The IRS spectrum, which covers the SED from 5 to 40 $\mu$m, is a particularly
useful constraint, since the depth of the silicate feature at 10 $\mu$m is connected
to the density and inclination angle of the envelope, while the slope between 15 and
40 $\mu$m helps to constrain the envelope shape, centrifugal radius and in part the
size of the cavity. 
Not all parameters can be uniquely defined by a model fit
to the IRS spectrum, but together with shorter- and longer-wavelength data,
the parameter space can be narrowed down considerably.  

We point out that, given the large number of parameters for each model, we 
did not determine the goodness of a model fit by ${\chi}^2$-minimization, 
but rather by judging by eye, based on our experience on how the different
parameters affect the resulting model SED.
However, we justify the choice of specific parameters 
and explain how the parameter space was narrowed. In addition, the more 
constraints a model has, in particular the availability of sub-mm and mm flux 
measurements and the source's luminosity, the more limited the choice of 
parameters becomes. The fitting by eye is much more efficient than 
${\chi}^2$-minimization, and in most cases we derive good fits that are
likely minima in ${\chi}^2$ space.
Typical uncertainties for our model parameters are listed in the footnote
of Table \ref{tab_model_fits}.

We also note that several assumptions made in the modeling code could be 
refined to better reproduce the observations: e.g., the assumption that the 
cavity follows the shape of the streamlines of infalling particles, the assumption 
of a spherically symmetric density distribution when computing the equilibrium 
temperature in the envelope, and the adoption of only the inner, optically thick 
regions of a flat accretion disk as the disk component. 
Some of our 22 models of Class I objects should be treated as preliminary, given 
the assumptions made in the modeling code and/or some weaker constraints
due to the lack of data, but our models already reveal a large range in parameters
necessary to obtain fits (see Table \ref{tab_model_fits} for a summary of
the model parameters).

We computed models for the majority of Class I objects in our sample; however,
we did not attempt to model objects with poor observational constraints for their
SEDs, or objects which are likely more evolved Class I objects and for which these 
envelope models might not apply. Thus, we did not attempt to model HH 30, 
whose spectrum is very noisy and whose overall SED is poorly constrained by 
the scarcity of measurements at other wavelengths. Also IRAS 04181+2654 B,
a member of a binary system, was not modeled, since it lacks long-wavelength
data (we modeled component A, which was observed at 1.3 mm). The three 
sources with decreasing SEDs over the IRS spectral range, IRAS 04278+2253, 
IC 2087 IR, and LkHa 358, were also not modeled. 
Finally, no model was produced for GV Tau, since its IRS spectrum is uncertain,
and the fact that the two components are separated by 1{\farcs}3 (which
corresponds to about 180 AU at the distance of the Taurus star-forming region)
likely introduces asymmetries in the envelope of the more embedded component.

Our envelope models indicate centrifugal radii, and thus disk sizes, that cover
a range from 10 AU up to 300 AU, with a median value of 60 AU. Since the 
centrifugal radius is thought to increase rapidly with time ($\propto t^3$), 
we could conclude that the objects in our sample span somewhat different 
ages (about a factor of 3). However, the main reason for the differences in 
centrifugal radii is likely different initial conditions of the collapse, since the 
initial centrifugal radius strongly depends on the mass of the central condensation 
and the initial angular velocity of the cloud core 
\citep[$R_c \propto M^3 {\Omega}^2$;][]{terebey84}. 
In addition, some of the more deeply embedded protostars that have been identified 
as Class 0/I objects, like IRAS 04166+2706 and 04368+2557, have large $R_c$ 
values, which would not be expected if the spread in centrifugal radii was only due 
to time evolution.
Since a disk size of a few 100 AU is comparable to the radii of accretion disks found 
around classical T Tauri stars \citep[e.g.,][]{dutrey96}, our median value for
$R_c$ (and thus outer disk radii) of 60 AU suggests that the accretion disks 
found around protostars will expand, as is expected due to angular momentum
transfer in the accretion process.

We note that the disk component dominates the flux in the 3-8 $\mu$m 
range for most of the models; for about 2/3 of these objects the fraction of 
the luminosity contributed by the star, $\eta_{star}$, is either 0.1 or 0.2, i.e., 
80-90\% of the luminosity is generated by accretion. About half of the objects that 
have no dominant disk contribution in the 3-8 $\mu$m range have luminosities 
dominated by the star ($\eta_{star}$=0.8), and the other half have smaller 
values for $\eta_{star}$, but generally large inclination angles. Thus, the 
Class I objects in our sample whose luminosity is dominated by accretion (and
which likely have large mass accretion rates through their disks) generate SEDs
that are dominated by the disk in the 3-8 $\mu$m range, except if the disk 
emission is diminished due to a high inclination angle.

The reference density $\rho_1$ varies from $4.0 \times 10^{-15}$ g 
cm$^{-3}$ to $7.0 \times 10^{-14}$ g cm$^{-3}$, with a median value
of $3.0 \times 10^{-14}$ g cm$^{-3}$, very similar to the value found by
KCH93. 
$\rho_1$ can be used to estimate the mass infall rate from the envelope onto
the disk by applying Equation 
\ref{rho1_equ}; for a 0.5 M$_{\odot}$ object, the mass infall rates for the 
objects in our Class I sample lie between $5.3 \times 10^{-7}$ and 
$9.3 \times 10^{-6}$ M$_{\odot}$ yr$^{-1}$, about two orders
of magnitude larger than the accretion rates of classical T Tauri stars \citep[e.g.,][]
{gullbring98}. This would confirm previous results that Class I objects are at
an earlier evolutionary stage than T Tauri stars and still accreting a substantial 
fraction of the final stellar mass \citep{muzerolle98}.

Even though the cavity semi-opening angle and, in some cases, the inclination 
angle are usually less well-constrained than the reference density and centrifugal
radius, we observe that our models indicate a large range of inclination
angles, covering values from 15\degr\ to 89\degr\ (with a median
value of 55\degr), and a smaller range of cavity semi-opening angles
(0.1\degr\ to 27\degr, with a median of 6\degr). 

In Table \ref{tab_lum_incl} we list the luminosities of the Class I objects
in our sample and their inclination angles. Both the values of bolometric luminosity
from the literature and those we derived by integrating under the SED are 
actually apparent luminosities, since envelopes are not spherically 
symmetric. Thus, we would expect the apparent bolometric luminosity to be larger 
than the system luminosity (which enters as a model parameter) at low inclination
angles, while at high inclination angles the opposite should be the case. 
The data in Table \ref{tab_lum_incl} shows that this trend is only weak for our
sample of Class I objects; in general, our model luminosities are higher than the
apparent bolometric luminosities.
This could suggest that we overestimated the luminosity of some of our
lower-inclination sources.

We note that most of our models do not require an initially flattened density
distribution ($\eta$ $\gtrsim$ 1); all but 5 of the 22 objects we modeled
can be fitted by TSC models, which assume an initially spherically symmetric
density distribution. However, we cannot exclude that sheet-collapse model 
fits could also be found for the remaining 17 Class I objects for which only 
TSC models have been considered so far. Thus, as a starting point for models,
it seems that star-forming cores in Taurus can be described by the approximation 
of collapsing, singular isothermal spheres.

In a few cases we found that an additional component of cold dust seems to be
present, since certain models are able to reproduce the 
sub-mm and mm fluxes measured in smaller apertures (10\arcsec-20\arcsec), 
but not the measurements at similar wavelengths derived from large apertures 
(40\arcsec-60\arcsec). In particular the 1.3 mm data points from \citet{motte01}, 
which correspond to fluxes integrated over a beam 60\arcsec\ in diameter, 
are often higher than the model prediction.
As suggested by \citet{jayawardhana01}, cold dust of roughly constant density 
outside of the infall region could contribute to the long-wavelength emission.
In these cases envelope models could aid in the distinction between emission from
infalling envelope material and emission from the surrounding dense cloud.

When comparing our modeling results with previous efforts, we note that our 
results often differ considerably. However, in this work we have the additional 
constraints through our IRS spectra, which, when combined with near-IR and
far-IR to mm data, allows us to determine the 9 parameters entering in the
envelope models. In particular the 10 $\mu$m silicate absorption feature
allows us to narrow down the parameter space. A next step in modeling 
would involve reproducing the images of the Class I sources at the same
time as the SEDs to better take into account the envelope geometry and 
asymmetries.

\section{Conclusions}
\label{conclusions}

After analyzing the IRS spectra and SEDs of 28 Class I objects in the Taurus 
star-forming region, and generating envelope models for 22 of these objects, 
we conclude the following:

${\bullet}$ Almost all protostars we observed display ice absorption features
in their mid-IR spectra, which originate in the envelopes around these young 
stars. In particular, the CO$_2$ ice feature at 15.2 $\mu$m is ubiquitous 
and strong among the Class I objects in our sample, signifying an origin in the 
cold outer envelope regions.
The envelope is also generally responsible for a deep silicate absorption feature 
at 10 $\mu$m. 
The three objects with silicate emission features in their spectra, IRAS 04248+2612, 
04264+2433, and CoKu Tau/1 are likely Class I objects seen through low-density 
envelopes, which are in an advanced stage of dispersal and thus considered to be
more evolved.

${\bullet}$ The SEDs of Class I objects peak in the mid- to far-IR, where 
the emission is dominated by the envelope. Objects with flat SEDs in the
infrared spectral range could be in transition between the Class I and II stage.
The inclination angle of protostellar systems strongly affects the appearance 
of their mid-IR SED; Class I objects seen pole-on could appear as Class II
objects with additional long-wavelength excess emission, while edge-on
Class I objects have SED shapes similar to Class 0 objects, but with less
flux in the sub-mm and mm wavelength region.

${\bullet}$ The IRS spectra are instrumental in constraining envelope model
parameters, in particular the reference density $\rho_1$ (which is linked to the
mass infall rate) and centrifugal radius $R_c$. The depth of the silicate feature
and the shape of the spectrum between 12 and 30 $\mu$m provide constraints
which can be further tightened by model fits to near-IR, far-IR, sub-mm, and 
mm observations. The median values for $\rho_1$ and $R_c$ for the Class I
objects we modeled are $3 \times 10^{-14}$ g cm$^{-3}$ and 60 AU,
respectively, with a relatively large range for both parameters. 
The disk contribution dominates the emission in the 3-8 $\mu$m range for 
most of the modeled Class I objects, confirming the importance of disk accretion
luminosity at this early evolutionary stage.
Most SEDs can be reproduced by TSC models, thus not requiring an initially 
flattened density distribution of the collapsing cloud core; only 5 of the 22 
objects we modeled needed sheet-collapse models for an adequate fit of 
their SEDs. 

The IRS spectra of Class I objects in Taurus display a variety in appearances
and add valuable constraints to envelope models. Fits to the
observed SEDs yield the parameters describing their physical structure,
and thus aid in confirming and refining our understanding of protostellar
cloud collapse and evolution.


\acknowledgments
Part of this work was carried out while the first author was supported by a 
NASA Postdoctoral Program Fellowship at the NASA Astrobiology Institute at 
the University of California, Los Angeles, administered by Oak Ridge Associated 
Universities through a contract with NASA.
This work is based on observations made with the {\it Spitzer Space Telescope}, 
which is operated by the Jet Propulsion Laboratory, California Institute of Technology, 
under NASA contract 1407. Support for this work was provided by NASA through 
contract number 1257184 issued by JPL/Caltech. N.C. and L.H. acknowledge support
from NASA Origins grants NAG5-13210 and NAG5-9670, and STScI grant AR-09524.01-A.
P.D. acknowledges grants from PAPIIT, UNAM, and CONACyT, Mexico.
This publication makes use of data products from the Two Micron All Sky Survey, 
which is a joint project of the University of Massachusetts and the Infrared Processing 
and Analysis Center/California Institute of Technology, funded by the National 
Aeronautics and Space Administration and the National Science Foundation. It has also 
made use of the SIMBAD and VizieR databases, operated at CDS (Strasbourg, France),
NASA's Astrophysics Data System Abstract Service, and of the NASA/ IPAC Infrared 
Science Archive operated by JPL, California Institute of Technology (Caltech), under 
contract with NASA. 

Facilities: \facility{Spitzer(IRS)}

\clearpage

\begin{deluxetable}{llcccc}
\tabletypesize{\scriptsize}   
\tablecaption{Properties of Observed Class I Objects \label{tab_properties}}
\tablehead{
\colhead{Name} & \colhead{Alt. name} & \colhead{Multiplicity$^a$} & \colhead{Spectral} &
\colhead{L$_{bol}$ } & \colhead{References}  \\
 & & & \colhead{Type}  & \colhead{($L_{\odot}$)} &  \\
\colhead{(1)} & \colhead{(2)} & \colhead{(3)} & \colhead{(4)} & 
\colhead{(5)} & \colhead{(6)} \\
}
\startdata
04016+2610    & L1489 IRS & s & K4 & 3.7 & 1, 2 \\
04108+2803 B$^b$  & L1495 IRS & 21.2\arcsec & \nodata & 0.62 &  2 \\ 
04154+2823    & \nodata & s & M0.5-M4.5 & 0.33 &  3, 4 \\
04158+2805    & \nodata & s & M5-M6 & 0.20  & 1, 5, 6 \\
04166+2706  & \nodata & s & \nodata & 0.5 & 7 \\
04169+2702    & \nodata & s & \nodata & 0.8 & 2 \\
04181+2654 A$^d$  & \nodata & 31.3\arcsec & \nodata & 0.26 &  2 \\  
04181+2654 B$^d$  & \nodata & 31.3\arcsec & \nodata & 0.25 &  2 \\ 
04239+2436 (A,B)   & \nodata  & 0.30\arcsec & \nodata & 1.27  &  8, 2 \\
04248+2612 (A,B,C)$^c$ & HH31 IRS2 & 0.16\arcsec, 4.55\arcsec & M4-M5 
& 0.36 & 9, 10, 1, 5, 2 \\
04264+2433    & Elias 6 & s & M1 & 0.37  & 1, 2 \\
04278+2253 (A,B)  & \nodata & 6.8\arcsec & G8+K7 &  7.2 & 1, 6 \\
04295+2251    & L1536 IRS & s & \nodata &  0.44 & 2 \\
04302+2247     & \nodata & s & \nodata & 0.34 & 2 \\
04325+2402 (A,B,C)$^e$ & L1535 IRS & 0.24\arcsec, 8.15\arcsec & \nodata & 0.9 & 11, 10, 12 \\
04361+2547 (A,B)  & TMR 1 & 0.31\arcsec & \nodata & 3.8 &  14, 6 \\
04365+2535    & TMC 1A & s & \nodata & 2.4 & 4 \\
04368+2557 (A,B)   & L1527 IRS & 0.17\arcsec & \nodata & 1.6 &  15, 6 \\
04381+2540 (A,B)   & TMC 1 & 0.6\arcsec & \nodata & 0.73 & 13, 4 \\
04489+3042    & \nodata & s & M3-M4 & 0.30 & 5, 1, 2 \\
CoKu Tau/1 (A,B) & \nodata & 0.24\arcsec & K7+M  &  $>$\,0.29 & 9, 1, 2 \\  
DG Tau B         & \nodata & s & \nodata & $>$\,0.02 & 2 \\
GV Tau (A,B)   & Haro 6-10 & 1.3\arcsec & K7 &  6.98 & 10, 1, 2 \\ 
HH 30             & \nodata & s & M0 & $>$\,0.1 & 1, 18 \\
HL Tau            & 04287+1807 & s & K5  & 6.60 & 1, 2 \\   
IC 2087 IR      & 04369+2539 & s & K4 & 3.80 & 1, 2 \\
L1551 IRS5 (A,B)    & 04287+1801 & 0.3\arcsec & G-K & 28  & 16, 17, 12 \\ 
LkHa 358        & CoKu Tau/2 & s & M5.5 & 0.59 & 3
\enddata

\tablecomments{
Column (1) gives the name of the object, column (2) other object names commonly
found in the literature, column (3) the multiplicity of the object (see note (a) below), 
column (4) the spectral type, column (5) the luminosity of the source taken from the
literature, and column (6) gives the references for the data listed in the previous columns. \\
$^a$ ``s'' means single star; for multiple systems, the separation between 
the components in arcseconds is listed. \\
$^b$ 04108+2803 B is separated by 21\arcsec from 04108+2803 A. \\
$^c$ 04248+2612 A and B are separated by 0.16\arcsec \citep{padgett99}; 
a third component was detected by \citet{duchene04} at a distance of 4.55\arcsec\
and a position angle of 15\degr. \\
$^d$ 31.3\arcsec\ is the separation between 04181+2654 A and B. \\
$^e$ 04325+2402 is a possible triple system; the sub-arsecond binary may just be
a single source surrounded by a complex reflection nebula \citep{hartmann99}.}

\tablerefs{
(1) \citet{white04};
(2) \citet{kenyon95};
(3) \citet{luhman00};
(4) \citet{myers87};
(5) \citet{luhman06a};
(6) \citet{kenyon90};
(7) \citet{young03};
(8) \citet{reipurth00};
(9) \citet{padgett99};
(10) \citet{duchene04};
(11) \citet{hartmann99};
(12) \citet{motte01};
(13) \citet{apai05};
(14) \citet{terebey98};
(15) \citet{loinard02};
(16) \citet{rodriguez98};
(17) \citet{kenyon98};
(18) \citet{reipurth93}
}
\end{deluxetable}

\clearpage

\LongTables
\begin{deluxetable}{lcccc}
\tabletypesize{\scriptsize}   
\tablecaption{Long-Wavelength Data of Class I Objects \label{tab_longwave}}
\tablehead{
\colhead{Name} & \colhead{$\lambda$} & \colhead{Flux} & 
\colhead{Aperture size} & \colhead{Reference} \\
 & \colhead{($\mu$m)} & \colhead{(Jy)} & \colhead{(FWHM)} &
}
\startdata
04016+2610    & 160 & 46.0 & 50\arcsec & 1 \\
                       & 350 & 12.48 & 9\arcsec & 2 \\
                       & 450 & 4.23 & 40\arcsec & 4 \\   
                       & 800 & 0.58 & 16{\farcs}8 & 5 \\
                       & 850 & 0.59 & 40\arcsec & 4 \\
                       & 1100 & 0.18 & 18{\farcs}5 & 5 \\
                       & 1300 & 0.15 & 60\arcsec & 6 \\
04108+2803 B & 450 & 1.13 & 40\arcsec & 4 \\
                      & 800 & 0.085 & 16{\farcs}8 & 5 \\
                      & 850 & 0.17 & 40\arcsec & 4 \\
                      & 1100 & $<$\,0.1 & 18{\farcs}5 & 5 \\
                      & 1300 & 0.04 & 60\arcsec & 6 \\
04154+2823    & 350 & 0.44 & 9\arcsec & 2 \\
                      & 450 & 0.495 & 9\arcsec & 2 \\
                      & 800 & $<$\,0.1 & 16{\farcs}8 & 5 \\
                      & 850 & 0.14 & 15\arcsec & 2 \\
                      & 1100 & $<$\,0.1 & 18{\farcs}5 & 5 \\
04158+2805   & 880 & 0.067 & $\sim$ 5\arcsec & 12 \\   
                   & 1300 & 0.11 & 60\arcsec & 6 \\
04166+2706  & 350 & 6.94 & 9\arcsec & 2 \\
                     & 450 & 4.2 & 40\arcsec & 4 \\
                     & 800 & 0.65 & 16{\farcs}8 & 3 \\
                     & 850 & 1.08 & 40\arcsec & 4 \\
                     & 1100 & 0.30 & 18{\farcs}5 & 3 \\
                     & 1300 & 0.80 & 60\arcsec & 6 \\
04169+2702    & 350 & 7.34 & 9\arcsec & 2 \\
                       & 450 & 6.09 & 40\arcsec & 4 \\
                       & 800 & 0.75 & 16{\farcs}8 & 5 \\
                       & 850 & 1.14 & 40\arcsec & 4 \\  
                       & 1100 & 0.28 & 18{\farcs}5 & 5 \\
                       & 1300 & 0.73 & 60\arcsec & 6 \\
04181+2654 A & 1300 & 0.23 & 60\arcsec & 6 \\
04181+2654 B & \nodata & \nodata & \nodata & \nodata\\
04239+2436 (A,B) & 350 & 1.14 & 9\arcsec & 2 \\
                       & 450 & $<$\,0.66 & 9\arcsec & 2 \\
                       & 800 & 0.33 & 16{\farcs}8 & 5 \\
                       & 850 & 0.21 & 15\arcsec & 2 \\
                       & 1100 & 0.11 & 18{\farcs}5 & 5 \\
                       & 1300 & 0.17 & 60\arcsec & 6 \\
04248+2612 (A,B,C) & 350 & 1.18 & 9\arcsec & 2 \\
                               & 450 & 2.96 & 40\arcsec & 4 \\
                               & 800 & 0.25 & 16{\farcs}8 & 5 \\
                               & 850 & 0.56 & 40\arcsec & 4 \\
                               & 1100 & 0.10 & 18{\farcs}5 & 5 \\
                               & 1300 & 0.45 & 60\arcsec & 6 \\
04264+2433    & 450 & 0.63 & 40\arcsec & 4 \\
                       & 850 & 0.13 & 40\arcsec & 4 \\
                       & 1300 & 0.03 & 60\arcsec & 6 \\
04278+2253 (A,B)  & 450 & $<$\,0.69 & 9\arcsec & 2 \\
                             & 850 & 0.04 & 15\arcsec & 2 \\
04295+2251    & 350 & 1.34 & 9\arcsec & 2 \\
                       & 450 & 2.66 & 40\arcsec & 4 \\
                       & 800 & 0.24 & 16{\farcs}8 & 5 \\
                       & 850 & 0.42 & 40\arcsec & 4 \\
                       & 1100 & 0.09 & 18{\farcs}5 & 5 \\
                       & 1300 & 0.115 & 60\arcsec & 6 \\
04302+2247    & 350 & 2.87 & 9\arcsec & 2 \\
                       & 450 & 2.09 & 40\arcsec & 4 \\
                       & 800 & 0.34 & 16{\farcs}8 & 5 \\
                       & 850 & 0.57 & 40\arcsec & 4 \\
                       & 1100 & 0.15 & 18{\farcs}5 & 5 \\
                       & 1300 & 0.18 & 60\arcsec & 6 \\
04325+2402 (A,B,C) & 160 & 38.0 & 50\arcsec & 1 \\
                               & 450 & 0.61 & 9\arcsec & 2 \\
                               & 800 & 0.30 & 16{\farcs}8 & 5 \\
                               & 850 & 0.19 & 15\arcsec & 2 \\
                               & 1100 & 0.07 & 18{\farcs}5 & 5 \\
                               & 1300 & 0.52 & 60\arcsec & 6 \\
04361+2547 (A,B)  & 450 & 2.35 & 40\arcsec & 4 \\ 
                       & 800 & 0.63 & 16{\farcs}8 & 5 \\
                      & 850 & 0.64 & 40\arcsec & 4 \\ 
                      & 1100 & 0.19 & 18{\farcs}5 & 5 \\
                      & 1300 & 0.44 & 60\arcsec & 6 \\
04365+2535   & 350 & 20.6 & 45\arcsec & 10 \\  
                      & 450 & 12.7 & 45\arcsec & 10 \\
                      & 800 & 1.01 & 16{\farcs}8 & 5 \\
                      & 850 & 1.80 & 45\arcsec & 10 \\
                      & 1100 & 0.44 & 18{\farcs}5 & 5 \\
                      & 1300 & 0.45 & 60\arcsec & 6 \\
04368+2557 (A,B)  & 160  & 69.0 & 50\arcsec & 1 \\
                       & 350  & 12.0 & 45\arcsec & 1 \\
                       & 450 & 2.85 & 9\arcsec & 2 \\
                       & 800 & 1.52 & 16{\farcs}8 & 5 \\
                       & 850 & 0.895 & 15\arcsec & 2 \\
                       & 1100 & 0.48 & 18{\farcs}5 & 5 \\
                       & 1300 & 1.50 & 60\arcsec & 6 \\
04381+2540 (A,B)  & 450 & 2.82 & 40\arcsec & 4 \\
                       & 800 & 0.29 & 16{\farcs}8 & 5 \\
                       & 850 & 0.56 & 40\arcsec & 4 \\  
                       & 1100 & 0.12 & 18{\farcs}5 & 5 \\
                       & 1300 & 0.30 & 60\arcsec & 6 \\
04489+3042    & 1300 & $\gtrsim$ 0.015 & 60\arcsec & 6 \\
CoKu Tau/1 (A,B) & 450 & $<$\,0.52 & 9\arcsec & 2 \\
                          & 850 & 0.035 & 15\arcsec & 2 \\
                          & 1300 & $<$\,0.012 & 11\arcsec & 9 \\
DG Tau B         & 1300 & 0.31 & \nodata & 11 \\
GV Tau (A,B)   & 350 & 1.68 & 9\arcsec & 2 \\
                      & 450 & 1.81 & 9\arcsec & 2 \\
                      & 800 & 0.57 & 16\arcsec & 8 \\
                      & 850 & 0.28 & 15\arcsec & 2 \\
                      & 1100 & 0.18 & 19\arcsec & 8 \\
                      & 1300 & 0.20 & 60\arcsec & 6 \\
HH 30             & 1300 & 0.035 & 60\arcsec & 6 \\
HL Tau            & 350  & 26.4 & 45\arcsec & 10 \\
                      & 450 & 16.8 & 45\arcsec & 10 \\
                      & 800 & 2.58 & 15{\farcs}8  & 7 \\
                      & 850 & 2.97 & 45\arcsec & 10 \\
                      & 1100 & 1.11 & 18{\farcs}4 & 7 \\
                      & 1300 & 1.20 & 60\arcsec & 6 \\
IC 2087 IR      & 450 & 1.365 & 9\arcsec & 2 \\
                      & 850 & 0.50 & 15\arcsec & 2 \\
L1551 IRS5 (A,B)  & 350  & 164 & 45\arcsec & 10 \\   
                      & 450 & 94 & 45\arcsec & 10 \\
                      & 800 & 8.05 & 16{\farcs}8 & 5 \\
                      & 850 & 12.1 & 45\arcsec & 10 \\
                      & 1100 & 2.77 & 18{\farcs}5 & 5 \\
                      & 1300 & 3.40 & 60\arcsec & 6 \\
LkHa 358        & 1300 & 0.032 & 11\arcsec & 9 
\enddata

\tablerefs{
(1) \citet{ladd91};
(2) \citet{andrews05};
(3) \citet{barsony92};
(4) \citet{young03};
(5) \citet{moriarty94};
(6) \citet{motte01};
(7) \citet{adams90};
(8) \citet{chandler98};
(9) \citet{osterloh95};
(10) \citet{chandler00};
(11) \citet{padgett99};
(12) \citet{andrews07}
}
\end{deluxetable}

\clearpage

\begin{deluxetable}{lccccccccccc}
\tabletypesize{\scriptsize}   
\tablecaption{Model Fits for Class I Objects \label{tab_model_fits}}
\tablehead{
\colhead{Name} & \colhead{L} & \colhead{$\rho_1$} & \colhead{$R_c$} &
\colhead{R$_{diskmin}$}& \colhead{R$_{max}$} & \colhead{$\eta$}  &  
\colhead{$\eta_{star}$} & \colhead{CO$_2$ ice} & \colhead{$\theta$} & 
\colhead{i}  \\
 & \colhead{($L_{\odot}$)} & \colhead{(g cm$^{-3}$)} & \colhead{(AU)} &
\colhead{(R$_{star}$)} & \colhead{(AU)} &  & & \colhead{abundance} & &  \\
\colhead{(1)} & \colhead{(2)} & \colhead{(3)} & \colhead{(4)} & 
\colhead{(5)} & \colhead{(6)} & \colhead{(7)} & \colhead{(8)} &
\colhead{(9)} & \colhead{(10)} & \colhead{(11)} 
}
\startdata
04016+2610  & 4.5 & $4.5 \times 10^{-14}$ & 100 & 1 & 6000 & 1.0 & 
0.1 & $1.5\times10^{-4}$ & 5\degr & 40\degr  \\
04108+2803 B  & 0.7 & $1.5 \times 10^{-14}$ & 40 & 3 & 6000 & TSC & 
0.3 & $1.0\times10^{-4}$ & 10\degr & 40\degr  \\
04154+2823  & 0.35 & $7.0 \times 10^{-15}$ & 10 & 1 & 5000 & TSC & 
0.1 & $1.0\times10^{-4}$ & 5\degr & 20\degr  \\
04158+2805  & 0.3 & $2.0 \times 10^{-14}$ & 60 & 3 & 5000 & TSC & 
0.7 & $4.0\times10^{-5}$ & 5\degr & 30\degr  \\
04166+2706  & 0.6 & $4.5 \times 10^{-14}$ & 300 & 5 & 10000 & TSC & 
0.2 & $2.0\times10^{-5}$ & 6\degr & 85\degr  \\
04169+2702  & 1.5 & $3.2 \times 10^{-14}$ & 100 & 1 & 10000 & 1.0 & 
0.8 & $6.0\times10^{-5}$ & 0.1\degr & 75\degr  \\
04181+2654 A  & 0.7 & $2.0 \times 10^{-14}$ & 50 & 5 & 10000 & TSC & 
0.2 & $7.0\times10^{-5}$ & 7\degr & 20\degr  \\   
04239+2436   & 1.5 & 1.3 $ \times 10^{-14}$  & 10 & 1 & 5000 & TSC & 
0.1 & $1.0\times10^{-4}$ & 5\degr & 15\degr  \\
04248+2612   & 0.4 & $4.0 \times 10^{-15}$ & 30 & 5 & 10000 & TSC & 
0.3 & $5.0\times10^{-5}$ &  15\degr & 70\degr  \\
04264+2433   & 0.7 & $5.0 \times 10^{-15}$ & 30 & 5 & 10000 & TSC & 
0.8 & $5.0\times10^{-5}$ &  13\degr & 87\degr  \\
04295+2251    & 0.8 & $8.0 \times 10^{-15}$ & 20 & 5 & 10000 & TSC & 
0.2 & $1.0\times10^{-4}$ &  5\degr & 70\degr  \\
04302+2247    & 1.0 & $3.0 \times 10^{-14}$ & 300 & 1 & 10000 & TSC & 
0.3 & $4.0\times10^{-5}$ &  22\degr & 89\degr  \\
04325+2402  & 0.9 & $3.0 \times 10^{-14} $ & 100 & 3 & 5000 & 1.0 & 
0.8 & $4.0\times10^{-5}$ & 15\degr & 80\degr  \\
04361+2547  & 4.0 & $2.0 \times 10^{-14}$ & 100 & 2 & 10000 & 1.5 &
0.9 & $7.0\times10^{-5}$ & 15\degr & 80\degr  \\
04365+2535  & 2.5 & $4.5 \times 10^{-14}$ & 50 & 1 & 10000 & TSC & 
0.1 & $7.0\times10^{-5}$ & 5\degr & 30\degr  \\
04368+2557    & 1.8 & $4.0 \times 10^{-14}$ & 200 & 1 & 10000 & TSC & 
0.5 & $1.0\times10^{-4}$ & 27\degr & 89\degr  \\
04381+2540  & 1.0 & $3.0 \times 10^{-14}$ & 70 & 5 & 10000 & TSC & 
0.1 & $1.2\times10^{-4}$ & 10\degr & 40\degr  \\
04489+3042  & 0.3 & $1.0 \times 10^{-14}$ & 15 & 1 & 1000 & TSC & 
0.7 & $4.0\times10^{-5}$ & 1\degr & 20\degr  \\
CoKu Tau/1    & 1.1 & $5.0 \times 10^{-15}$ & 40 & 7 & 5000 & TSC & 
0.8 & $5.0\times10^{-5}$ & 5\degr & 80\degr  \\
DG Tau B    & 2.5 & $3.5 \times 10^{-14}$ & 60 & 1 & 10000 & TSC & 
0.1 & $3.0\times10^{-5}$ & 10\degr & 55\degr  \\
HL Tau       & 8.0 & $4.5 \times 10^{-14}$ & 100 & 2 & 10000 & 1.0 & 
0.2 & $4.0\times10^{-5}$ & 1\degr & 15\degr  \\  
L1551 IRS5  & 25.0 & $7.0 \times 10^{-14}$ & 100 & 5 & 10000 & TSC & 
0.3 & $3.0\times10^{-4}$ & 5\degr & 45\degr  
\enddata

\tablecomments{
Column (1) gives the name of the object, column (2) the total luminosity 
($L_{star}+L_{disk}$) of the system, column (3) the reference density
$\rho_1$, column (4) the centrifugal radius, column (5) the inner disk radius
(1 R$_{star}$ = 2 R$_{\odot}$),
column (6) the outer radius of the envelope, column (7) gives the flattening 
parameter $\eta$, column (8) the fraction of the luminosity arising from the star, 
column (9) the CO$_2$ ice abundance, column (10) the semi-opening angle of 
the cavity, and column (11) the inlincation angle. \\
Typical uncertainties for our model parameters are as follows: 
$L$ $\pm$ 10\%, $\rho_1$ $\pm$ 15\%, $R_c$ $\pm$ 20\%, 
R$_{diskmin}$ $\pm$ 10\%, R$_{max}$ $\pm$ 25\%, $\eta$ $\pm$ 0.25
(if not TSC model), $\eta_{star}$ $\pm$ 0.05, CO$_2$ ice abundance $\pm$ 10\%,
 $\theta$ $\pm$ 2\degr, $i$ $\pm$ 5\degr.  
Note that most uncertainties are correlated; e.g., a decrease in $\rho_1$ should
be accompanied by a decrease in $R_c$ and/or increase in $i$ to still result in 
a comparable fit.
}

\end{deluxetable}

\clearpage

\begin{deluxetable}{lccccccc}
\tabletypesize{\scriptsize}   
\tablecaption{Comparison of Luminosities and Inclination Angles \label{tab_lum_incl}}
\tablehead{
\colhead{Name} & \colhead{L$_{bol}$ ($L_{\odot}$)} & \colhead{Ref.} & 
\colhead{L$_{bol}$ ($L_{\odot}$)} & \colhead{L$_{bol}$ ($L_{\odot}$)} &
\colhead{i (deg)} & \colhead{Ref.} & \colhead{i (deg)} \\
 & \colhead{(literature)} & & \colhead{(measured)} & \colhead{(model)} &
 \colhead{(literature)} & & \colhead{(model)}  \\
\colhead{(1)} & \colhead{(2)} & \colhead{(3)} & \colhead{(4)} &
\colhead{(5)} & \colhead{(6)} & \colhead{(7)} & \colhead{(8)} \\
}
\startdata
04016+2610    & 3.70 & 1 & 3.6 & 4.5 & 60 & 7 & 40 \\
04108+2803 B & 0.62 & 1 & 0.5 & 0.7 & \nodata & \nodata & 40 \\ 
04154+2823    &  0.33 & 2 & 0.4 & 0.35 & \nodata & \nodata & 20 \\
04158+2805    & 0.20  & 3 & 0.2 & 0.3 & \nodata & \nodata & 30 \\
04166+2706  & 0.5 (0.40) & 4 (3) & 0.4 & 0.6 & \nodata & \nodata & 85 \\
04169+2702    & 0.8 (1.4) & 1 (3) & 1.2 & 1.5 & 60 & 10 & 75\\
04181+2654 A  & 0.26 (0.70) & 1 (3) & 0.7 & 0.7 & \nodata & \nodata & 20 \\  
04181+2654 B  & 0.25 &  1 & 0.3 & \nodata & \nodata & \nodata & \nodata \\ 
04239+2436 (A,B)   & 1.27  & 1 & 1.5 & 1.5 & \nodata & \nodata & 15 \\
04248+2612 (A,B,C) & 0.36 & 1 & 0.2 & 0.4 & 78 & 8 & 70 \\
04264+2433    & 0.37  & 1 & 0.5 & 0.7 & \nodata & \nodata & 87 \\
04278+2253 (A,B)  & 7.2 & 3 & 5.7 & \nodata & \nodata & \nodata & \nodata \\
04295+2251    & 0.44 (0.64) & 1 (2) & 0.5 & 0.8 & \nodata & \nodata & 70 \\
04302+2247     & 0.34 & 1 & 0.3 & 1.0 & 90 & 8 & 89 \\
04325+2402 (A,B,C) & 0.9 (0.70) & 5 (1) & 1.2 & 0.9 & 60 & 7 & 80 \\
04361+2547 (A,B)  & 3.8 (2.90) & 3 (1) & 2.5 & 4.0 & 60 & 7 & 80 \\
04365+2535    & 2.4 & 2 & 2.8 & 2.5 & 55 & 7 & 30 \\
04368+2557 (A,B)   & 1.6 & 3 & 1.9 & 1.8 & 75 & 7 & 89 \\
04381+2540 (A,B)   & 0.73 & 2 & 0.7 & 1.0 & 55 & 7 & 40 \\
04489+3042    & 0.30 & 1 & 0.3 & 0.3 &\nodata & \nodata & 20 \\
CoKu Tau/1 (A,B) & $>$\,0.29 & 1 & 1.0 & 1.1 & $\sim$\,90 & 9 & 80\\  
DG Tau B         & $>$\,0.02 & 1 & 1.8 & 2.5 & $\sim$\,90 & 9 & 55 \\
GV Tau (A,B)   & 6.98 & 1 & 9.0 & \nodata & 30 & 7 & \nodata \\ 
HH 30             & $>$\,0.1 & 6 & 0.02 & \nodata & $\sim$\,90 & 12 & \nodata \\
HL Tau            & 6.60 (7.1) & 1 (6) & 6.9 & 8.0 & 67 & 11 & 15 \\   
IC 2087 IR      & 3.80 & 1 & 4.9 & \nodata & \nodata & \nodata & \nodata \\
L1551 IRS5 (A,B)    & 28 (21.90) & 5 (1) & 23.2 & 25.0 & 65 (56) & 7 (8) & 45 \\ 
LkHa 358        & 0.59 & 13 & 0.3 & \nodata & \nodata & \nodata & \nodata 
\enddata

\tablecomments{
Column (1) gives the name of the object, columns (2) and (3) the bolometric 
luminosity of the source taken from the literature and its reference, respectively, 
column (4) gives the bolometric luminosity measured by integrating under the SED, 
column (5) the luminosity used in the models, columns (6) and (7) give the 
inclination angle taken from the literature and its reference, respectively, and
column (8) gives the inclination angle used in the models.
}

\tablerefs{
(1) \citet{kenyon95};
(2) \citet{myers87};
(3) \citet{kenyon90}
(4) \citet{young03};
(5) \citet{motte01};
(6) \citet{reipurth93};
(7) \citet{hogerheijde98};
(8) \citet{lucas97};
(9) \citet{padgett99};
(10) \citet{ohashi97b};
(11) \citet{close97};
(12) \citet{burrows96};
(13) \citet{luhman00}
}
\end{deluxetable}

\clearpage

\clearpage
\centerline{\includegraphics[angle=90, scale=0.62]{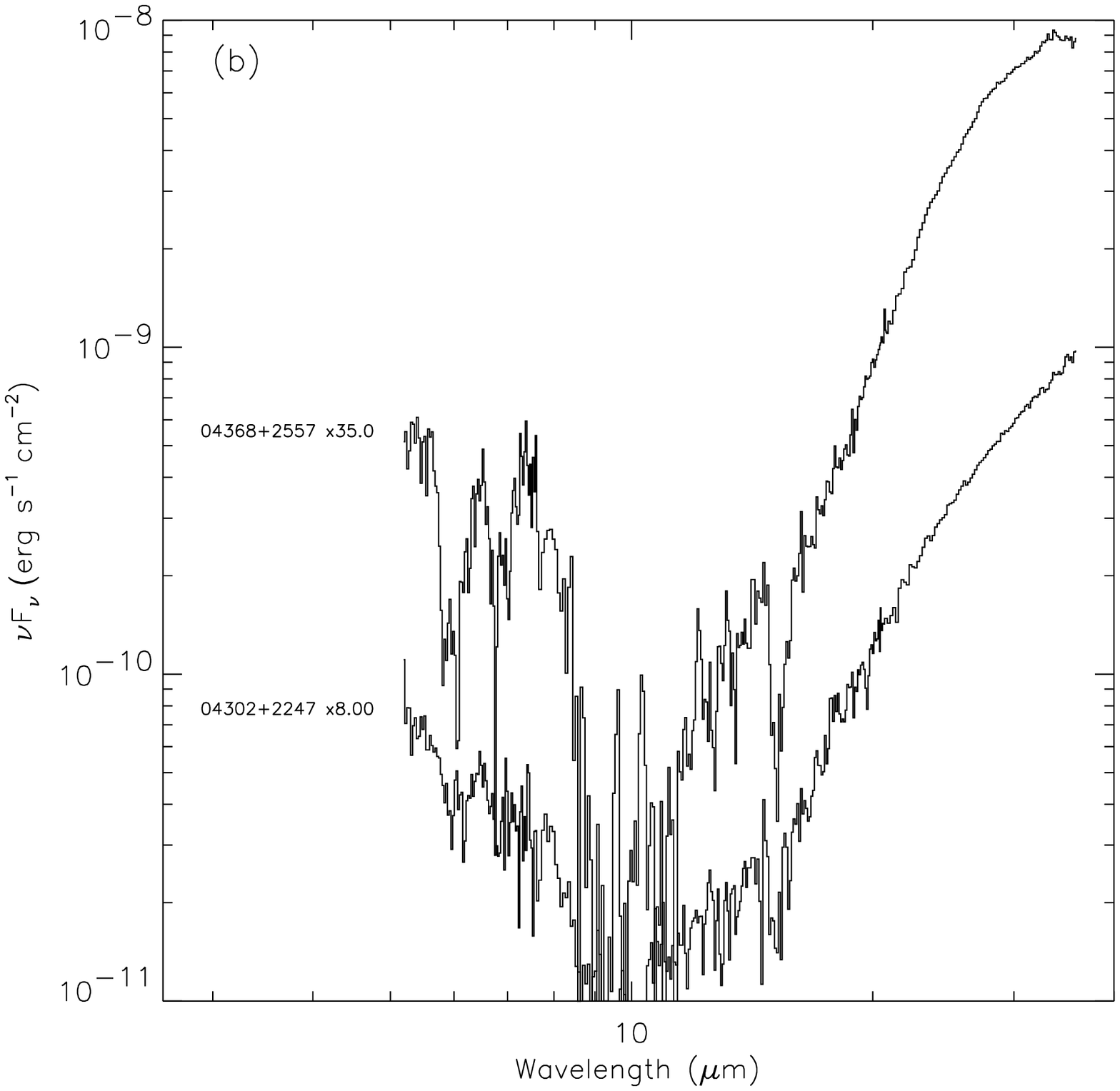}}
\centerline{Fig. 2. --- continued.}
\centerline{\includegraphics[angle=90, scale=0.62]{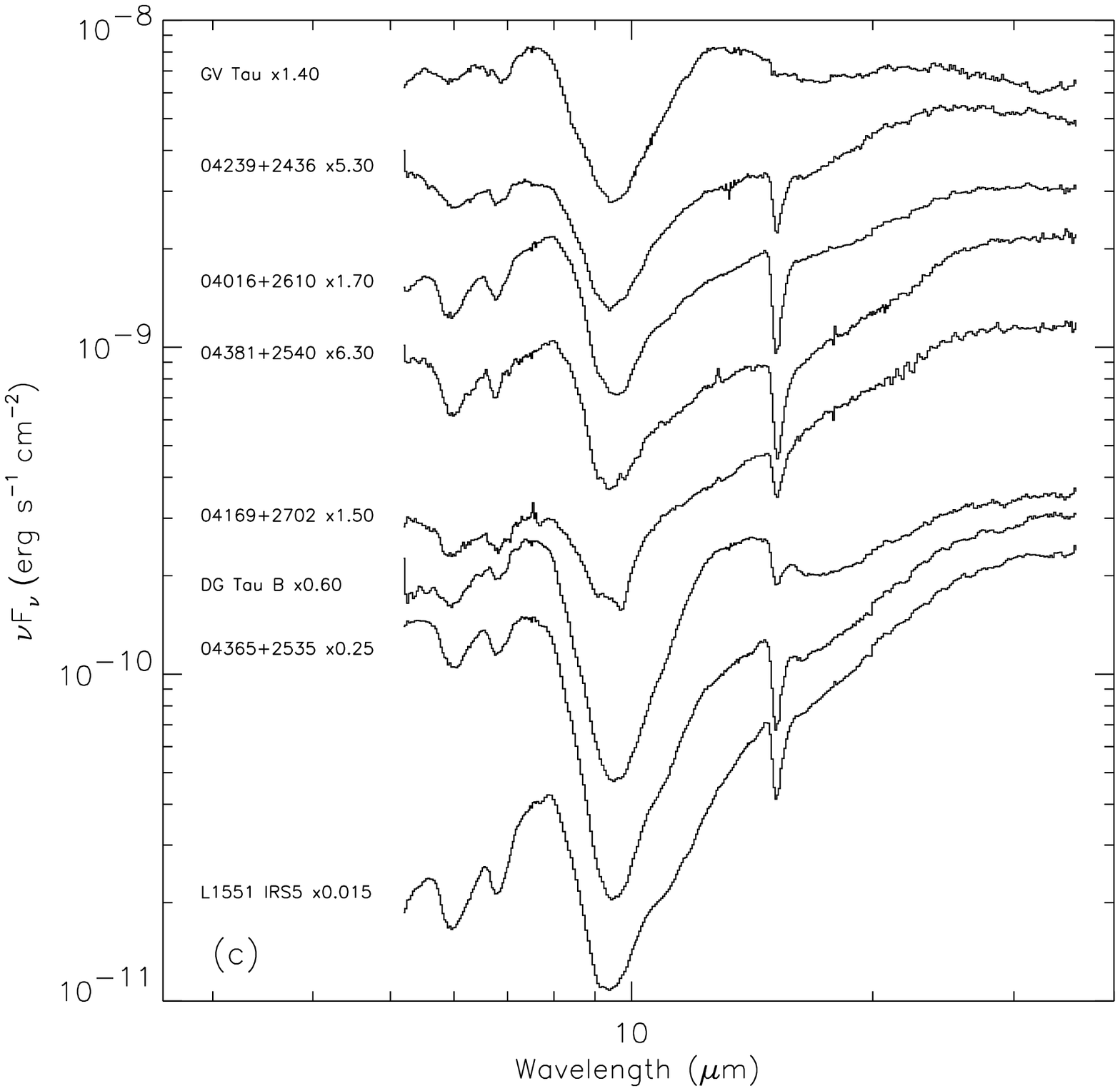}}
\centerline{Fig. 2. --- continued.}
\centerline{\includegraphics[angle=90, scale=0.62]{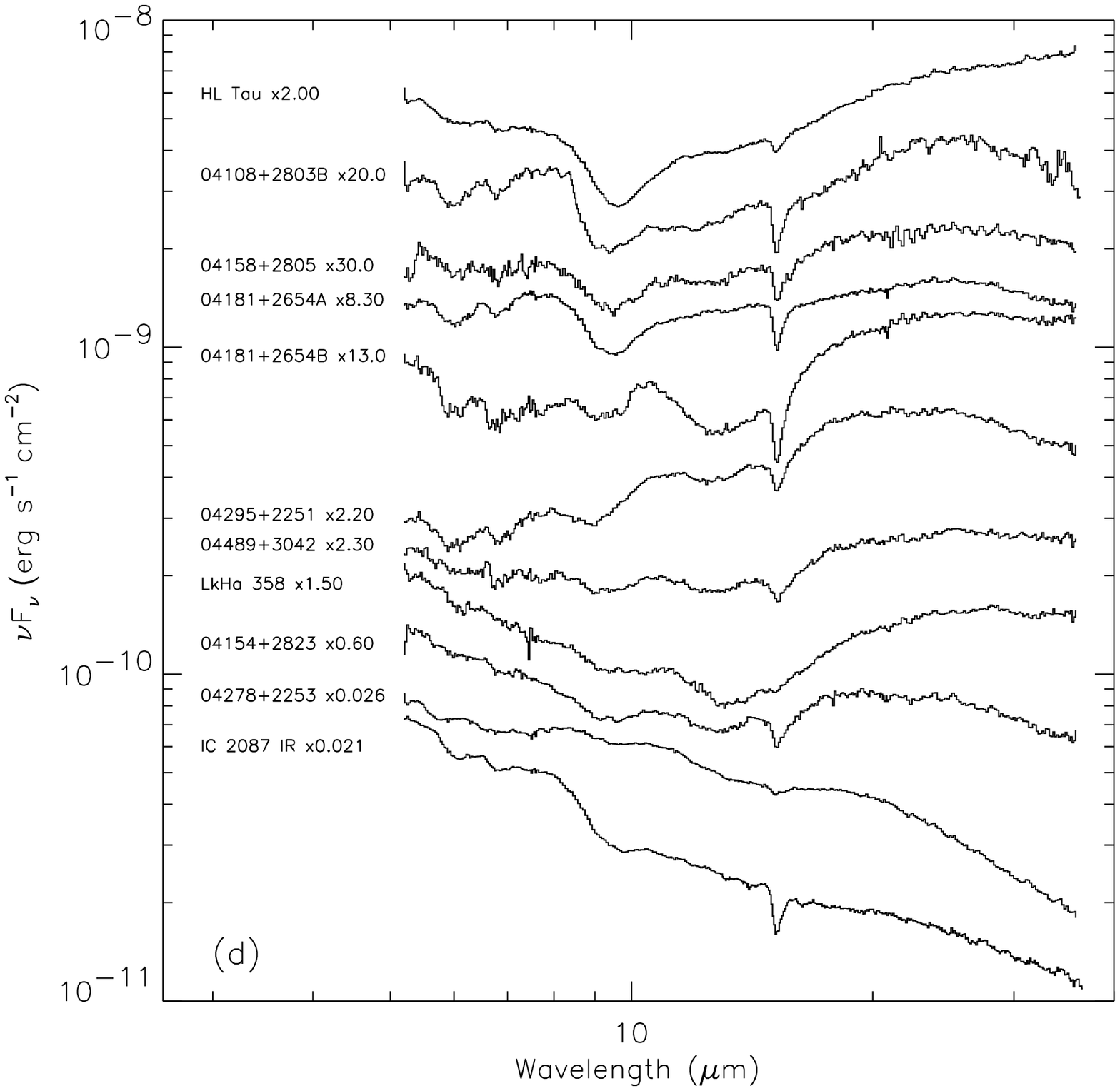}}
\centerline{Fig. 2. --- continued.}
\centerline{\includegraphics[angle=90, scale=0.62]{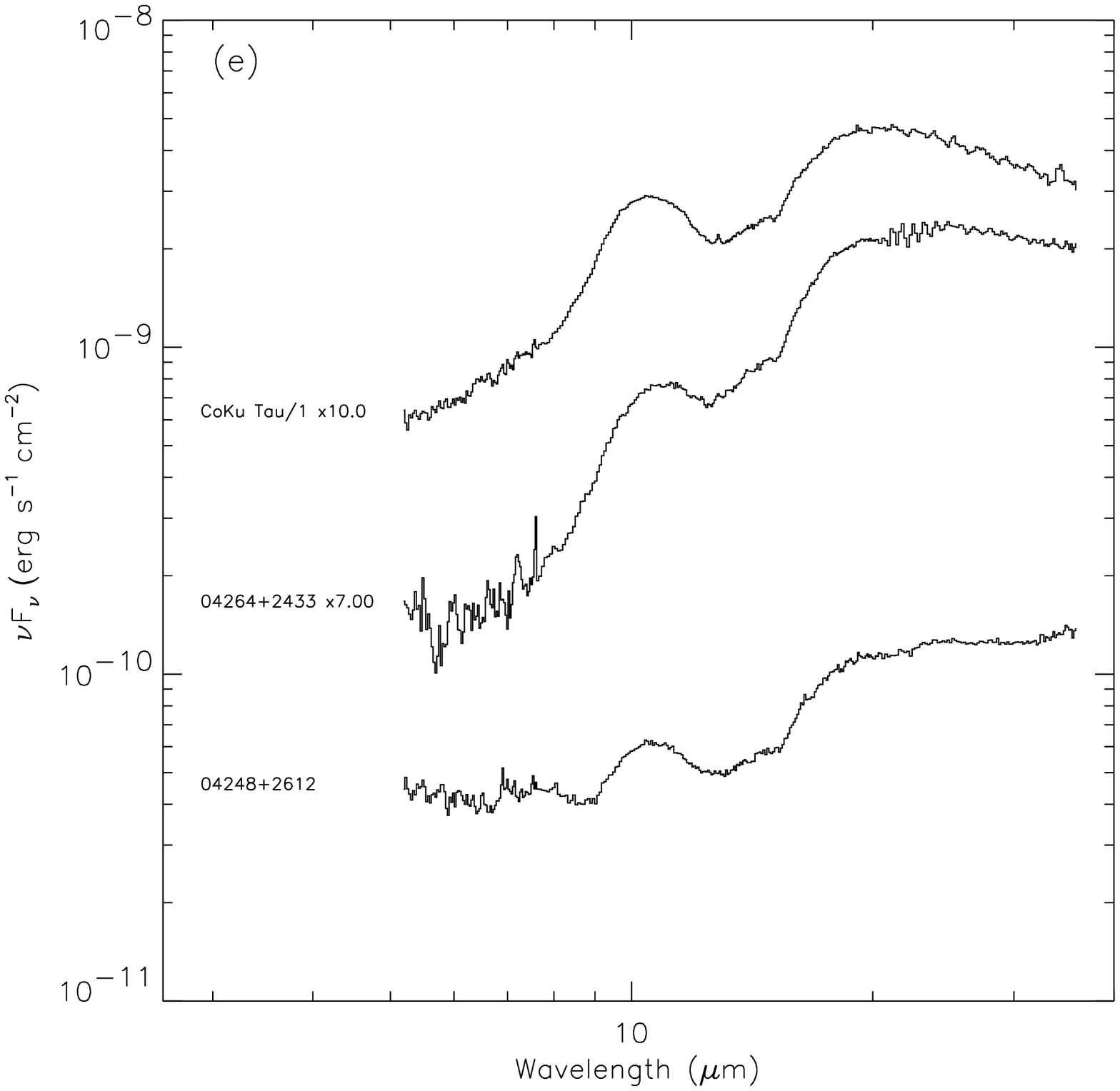}}
\centerline{Fig. 2. --- continued.}

\clearpage
{\plotone{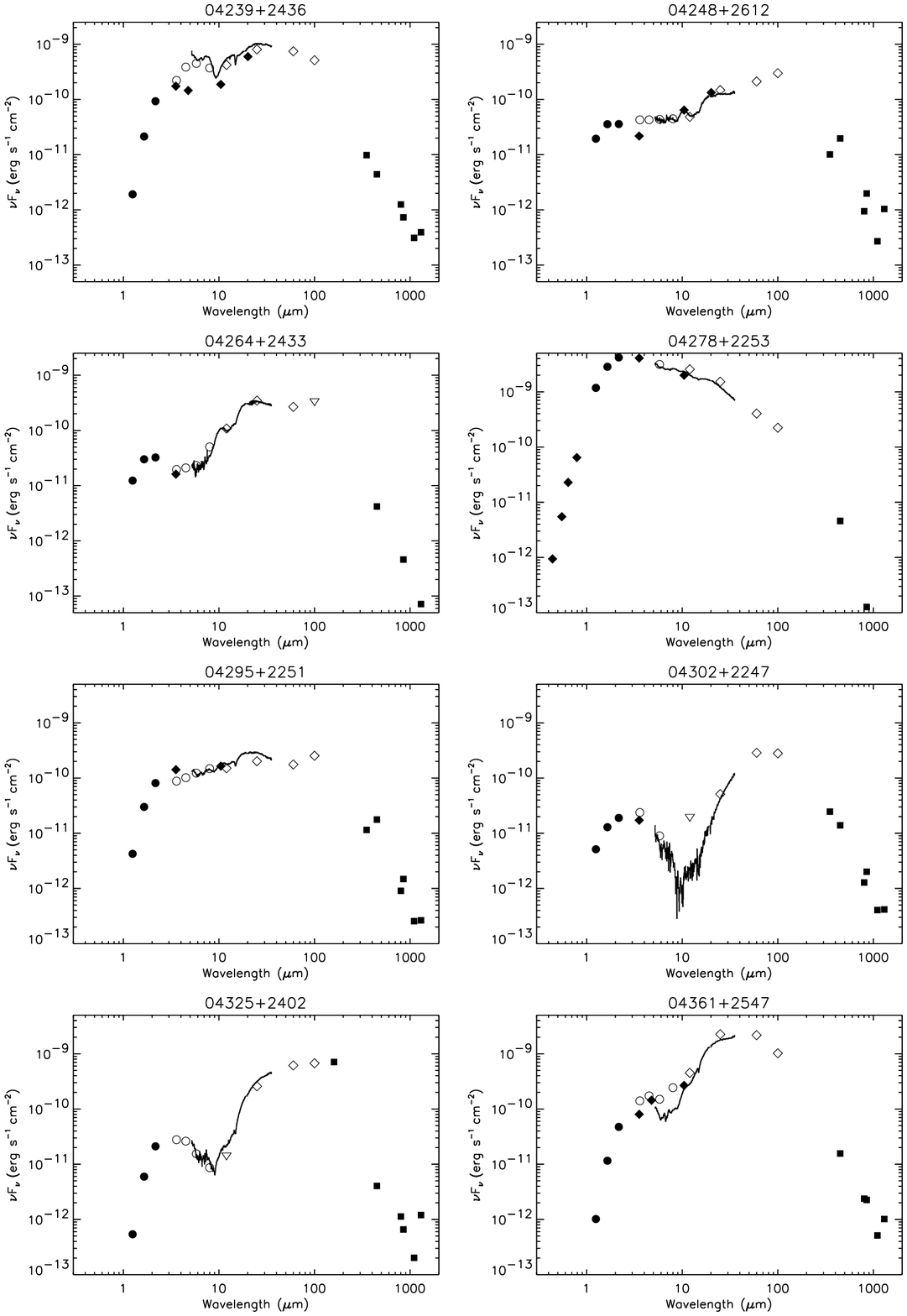}}\\
\centerline{Fig. 3. --- continued.}
\clearpage
{\plotone{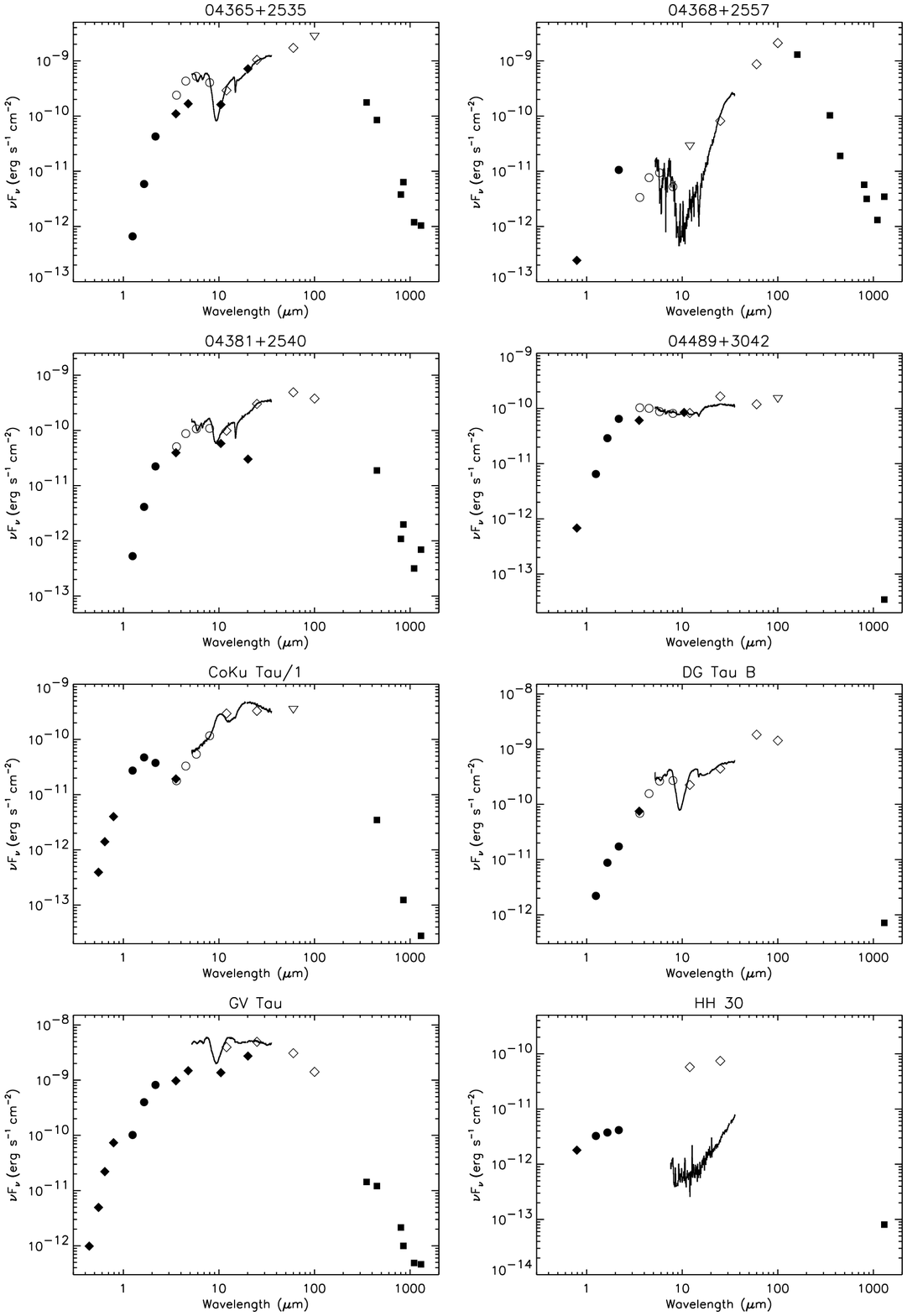}}\\
\centerline{Fig. 3. --- continued.}
\clearpage
{\plotone{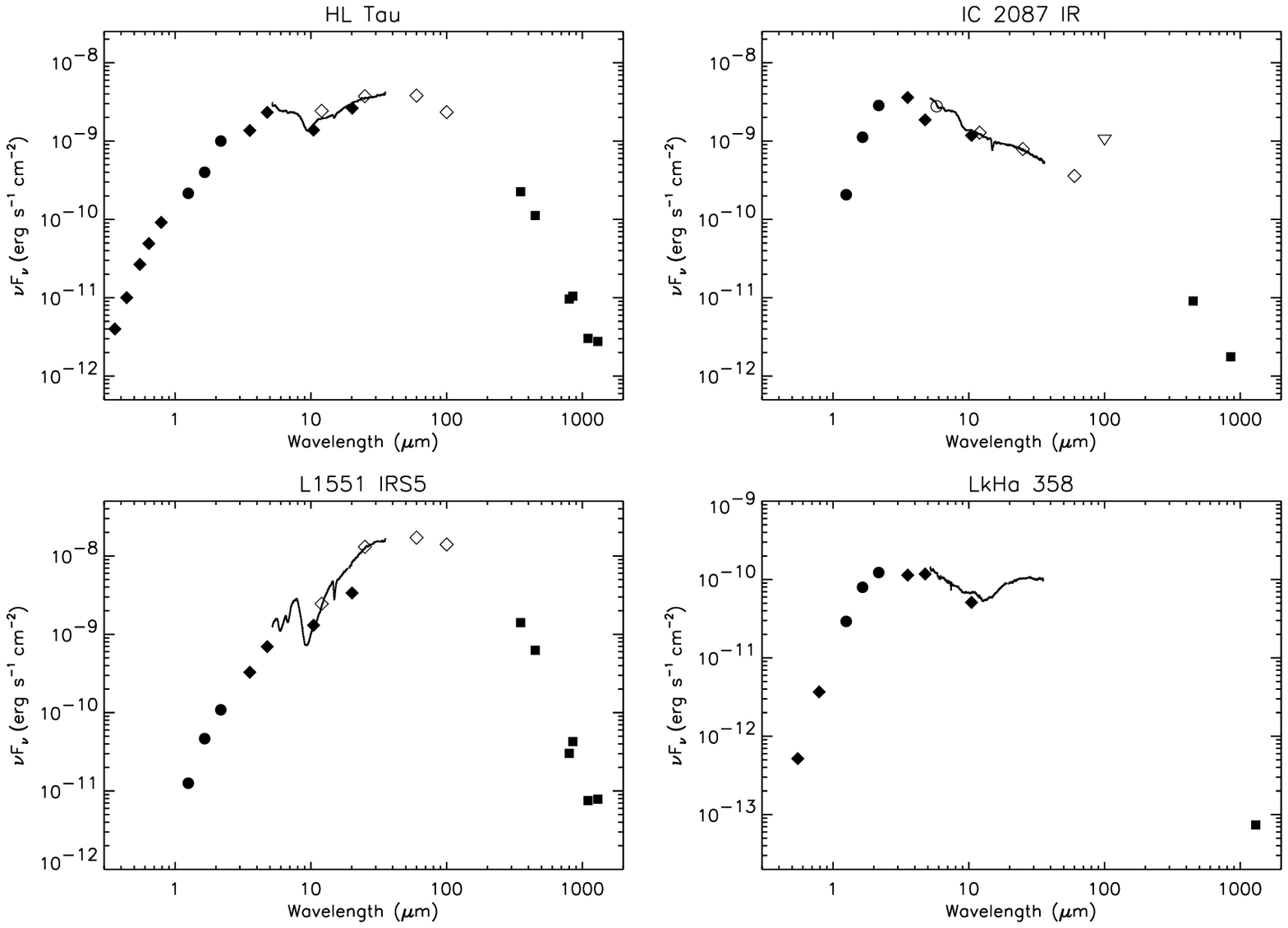}}\\
\centerline{Fig. 3. --- continued.}

\clearpage 

\end{document}